\documentclass[12pt,preprint]{aastex}

\begin{document}
\title{Seyfert's Sextet: A Slowly Dissolving Stephan's Quintet?}

\author {
A. Durbala\altaffilmark{1}, A. del Olmo\altaffilmark{2}, M. S.
Yun\altaffilmark{3}, M. Rosado\altaffilmark{4}, J. W.
Sulentic\altaffilmark{1}, H. Plana\altaffilmark{5}, A.
Iovino\altaffilmark{6}, J. Perea\altaffilmark{2}, L.
Verdes-Montenegro\altaffilmark{2}, I.
Fuentes-Carrera\altaffilmark{7}}

\altaffiltext{1}{Department of Physics and Astronomy, University of
Alabama, Box 870324, Tuscaloosa, AL 35487-0324, USA;
adriana.durbala@ua.edu}

\altaffiltext{2}{Instituto de Astrof\'{\i}sica de Andaluc\'{\i}a,
CSIC, Apdo. 3004, 18080 Granada, Spain}

\altaffiltext{3}{Department of Astronomy, University of
Massachusetts, Amherst, MA 01003, USA}

\altaffiltext{4}{Instituto de Astronom\'{\i}a, Universidad Nacional
Aut\'{o}noma de M\'{e}xico (UNAM), Apdo. Postal 70-264, 04510,
M\'{e}xico, D.F., Mexico}

\altaffiltext{5}{Laboratorio de Astrofisica Teorica e Observacional,
Universidade Estadual de Santa Cruz, Brazil}

\altaffiltext{6}{INAF--Osservatorio Astronomico di Brera, via Brera
28, 20121 Milano, Italy}

\altaffiltext{7}{GEPI, Observatoire de Paris, CNRS, Universit\'{e}
Paris Diderot, Place Jules Janssen 92190, Meudon, France}

\begin{abstract}

We present a multiwavelength study of the highly evolved compact
galaxy group known as Seyfert's Sextet (HCG79: SS). We interpret SS
as a 2-3 Gyr more evolved analog of Stephan's Quintet (HCG92: SQ).
We postulate that SS formed by sequential acquisition of 4-5
primarily late-type field galaxies. Four of the five galaxies show
an early-type morphology which is likely the result of secular
evolution driven by gas stripping. Stellar stripping has produced a
massive/luminous halo and embedded galaxies that are overluminous
for their size. These are interpreted as remnant bulges of the
accreted spirals. H79d could be interpreted as the most recent
intruder being the only galaxy with an intact ISM and uncertain
evidence for tidal perturbation. In addition to stripping activity
we find evidence for past accretion events. H79b (NGC6027) shows a
strong counter-rotating emission line component interpreted as an
accreted dwarf spiral. H79a shows evidence for an infalling
component of gas representing feedback or possible cross fueling by
H79d. The biggest challenge to this scenario involves the low gas
fraction in the group. If SS formed from normal field spirals then
much of the gas is missing. Finally, despite its advanced stage of
evolution, we find no evidence for major mergers and infer that SS
(and SQ) are telling us that such groups coalesce via slow
dissolution.

\end{abstract}


\section{Introduction}

Compact groups (CG) are an intriguing component of the large scale
structure \citep[e.g.][]{hic97}. They can be described as
contradictions in terms involving very high-density galaxy
aggregates (typical projected separation $\sim$ 30-40 kpc) found in
some of the lowest galaxy density environments. In some ways they
mimic a cluster-like environment (galaxy
harassment/stripping/secular evolution). Early models
\citep[e.g.][]{mam87,bar89} found them to be gravitationally
unstable and therefore candidates for rapid collapse into ``fossil''
ellipticals. Unfortunately very few fossil elliptical candidates
have been found \citep{sul94,zab98} except in denser and,
especially, cluster environments. Very few CG show signs of advanced
merging \citep[only $\sim$10\% contain a first ranked
elliptical;][]{sul97}. Dynamical models, taking into account massive
dark matter (DM) haloes \citep{ath97} or formation from a diffuse
configuration \citep{ace02} now suggest that such groups can survive
for even a Hubble time.

One can study CG both statistically and individually. Each approach
has strengths and weaknesses. In either case, but especially the
former, this requires a clear definition of the compact group
phenomenon. The most local samples \citep[hereafter
H1982]{iov02,hic82} suggest that CG can be defined as physically
dense aggregates of 4-8 galaxies with separations on the order of a
few component diameters. Quartets are by far the most common while
systems of 6-8 galaxies are rarely found. It is of course easier to
identify numerical populations of such aggregates than to establish
true physical density, however first attempts based on redshift
concordance and evidence for interaction \citep{men94} suggest that
60/100 of the aggregates in the Hickson Compact Groups (HCG) are
{\it bona fide} compact groups \citep[see e.g. table 2 in][]{sul97}.
It is assumed that triplets are dynamically distinct (inherently
unstable) entities and should not be included in a compact group
sample.

Above considerations suggest that CG will be found in a wide range
of evolutionary states \citep[e.g.][]{rib98,ver01} and that detailed
study of individual CG might be best accomplished within the context
of an assumed evolutionary stage. Relatively unstripped, but
interacting groups should be very young while aggregates of
early-type galaxies embedded in a common stellar halo should be
highly evolved. All or most of the CG that have formed in the local
Universe may still exist as CG. SQ and SS, two of the densest and
best known groups, might be viewed as young/middle aged (few Gyr)
and highly evolved (many Gyr) examples, respectively. We considered
SQ earlier \citep{sul01} and focus on SS in this paper. The best
examples for detailed study involve those compact groups with
abundant multiwavelength data and SS certainly qualifies by that
criterion. As one of the most evolved groups  we have a further
motivation for detailed study.

Seyfert's Sextet (HCG 79, VV 115) is probably the densest (most
compact) galaxy aggregate in the local Universe (the systemic
recession velocity is v$_R$$ \sim$ 4400 km s$^{-1}$ implying a
distance D $\sim$ 60 Mpc assuming H$_0$ = 75 km s$^{-1}$
Mpc$^{-1}$). It is also one of the most isolated systems
\citep[H1982;][]{sul87,iov02}. In this paper we present new
multiwavelength data (optical, IR and radio) including line and
continuum imagery, Fabry-P\'{e}rot and slit spectroscopy. In
addition to these datasets the paper harvests previously published
multiwavelength data for the group in an attempt to give a clear and
coherent picture of the phenomena involved and the current
evolutionary state. Much of our interpretation of SS will be within
the context of our observations and ideas about SQ.

The paper is organized as follows: \S~2 basic observations and
reduction information; \S~3 data analysis and presentation with
minimal interpretation and \S~4: Inferences about the evolutionary
history of SS using new and old observations and interpreted in the
light of what we have learned about SQ. Within \S~2 and \S~3 we
organize into subsections according to wavelength and/or date type
(e.g. IR, radio, optical and imaging, spectroscopy, line,
continuum). Discussion in \S~4 contains frequent citations of the
relevant parts of \S~3 that support specific inferences.

\section{New Observations of SS}

\subsection{Broad-band Optical Images}

SDSS: Optical properties (e.g. magnitudes, morphologies, geometry)
of SS and its neighbors were derived from Sloan Digital Sky Survey
(Data Release 5: SDSS DR5 -- \citealt{ade07}) images\footnote{This
study has made use of SDSS Data Release 5. Funding for the SDSS and
SDSS-II has been provided by the Alfred P. Sloan Foundation, the
Participating Institutions, the National Science Foundation, the
U.S. Department of Energy, the National Aeronautics and Space
Administration, the Japanese Monbukagakusho, the Max Planck Society,
and the Higher Education Funding Council for England. The SDSS Web
Site is http://www.sdss.org/. The SDSS is managed by the
Astrophysical Research Consortium for the Participating
Institutions. The Participating Institutions are the American Museum
of Natural History, Astrophysical Institute Potsdam, University of
Basel, University of Cambridge, Case Western Reserve University,
University of Chicago, Drexel University, Fermilab, the Institute
for Advanced Study, the Japan Participation Group, Johns Hopkins
University, the Joint Institute for Nuclear Astrophysics, the Kavli
Institute for Particle Astrophysics and Cosmology, the Korean
Scientist Group, the Chinese Academy of Sciences (LAMOST), Los
Alamos National Laboratory, the Max-Planck-Institute for Astronomy
(MPIA), the Max-Planck-Institute for Astrophysics (MPA), New Mexico
State University, Ohio State University, University of Pittsburgh,
University of Portsmouth, Princeton University, the United States
Naval Observatory, and the University of Washington. }. We
downloaded g and r band FITS format images of night sky
(``corrected'') frames using the Catalog Archive
Server\footnote{http://www.sdss.org/dr5/access/index.html}. The
corrected frames are  flat-field, bias, cosmic-ray, and pixel-defect
corrected \citep{sto02}. Sky background fitting and subtraction were
performed using IRAF\footnote{Image Reduction and Analysis Facility
(IRAF) is distributed by the National Optical Astronomy
Observatories, which are operated by the Association of Universities
for Research in Astronomy, Inc., under cooperative agreement with
the National Science Foundation - http://iraf.noao.edu/} task
IMSURFIT. IRAF task ELLIPSE was used to derive surface brightness,
deviation of isophotes from pure ellipses (Fourier coefficient b4 -
fourth cosine coefficient of the Fourier expansion), position angle
(PAs) and ellipticity as functions of the semimajor axis. Positive
and negative values of b4 indicates disky and boxy isophotes
respectively. Photometric
calibration\footnote{http://www.sdss.org/dr5/algorithms/fluxcal.html}
of SDSS images was accomplished using the \emph{aa}, \emph{kk} and
\emph{airmass} coefficients (zeropoint, extinction coefficient and
airmass) from the TsField files. The surface brightness zeropoint
was calculated as:
$2.5\times\log(exptime\times0.396^{2})-2.5\times0.4\times(aa+kk\times
airmass)$, using the exposure time \emph{exptime} of 53.907456
seconds and pixel size of 0\farcs396. Our magnitudes are
conventional (``Pogson'') magnitudes.

NOT: We also obtained deep B and R filter CCD images of SS with the
2.5 m Nordic Optical Telescope (NOT) of the Observatorio del Roque
de los Muchachos (La Palma), using the ALFOSC\footnote{The data
presented here have been taken using ALFOSC, which is owned by the
Instituto de Astrofisica de Andalucia (IAA) and operated at the
Nordic Optical Telescope under agreement between IAA and the NBIfAFG
of the Astronomical Observatory of Copenhagen} spectrograph. The
detector was a Loral/Lesser $2048\times2048$ pixels with a spatial
scale of 0\farcs19 which gives a field of view of $6\farcm5
\times6\farcm5$. The broad band images of SS were obtained on a
single night of a three night observing run with ALFOSC. A summary
of the main characteristics of the images can be found in Table 1.
Atmospheric conditions were photometric. Photometric calibration was
accomplished with observations of 10 different Landolt standard
stars (from \citealt{lan83,lan92} list) in the fields SA110 and SA92
obtained on the same night that we observed SS. Seven of these stars
were observed 4 times during the night at 4 different airmasses and
the other three stars were observed 3 different times at 3 different
airmasses. We have in total 37 star flux measures during the night
and could obtain with good reliability the extinction during that
night allowing us to quantify the photometric conditions. The seeing
was also good, as indicated in Table 1. Reduction and calibration of
the images were carried out using standard techniques. Bias
exposures were used to construct average bias images that were
subtracted from the images of SS. Pixel-to-pixel variations were
evaluated with a median normalized sky flat-field in each filter.
Flux calibration was carried out following the method developed by
\citet{you74}. Standard star residuals in the final calibration were
always smaller than 0.03 mag. The errors due to variations in the
sky were smaller than $1\%$.

HST: We downloaded seven HST/WFPC2 archival images\footnote{This
research is based on observations made with the NASA/ESA Hubble
Space Telescope, obtained from the Data Archive at the Space
Telescope Science Institute, which is operated by the Association of
Universities for Research in Astronomy, Inc., under NASA contract
NAS 5-26555. These observations are associated with program \#
8717.} (4-F555W and 3-F814W) of SS all with exposure times of 500 s.
We used IRAF to reject cosmic rays and to combine images. The left
panel of Figure 1 presents a 31$\times$31 pixels high-pass median
filtered image of the average of all seven images. Two contours are
superimposed indicating the maximum extent of the luminous halo on
HST/SDSS (solid) and NOT images (dotted). The inner contour
corresponds to the maximum halo extent on a 51$\times$51 pixels
low-pass filtered HST average. The outer contour shows a circle best
representing the halo extent on the NOT B-R image shown later in
Figure 7. We also downloaded the flux-calibrated archival HST/WFPC2
F555W image of SS with total exposure time of 2000 s. Figure 1 upper
and lower right panels show 15$\times$15 pixels low-pass median
filtered images of H79b and c respectively. Images were extracted
and enlarged by a factor of 5.5.

\subsection{Narrow-band Optical--H$\alpha$ Images}

H$\alpha$ interference filter images (0\farcs533~pixels) were
obtained with the Calar Alto (Centro Hispano Aleman) 2.2m telescope
in June 1997 \citep[see also][]{xus99}. H$\alpha$ interference
filter 667/8 centered at 6667 {\AA} (FWHM = 76 {\AA}) and Johnson R
641/158 centered at 6412 {\AA} (FWHM = 157.5 {\AA}) were used. Three
different exposures were obtained with each filter(600s/exposure and
300s/exposure for 667/8 and Johnson R filters, respectively). The
IRAF package was used for the H$\alpha$ reduction. The images were
bias, flat field, cosmic ray and sky corrected. Individual images
were combined into average line and continuum images after
centering, equalizing the psf and rescaling. H$\alpha$+[NII]
$\lambda\lambda$6548,83 equivalent widths were calculated following
the relation from \citet{igl99}:
$$EW(H_{\alpha}+[NII])=\frac{C_{\alpha}}{C_{cont}}~W_{f}~,$$ where
$C_{\alpha}$ is the number of galaxy counts in the net $H_{\alpha}$
image, $C_{cont}$ is the number of counts from the scaled continuum
image and $W_{f}$ is the FWHM of the filter in {\AA}.

\subsection{Optical Spectroscopy (2D)}

Long-slit spectra were obtained with ALFOSC at the NOT telescope
with the same detector used for NOT images. Table 2 contains a
summary of the long-slit observations with format as follows: Col. 1
spectrum identification with the galaxy or the direction of the
slit, Col. 2 position angle, Col. 3 grism used, Col.4 number of
exposures in each slit, Col. 5 total exposure time for each slit,
Col. 6 resolution in the Spatial scale, Col. 7 spectral resolution
in {\AA}/pix and Col. 8 spectral range.

The directions of the slits are marked in Figure 2. The spectra were
reduced according to the usual techniques, including subtraction of
a mean bias calculated for each night and division by a median
flat-field obtained for each configuration. Wavelength calibration
was performed in a standard way using He-Ne lamps in each position.
The rms of the bidimensional wavelength calibration was found to be
of 0.06\AA\ for grism\#8 and 0.4\AA\ for grism\#4. In all the cases
several exposures were taken in order to increase the S/N ratio and
to remove cosmic rays. Spectra of the spectrophotometric standard
stars BD+26\ 2606, HZ44 and BD+33\ 2642 were observed for flux
calibration. Observations with Grism\#4 were performed in order to
derive some general spectral characteristics of the galaxies. To
obtain the kinematics of the galaxies we used the GR8 spectra with a
spectral coverage from 5800\AA\ to 8300\AA\ and a spectral
resolution of 1.24\AA/pixel. This setup includes the emission lines
H$\alpha$, {[}NII{]}, and {[}SII{]} together with the interstellar
Na absorption.

In order to obtain redshifts and rotation curves we used the
cross-correlation technique developed by \citet{ton79}. For the
emission line spectra we used as templates the brightest spatial
section of galaxies H79b and H79d, together with a synthetic
spectrum built from the rest frame wavelengths of the emission
lines. In the case of the absorption line spectra, we used as
template also a synthetic spectrum and exposures of three radial
velocity standard giant star observed with the same setup.

\subsection{H$\alpha$ Fabry-P\'{e}rot Spectroscopy (3D)}

SPM: Scanning Fabry-P\'{e}rot (FP) H$\alpha$ interferometry was
carried out during the nights of March, 7 and 8, 2000 and March 24,
2001 using the FP interferometer PUMA \citep{ros95} attached to the
f/7.9 Ritchey-Chretien focus of the 2.1m telescope at the
Observatorio Astronomico Nacional (San Pedro Martir - SPM, Baja
California). PUMA involves a scanning FP interferometer, a focal
reducer with an f/3.95 camera, a filter wheel, a calibration system
and a Site 1024$\times$1024 CCD detector. CCD readout was binned
2$\times$2 resulting in a pixel size of 1\farcs16 with a
10$\times$10 arcmin FOV. The FP has an interference order of 330 at
6563{\AA}. The free spectral range of 934 km s$^{-1}$ was scanned in
48 steps with a sampling resolution of 19 km s$^{-1}$.

We have obtained five data cubes at 6658 \AA, each one with a
exposure time of 120 s per channel (implying a total exposure time
of 96 min per data cube), during our  observing runs. We have
selected the best data cube (the data cube with better seeing and
transparency conditions) in order to do our analysis. The best data
cube was from 2001. A log of FP observations is presented in Table
3.

The calibration of the data cubes was carried out by obtaining,
under the same observing conditions, calibration cubes using the
line at $\lambda$6598.95 \AA ~of a diffuse neon lamp. The
calibration cubes were obtained before and after SS observations in
order to check for possible flexures of the instrument.

Reduction of the data cubes was carried out using the CIGALE/ADHOCw
software \citep{bou93,lec93}. The data reduction procedure has been
extensively described in \citet{amr91} and \citet{fue04}. The
accuracy of the zero point for the wavelength calibration is a
fraction of a channel width ($\leq$ 3 km s$^{-1}$) over the whole
field. OH night-sky lines passing through the filter were subtracted
by determining the emission in the field outside SS \citep{lav87}.

From PUMA observations we can extract, in addition to the velocity
cubes, velocity profiles and the 2D velocity fields of the objects
as well as purely monochromatic and continuum images obtained by
integrating the intensities of the radial velocity profiles, pixel
per pixel, up or down to a certain value of the velocity peak,
respectively.

CFHT: The Canada-France-Hawaii Telescope (CFHT) data reduction
procedure has been extensively described in \citet{amr92} and
references therein. A summary of the FP observations can be found in
Table 3.

Wavelength calibration was obtained by scanning the narrow Ne 6599
\AA\ line under the same conditions as the observations. Velocities
measured relative to the systemic velocity are very accurate, with
an error of a fraction of a channel width (${\rm <~3 \,~km~s^{-1}}$)
over the whole field.

Subtraction of bias, flat fielding of the data and cosmic-ray
removal have been performed for each image of the data cube for the
CFHT observations. To minimize seeing variation, each scan image was
smoothed with a gaussian function of full-width at half maximum
equal to the worse-seeing data of the data cube. Transparency and
sky foreground fluctuations have also been corrected using field
star fluxes and galaxy-free windows for the observations.

The signal measured along the scanning sequence was separated into
two parts: (1) an almost constant level produced by the continuum
light in a narrow passband around H$\alpha $ (continuum map), and
(2) a varying part produced by the H$\alpha $ line (H$\alpha$
integrated flux map). The continuum level was taken to be the mean
of the three faintest channels, to avoid channel noise effects.  The
H$\alpha$ integrated flux map was obtained by integrating the
monochromatic profile in each pixel. The velocity sampling was
11~km~s$^{-1}$ at CFHT. Profiles were spatially binned to 3$\times$3
or 5$\times$5 pixels in the outer parts, in order to increase the
signal-to-noise ratio. Strong OH night sky lines passing through the
filters were subtracted by determining the level of emission from
extended regions away from the galaxies \citep{lav87}.

\subsection{Mid- and Far-Infrared}

Mid Infrared (MIR) observations for SS were obtained in the raster
imaging mode (AOT:ISOCAM01) with the mid-infrared camera ISOCAM
\citep{ces96} on board the satellite ISO \citep[Infrared Space
Observatory,][]{kes96}.

We observed our group using the filter LW10, centered at 11.5
\micron, and the configuration with a PFOV = 6\arcsec~for the
32$\times$32 infrared array. We adopted a raster mode 2$\times$2,
with a raster step of 9\arcsec, for a total exposure time of 450
seconds on target (each single exposure had 10 sec integration).

The data were reduced using the software CIA (CAM Interactive
Analysis, v. 3) and included standard steps like dark subtraction,
removal of cosmic rays events (deglitching), correction for the
transient of the detector (stabilization), averaging the frames at
each raster position and flat-field correction and, finally
combining the frames at each position into one mosaic image taking
into account the instrument distortions \citep[projecting; for a
detailed description of these steps see][]{sie96}.

Figure 15 shows the isophotal contours of the 12 $\micron$ image
overlapped on a DPOSS image of the same group.

In ISOCAM images a major source of photometric error are due to
variation of the sky background, correction for the transient
(memory) effect and undersampling of the objects. In our case the
near blending of galaxies H79b and e is a further source of error. A
conservative estimate for the accuracy of our measured fluxes is
$\sim$25\% \citep[see also][]{sie96,biv98}.

\subsection{Radio Line and Continuum}

Observations of SS using the Very Large Array (VLA\footnote{The Very
Large Array is a facility of the National Radio Astronomy
Observatory which is operated by Associated Universities, Inc. under
cooperative agreement with the National Science Foundation.}) were
conducted in its 3 km (C) and 1 km (D) configuration on August 2,
1997 and February 6, 1998, respectively, using all 27 telescopes.
Shortest baselines at the shadowing limit of 25 meter are present,
and structures as large as 15\arcmin\ in size should be visible in
each channel maps. The spectrometer was configured in the dual
polarization 2IF mode to have 64 spectral channels at a frequency
resolution of 48.8 kHz ($\sim$ 10.5 km s$^{-1}$) to cover a total
bandwidth of 3.125 MHz ($\sim 660$ km s$^{-1}$). All of the data are
calibrated following the standard VLA calibration procedure in AIPS
and imaged using IMAGR. Absolute uncertainty in the resulting flux
scaling is about 15\%, and this is the formal uncertainty we quote
for all physical parameters derived from the flux density.

The synthesized beam produced using a robust weight of $R=1$ is
27\farcs0 $\times$ 16\farcs4. The resulting spectral-line maps have
an rms noise level of 0.5 mJy beam$^{-1}$. The 3$\sigma$ HI flux
limit in each map is about 0.016 Jy km s$^{-1}$, corresponding to an
HI column density limit of $4\times 10^{19}$ atoms cm$^{-2}$. At the
adopted distance of the compact group (60 Mpc), the corresponding
mass detection limit is $1.4\times 10^7 M_\odot$.\footnote{The
atomic gas mass has been calculated as $M_{HI} = 2.36 \times
10^5~D^2_{Mpc} \int S_V dV$, where $S_V dV$ is in Jy km s$^{-1}$.}

A line-free continuum image constructed by averaging the 14
line-free channels is shown in Figure 16. This continuum image has
an effective bandwidth of 1.37 MHz centered on an effective
frequency of 1399 MHz. The rms noise level achieved in Figure 16 is
0.20 mJy beam$^{-1}$.

In addition, we also produced a higher resolution 1.4 GHz continuum
image of SS using the data from the archive. The B-configuration
snapshot image with $\sim5\arcsec$ resolution is obtained in the
continuum mode with a total bandwidth of 200 MHz, and the rms noise
in the image is $\sim 0.06$ mJy beam$^{-1}$.

\section{Data Analysis}

\subsection{Broad Band Optical}

\subsubsection{SS Environment}

One of the characteristics of the best known local examples of
compact groups \citep[H1982;][]{iov02} involves the low galaxy
surface density environments in which they are found. Since they are
usually selected with an isolation criterion it is not surprising
that they are at least locally isolated but they are often so
compact that this does not necessarily mean very much in the context
of loose group scales. Attempts have been made to estimate the
galaxy surface density near H1982 groups over a larger area/volume
\citep{sul87,roo89} and a redshift survey has also identified
accordant redshift neighbors around 17 Hickson groups \citep{dec97}.
Many were found to be quite isolated even on Mpc scales where they
are often found on the periphery of nearby superclusters in stark
contrast to their implied space densities that can exceed those of
cluster cores. SS is arguably the most compact group in the local
Universe showing the largest surface density enhancement of any
Hickson group. A little reconnaissance usually allows one to
identify the component of large scale structure with which a compact
group is associated. SS can be assigned to group LGG403
\citep{gar93,roo94} whose identified members span several Mpc.
NED\footnote{This research has made use of the NASA/IPAC
Extragalactic Database (NED) which is operated by the Jet Propulsion
Laboratory, California Institute of Technology, under contract with
the National Aeronautics and Space Administration.} reveals only one
accordant redshift (3400-5400km/s) neighbor  (UGC10127) within
$30\arcmin$ ($\sim0.5$ Mpc) radius. Only four additional accordant
(within $\pm$1000 km s$^{-1}$) galaxies are found within $60\arcmin$
(CGCG137-004, UGC 10117, CGCG137-019, NGC6032). This radius
corresponds to crossing time t$_c$$\sim$ 5 Gyr assuming a random
velocity on the plane of sky of 200 km s$^{-1}$.

Table 4 provides a census of the 17 known accordant redshift
neighbors within 2 Mpc (an Abell radius) of SS and within 1000 km
s$^{-1}$ of the mean group recession velocity. This census is based
upon SDSS images of the field and includes all galaxies brighter
than g magnitude of the faintest member of SS (H79d with
g$\sim$15.5; redshift implied luminosity similar to LMC). Redshifts
were taken from NED because SDSS does not provide spectroscopy for
this field. The g band survey also found twelve galaxies beyond 5500
km s$^{-1}$ (many near 10-11000 km s$^{-1}$ associated with the
Hercules Cluster), one galaxy near 2000 km s$^{-1}$ and three
without redshift measures. An equivalent r-band census to r=15.8
(r-band magnitude of H79d) yields many more galaxies (without
redshifts) but visual inspection suggest that all/most of these are
likely to be much more distant. We adopted g-band census because: 1)
it shows higher redshift completeness, 2) we wanted to minimize
background contamination and 3) we regard late-type field galaxies
as the principal source of infalling intruders as suggested by
NGC7318b in SQ and H79d in SS. The census yields a surface density
of 1.3 galaxies Mpc$^{-2}$, or 1.6 galaxies Mpc$^{-2}$ if all three
bright neighbors with unmeasured redshift show accordant measures.
Some recent surface density estimates range from $\sim$ 1 Mpc$^{-2}$
for the sparsest environments up to $\sim$ 6 Mpc$^{-2}$ in the
poorer cluster cores \citep{got03a}. SS shows a surface density of
$\sim$7-8000 galaxies Mpc$^{-2}$ and a surface density enhancement
of $\sim$5-6000 in agreement with \citet{sul87}. SS shows the
highest density enhancement of any compact group in the local
Universe and one of the lowest galaxy surface densities in its
Mpc-scale environment.

\subsubsection{SS Component/Neighbor Morphologies and Geometries}

SS involves four or five accordant redshift (H79abcdf) galaxies plus
one discordant redshift galaxy (H79e). H79f on the NE edge of the
group can be described either as a tidal tail or as the remnants of
a tidally stripped galaxy. The mean separation between the galaxies
is about 7.2 kpc and the mean velocity dispersion $\sigma_V$= 121 km
s$^{-1}$ \citep[hereafter D2005]{dar05}. Figure 1 identifies
specific components of the group following designations in
\citet{hic82,hic93a} with the addition of H79f.

Table 4 gives estimated Hubble types for assumed members of SS as
well as all neighbors within a radius of $\sim$2 Mpc. Positions and
recession velocities are given for all galaxies along with projected
separations in arcmin and kpc for the neighbors. SS members show
unusual and sometimes ambiguous structure. Figures 3 and 4 present
the most interesting standard geometric profiles for the brightest
two member galaxies (H79ab). We derived these measures from r-band
SDSS images. Results for g-band were very similar. Several sets of
geometric profiles and derived parameters have already been given
for galaxies in SS: \citet[hereafter H1989a]{hic89a};
\citet[hereafter R1991]{rub91}; \citet[hereafter B1993]{bet93};
\citet[hereafter N2000a]{nis00}. A serious problem with deriving
standard parameters for SS components involves the fact that they
are embedded in a luminous halo. For example the r-band surface
brightness of the last concentric isophotes in H79abc are 21.7, 21.0
and 21.2 mag arcsec$^{-2}$ respectively.

The three brightest members (H79abc) all show early-type
morphologies in the simple sense that they are axially symmetric
with no obvious spiral arms or visible emission regions. H79a
(NGC6027a) shows a very smooth light distribution and an ellipticity
consistent with Hubble type E3-4 although it has been classified E0,
S0 and Sa in the past. It is bisected by a prominent dust lane
adding to the confusion in assigning a type and in determining the
position of the nucleus. The relaxed appearance (i.e. flatness) of
the dust lane forces us to consider the possibility that H79a might
be an edge-on S0 with a weak stellar disk component. The alternative
interpretations are that the dust lane: 1) hints that H79a was
originally a spiral galaxy or 2) is the relaxed signature of a past
accretion event. ``Relaxed'' is perhaps an overstatement because the
dust lane shows a pronounced bend (see Figure 1) in the direction
towards the center of the group. Figure 3 shows azimuthally averaged
surface brightness and b4 profiles for this galaxy derived with
fixed position angle (68\arcdeg) and ellipticity ($\epsilon$=0.36).
In the figures a$_{max}$ indicates the semimajor axis of the last
concentric isophote. Isophotal boxiness is observed in the center
and is probably induced by the dust lane (see also B1993). No
surface photometric study of this galaxy has revealed a stellar disk
component that might be a counterpart to the dust disk.

The unusual shape and internal structure of H79b revealed by HST
images (Figure 1) do not lend themselves to a standard geometric
analysis. Figure 4 shows evidence for two disk components (centered
at $\approx$ 2-3\arcsec and 5-8\arcsec) in both surface brightness
and b4 parameter. Seeing in the SDSS r-band image is $\sim$2.5
pixels corresponding to $\sim$1\arcsec. The above mentioned WFPC2
images show signs of highly inclined internal (spiral?) structure
(see Figure 1 upper right where a tightly wrapped spiral dust lane
is seen). Earlier WFPC1 images revealed the presence of a twisted
dust lane oriented approximately diagonally across the galaxy
\citep{sra94}. The axial ratios of H79b and c suggest that both are
highly inclined to our line of sight. The surface brightness and b4
profiles for H79c also indicate considerable complexity. The WFPC2
images suggest that the tidal tail extending to the NW originates
near this galaxy. It crosses the disk at an angle of about
20\arcdeg\ forming an apparent ``X'' structure (see Figure 1 lower
right).

H79d is the only member that shows a late-type (spiral) morphology
with numerous condensations resembling HII regions detected on many
images but most dramatically with WFPC2. It is highly inclined with
an axial ratio of 0.18 suggesting an inclination of 80\arcdeg. H79d
appears to be projected on H79a because a few dusty condensations
can be seen in silhouette against the halo light of that galaxy on
the WFPC2 images \citep{pal02}. Some published studies suggest a
link between galaxies H79ad \citep[R1991;][]{pla02}.

H79e shows a much higher redshift and is assumed to be a near
face-on high luminosity ScI spiral at approximately 4.5$\times$
greater distance. H79f lies to the NE of H79b and is either a tidal
filament or another early-type member. It was not included as a
member by \citet{hic82,hic93a} yet the RC3 catalog \citep[herafter
RC3]{dev91} lists it as a galaxy with designation NGC6027e. The
surface brightness profile for this galaxy follows an exponential
law. The position angle and ellipticity profiles in the outer part
of H79f are constant ($\sim$50\arcdeg\ and $\sim$0.57,
respectively). The surface brightness of the last concentric
isophote in H79f is 21.6 mag arcsec$^{-2}$ in r-band and 22.3 mag
arcsec$^{-2}$ in g-band. The central concentration and elliptical
isophotes argue that it is a member galaxy however in this
interpretation it has likely been heavily stripped given the
weakness of the central concentration for an apparently early-type
morphology (lacking any signs of gas or dust).

Perhaps more unusual than their internal peculiarities are the small
measured sizes of all members of SS. Table 5 presents different
estimates for the sizes of the galaxies in SS. We compare our SDSS
based measurements with four previous studies (H1989a, B1993, RC3
and N2000a). We find a large scatter among the measures consistent
with the difficult task of extracting discrete diameters from
galaxies embedded in a luminous halo. Measured diameters exceed 10
kpc only if one attempts to apply a standard model of the galaxy
that extends beyond the last concentric isophote. We used SDSS g and
r band images to determine the last concentric major/minor axis
isophotes in each galaxy. We think that attempts to measure standard
diameters (e.g. 25 mag arcsec$^{-2}$) have little or no meaning in
the context of individual galaxy properties since this level is
three magnitudes below the surface brightness level of the last
concentric isophote. The luminous halo of SS likely contains a
significant fraction of the stellar mass of galaxies H79abc and
especially H79f. Attempts at model-based (e.g. exponential and
r$^{1/4}$) galaxy subtraction of H79abcdf suggest that the halo is
not a product of isophotal overlap \citep[][N2000a]{sul83}. Galaxy
H79d is the only component of SS that shows little sign of tidal
stripping and appears to be a recently arrived low luminosity late
type spiral.

The shape and extent of the luminous halo in SS is indicated by two
contours in Figure 1. The inner very irregular contour shows the
extent of the halo on HST and SDSS images. In the latter case this
correspond to 24.7 and 24.2 mag arcsec$^{-2}$ in g and r-bands
respectively. The outer more symmetric contour shows the the circle
best fitting the halo extent on more sensitive NOT B-band images
(see also Figure 7) and corresponds to 27 mag arcsec$^{-2}$. The
radius of the outer circle is 1.3 arcmin ($\sim$ 23 kpc). The SDSS
sky levels at that radius are 26.4 and 26.2 arcsec$^{-2}$ for g and
r-bands respectively.

Seventeen accordant redshift neighbors brighter than g=15.5 lie
within $\pm1000$ km s$^{-1}$ and r=2 Mpc of SS. Table 4 lists the
assigned Hubble types for the galaxies with 75-80\% of SS members
(3/4 or 4/5 depending on whether H79f is considered) and 30-40\% of
neighbors showing early-type (E/S0 or E/S0/Sa) morphologies. Recent
morphological reevaluation of a sample of about 1000 very isolated
field galaxies \citep{sul06} leads us to predict a $\sim$14\%
early-type fraction in SS and in its neighborhood. The SS
environment shows a 2-3$\times$ higher early-type fraction so this
prediction is likely too low. A recent quantification of the
morphology-density relation \citep{got03a} finds 16\% E plus 30\% S0
for environments with surface density similar to the SS neighborhood
suggesting that the observed early-type galaxies are overrepresented
in SS if we accept H79abc, and possibly f, as {\it bona fide} E/S0
galaxies.

Table 6 provides g-band major axis diameters at the 25 mag
arcsec$^{-2}$ isophote derived from the SDSS images for thirteen
accordant redshift neighbors. Comparison of these diameters with
those for SS components leads to the conclusion that the galaxies in
SS are 3-4 times smaller. A similar conclusion was found by
\citet{wil91} where they noted that the individual galaxies in SS
are on average one third the size of a typical normal galaxy. We
estimated the diameters of neighboring S0-Sb galaxies truncated to
the same surface brightness level as the last concentric isophotes
in H79abc. We find neighbor diameters in the range 4-7 kpc at
$\mu_g$=21.5 mag arcsec$^{-2}$ and 4-12 kpc at $\mu_g$=21.8 mag
arcsec$^{-2}$ compared to 5-8 kpc for H79abc at their last
concentric isophotes. The SS members are therefore more similar in
size to the bulge components of neighboring spiral galaxies. The
galaxies in SS are either: a) intrinsically dwarf or b) normal
galaxies that have undergone severe tidal stripping. The angular
size of the entire group is $\sim$1.3 arcmin (H1982) $\sim$23 kpc
which is comparable to the size of some of the neighboring disk
galaxies. The massive common halo in SS implies significant
stripping and disfavors the hypothesis that most SS members are
intrinsically dwarf galaxies.

\subsubsection{SS Component and Neighbor Luminosities}

Table 7 presents apparent magnitudes for members of SS in different
filters. We compare our g and r-band SDSS based measurements to
H1989a, R1991, RC3 and N2000a. Our apparent magnitudes measure the
light within the last concentric isophote corrected for galactic and
internal extinction as well as K-corrected. Extinction corrections
were performed using The York Extinction Solver (YES)\footnote{The
York Extinction Solver (YES) at http://cadcwww.hia.nrc.ca/yes}
\citep{mcc04}. YES allows a user to determine the optical depth at
1$\mu$m from an estimate of the color excess, and then to determine
the extinction of the target from the optical depth. Color excess is
estimated using \citet{sch98} extinction maps employing a
\citet{fit99} reddening law.

Table 8 presents our g and r-band absolute magnitudes and
luminosities for SS galaxies. As noted earlier the galaxies appear
to be remarkably luminous for their small sizes which are more
similar to the size of the bulges of late-type neighbors. Along with
diameters, Table 6 presents absolute magnitudes, total luminosities
and bulge luminosities for the accordant redshift neighboring
galaxies. Overlapping structure in the close spiral pair NGC6052
prevents precise determination of their sizes and magnitudes. Other
late-type neighbors (NGC6028 and CGCG108-085) lack any SDSS imaging
data. Apparent magnitudes are corrected model magnitudes (modelMag-
better of exponential/deVaucouleurs fit) given by the SDSS DR5
pipeline. The total luminosities in each filter were calculated
using the formulae:
$$L_{gT}=10^{0.4(M_{g\sun}-g_{TC}+5logD-5)} (L_{g\sun})$$
$$L_{rT}=10^{0.4(M_{r\sun}-r_{TC}+5logD-5)} (L_{r\sun})$$
where $M_{g\sun}=5.12$ and $M_{r\sun}=4.68$ are adopted for g and r
absolute magnitudes of the Sun respectively. To express the
luminosities in terms of bolometric solar luminosity we used
$M_{\sun}=4.76$ which implies ratios $L_{\sun}/L_{g\sun}=1.39$ and
$L_{\sun}/L_{r\sun}=0.93$. The assumed distance is D= 60 Mpc. Bulge
luminosities are estimated considering a bulge to disk ratio
$(B/D)\sim 1$ for Sa, $\sim0.65$ for Sab, $\sim0.4$ for Sb, $\sim
0.1$ for Sc and $\sim 0.01$ for Sd morphological type
\citep{ken85,kop90} and the equation: $B/T=1/(1+D/B)$, where T is
total luminosity. $B/T$ is assumed to be 0.68 for SO morphological
type \citep{ken85}. Inspection of Tables 6 and 8 shows that the
total luminosities of galaxies H79abc are comparable to the bulge
luminosities of the Sb-Sc and SO neighbors.

Figure 5 plots apparent diameter versus g-band magnitude for SS
galaxies (open circles) and thirteen accordant redshift neighbors
(filled squares): 1) within $\pm1000$ km s$^{-1}$, 2) within 2 Mpc
radius and 3) brighter than g=15.5. Measured diameters for the
neighbors correspond to the 25 mag arcsec$^{-2}$ g-band isophote. We
used the corrected modelMag values from SDSS. We also plot the
bulges for SO and Sb-Sc neighbors with sizes and magnitudes
corresponding to the 21.8 mag arcsec$^{-2}$ g-band isophote.
Diameters and magnitudes for SS components correspond to the last
concentric g-band isophote. The plot shows the expected correlation
between angular diameter and apparent magnitude for the accordant
neighbors. It is reasonably well fit by the indicated linear
regression line. The brightest neighbor NGC6060 is similar in
luminosity to M31 ($M_{g}\sim-21.5$) while H79d is similar to the
LMC ($M_{g}\sim-18.2$). H79f, considered as a galaxy, would be 0.4
magnitudes fainter than the SMC ($M_{g}\sim-16.9$). Apparent g-band
magnitudes for M31, LMC and SMC were estimated from V magnitude and
B-V color using the transformation equation: $g=V+0.6(B-V)-0.12$
\citep{jes05}. Conversion to absolute magnitudes assumed distances
of $\sim$ 750 kpc for M31 \citep{rib05}, $\sim$ 50 kpc for LMC and
$\sim$ 60 kpc for SMC \citep{kel06}. SS members are smaller than all
of the galaxies in our neighborhood sample.

Figure 5 shows that H79abc are similar in apparent brightness to the
faintest neighbors, but they show much higher mean surface
brightness. H79abc show g-band mean surface brightness within the
last concentric isophote $\mu_g$=19.9-21.1 mag arcsec$^{-2}$
compared to $\mu_g$=22.5-23.5 mag arcsec$^{-2}$ calculated within
the 25 mag arcsec$^{-2}$ isophote for neighbors. The two neighbors
with most similar size and apparent brightness show
$\mu_g$=22.5-22.9 mag arcsec$^{-2}$. Galaxies in SS show much higher
mean surface brightness and are embedded in a luminous halo. Their
sizes and surface brightness are more similar to the bulge
components of neighboring spiral and SO neighbors. The mean surface
brightness of one of the brightest neighboring Sb galaxies (NGC6008)
would increase from 22.9 to 20.8 mag arcsec$^{-2}$, while the
diameter would decrease from 1.5 to 0.2 arcmin if we consider only
the central bulge. The overall mean surface brightness of SS is also
high. It is by far the highest observed (20.5 mag arcsec$^{-2}$) for
any H1982 group. Only three other groups (HCG8, 40 and 95) shows
values within one magnitude (21.3-21.4 mag arcsec$^{-2}$). Allowing
for a different zero point in the surface brightness scale suggests
that none of the 121 southern compact groups \citep{iov02}
approaches the surface brightness of SS.

Figure 6 is a composite of the B and R images showing the extension
of the diffuse halo. According to D2005 the halo contributes
$\sim46\%$ of the total light in B band \citep[see also][]{sul83}.
We get a similar result with g and r band SDSS images. The halo is a
fraction of $46\pm10\%$ and $45\pm10\%$ of the total light in g and
r band, respectively, if we don't include galaxy H79f as part of the
halo. This corresponds to a corrected apparent magnitude of
$g=13.4\pm0.1$ and $r=12.9\pm0.1$. If we included H79f, the halo
contribution would raise by $2\%$ in both g and r band. We estimated
the diffuse light in SQ using the magnitudes of the halo and
component galaxies given in \citet{mol98}. The diffuse light
component represents $\sim13\%$ of the total light, suggesting that
SQ is a less evolved compact group, younger than SS. The early-type
fraction in SQ is about $40-50\%$.

A diffuse light image was derived for SS using the wavelet technique
(Figure 6 in D2005). The symmetry and smoothness of the halo light
distribution in Figure 6 suggests a reasonable degree of relaxation
compared to e.g. SQ. There are two peaks in the halo light
distribution of SS. One is coincident with the NW tidal tail marked
in that figure. The other, which does not coincide with a galaxy or
tail, lies much closer to the center of the diffuse light
distribution. We attempted to model the radial profile of the CFHT
B-band halo image (D2005) and find that an exponential yields the
best fit. We get disk scale lengths R$_d$= 23 and 28 kpc, centered,
respectively, on an approximate outer halo contour (circle in Figure
1 left) and on the SE condensation. We also attempted de Vaucouleurs
and S\'{e}rsic fits on the surface brightness profile of the halo.
However, they could be applied only for the outer part and even so
we obtained unphysical results, namely extreme values for the
effective radii. While the overall halo is smooth and roughly
circular, consistent with relaxation, the condensation connected
with the tidal tail indicates that the halo is still growing.

\subsubsection{Group Morphology and Colors}

We supplement published color information on SS with our SDSS g-r
and our NOT B-R measures. Table 9 lists new and old color measures
for group members as well as the halo while Figure 7 shows our best
attempt at a 2D B-R color image. Tabular results are consistent with
Figure 7. The central region of H79a shows the reddest color
(B-R$\sim$1.6-1.8) due to the presence of the strong dust lane. The
inner parts of galaxies H79a/b show B-R$\sim$1.5 which is typical of
an early-type or bulge stellar population. The outer parts of
galaxies H79a/b as well as H79f show B-R$\sim$1.3-1.4. The color of
galaxy H79f is in good agreement with \citet{nis02}. Galaxy H79c,
much of the diffuse halo as the NW tidal tail show B-R$\sim$1.2-1.3
while the outer halo is distinctly bluer at B-R$\sim$0.8-1.1. All
results indicate that galaxy H79c is bluer than galaxies H79a/b with
complex color structure. It is bluest on the side towards the center
of the group. The bluest colors are seen in galaxy H79d with
B-R$\sim$0.7-0.9 which is typical of a late type spiral. We see a
slight color difference $\Delta(B-R)$=0.3 along the major axis with
reddest color in the direction of galaxy H79a.

Color measures for the halo component are the most complex. We find
a mean halo color that is bluer than galaxies H79a/b and more
similar to galaxies H79c/f. D2005 also reported a much bluer color
for the halo B-R=0.9 compared to B-R=1.5 for the galaxies. We
derived halo properties after subtracting galaxies at their last
concentric isophote which might leave a significant red galaxy
component in our halo measures. The wavelet technique employed for
the D2005 estimates more effectively removes the galaxies leaving a
much flatter halo. The resultant galaxy color will be dominated by
red galaxies H79a/b. The wavelet derived halo color is similar to
galaxy H79d. Figure 7 suggests that the halo shows a significant
color gradient and that the outer parts are as blue as those of a
late-type spiral. If the mean color of H79d is B-R=0.8 then we can
adopt a mean halo color of B-R=1.0$\pm$0.2 as a best estimate. B-R
colors were only corrected for galactic extinction using
\citet{bur82}. Other corrections are very dependent on the
morphological type, which is uncertain for some members of the
group. An internal extinction correction for galaxy H79d
(morphological type Sd) would make it 0.2 B-R magnitudes brighter.
The color of the outermost halo isophotes will be sensitive to the
S/N match between the B and R frames; our estimate based on the
intermediate halo color is likely to be more robust.

\subsection{Optical Line (H$\alpha$ Emission)}

Figure 8 presents a continuum subtracted H$_{\alpha}$ image derived
from the CFHT Fabry-P\'{e}rot observations. Figure 9 shows a similar
continuum subtracted interference filter (IF) image obtained with
the Calar Alto 2.2m telescope that confirms all features seen in
Figure 8. The former provides higher sensitivity and resolution
while the latter image gives a much larger field of view that allows
a search to be made for stripped warm gas or gas rich companions. In
this section we use the 3D CFHT data only as the source of an
additional 2D H$_{\alpha}$ map. H$_{\alpha}$ emission is detected
from galaxies H79a, b and d. Emission is detected over the full
optical extent of galaxy H79d while emission from H79a and b is
detected only in the central regions. The weak emission signature
associated with H79c is not confirmed with our slit spectra and is
likely an artifact of the strong Balmer absorption detected in that
galaxy.

Table 10 summarizes new and published H$_{\alpha}$ fluxes for these
galaxies including both IF and slit spectral measures. The former
include a contribution from [NII] $\lambda$6548,83 emission. Our
flux measures have an uncertainty lower than $\pm$10\%, computed as
Poissonian error. We find a large scatter among published EW
H$_{\alpha}$(+[NII]$\lambda$6548,83) measures: 1-7{\AA}, 4-8{\AA}
and 13-116{\AA} \citep[][N2000a]{igl99,coz04} for galaxies H79a,b
and d respectively. We measure 10{\AA}, 9{\AA} and 36{\AA},
respectively, for the three galaxies. [NII] contamination could not
be avoided with the IF filter employed (FWHM= 76{\AA}). Typical EW
for E/SO galaxies range from -3 to 4{\AA} \citep{ken98}. Our EW
measure for H79d is similar to published values for normal Sd
galaxies \citep{ken98,jam04}. A mean EW H$_{\alpha}$=36{\AA} was
found in a recent study involving Sd galaxies \citep{jam04}.
H$_{\alpha}$ flux measures for H79d tabulated in Table 10 show much
less scatter than EW estimates reflecting the uncertainty of the
continuum normalization for this late-type edge-on galaxy. We
estimate a star formation rate (SFR= 0.07 M$_{\sun}$ $yr^{-1}$)
which is smaller than the mean value (0.6 M$_{\sun}$ $yr^{-1}$)
found for normal Sd spirals \citep{ken83,jam04}. SFR is computed
using the formula derived for normal disk galaxies:
$$SFR(M_{\sun}~yr^{-1})=7.9\times10^{-42}~L(H_{\alpha})~(ergs
~s^{-1})$$ assuming a \citet{sal55} initial mass function (IMF) with
mass limits 0.1 and 100 M$_{\sun}$ \citep{ken98}. We derived the net
H$_{\alpha}$ luminosities using the [NII]$\lambda$6583/H$_{\alpha}$
line ratios from N2000a. The SFR for H79d increases to $\sim$0.11 if
we apply the same (1.1 magnitude) extinction correction as used in
\citet{jam04}. H$\alpha$ flux and derived SFR values for H79d
therefore show no evidence for an interaction induced enhancement.

Using the same formula we estimate SFR $\sim$ 0.05 and 0.06
M$_{\sun}$ $yr^{-1}$ for galaxies H79a and b, respectively. The
[NII]$\lambda$6583/H$_{\alpha}$ line ratio is $\sim$ 0.5 for both
galaxies (N2000a) which is at the lower limit of the typical values
(0.5-3) found for early-type galaxies \citep{phi86}. H$_{\alpha}$
emission in galaxies H79a and b shows an extension on a scale of 3-4
kpc, larger than the typical size of H$_{\alpha}$ emission regions
in early type galaxies($<$ 3 kpc) \citep{phi86}. Apparently there is
too much gas in H79a/b if they are interpreted as normal SO
galaxies. The emission in H79a shows two condensations separated by
3\arcsec\ superimposed on weaker more extended emission. One
component (NE) is coincident with or slightly E of the position of
the optical nucleus and lies at the end of an apparent emission
``bridge'' between H79d and H79a that can be seen in Figure 8. The
other component (SW) is about 0.74$\times$ the intensity of the NE
component. Galaxy H79b shows a compact central emission component
with diameter D $\sim$ 5\arcsec\ superimposed on a weak elongated
diffuse component with major axis diameter D $\sim$ 15\arcsec. The
latter component shows distinct curvature or warping. The central
component appears to be slightly offset towards the N on Figure 8.

We estimated the mass of ionized gas following \citet{phi86} where:
$$M_{ionized~gas}=(L_{H_{\alpha}}m_{H}/N_{e})/(4\pi
j_{H_{\alpha}}/N_{e} N_{p})$$ where L$_{H_{\alpha}}$ is the
H$_{\alpha}$ luminosity, m$_{H}$ is the mass of the hydrogen atom,
j$_{H_{\alpha}}$ is the H$_{\alpha}$ emissivity, N$_{e}$ and N$_{p}$
are the electron and proton densities, respectively. The assumed
electron temperature was 10$^{4}$ K. [SII] $\lambda6717/\lambda6731$
line ratio was used as an electron density indicator \citep{ost06}.
[SII] $\lambda6717/\lambda6731$ line ratios for galaxies H79b and d
are estimated from our slit spectra and are tabulated in Table 11
along with other line ratios. No [SII] $\lambda6717/\lambda6731$
line ratio exists for galaxy H79a so we adopted the same ratio as
galaxy H79b. The value for $4\pi j_{H_{\alpha}}/N_{e}N_{p}$ was
derived using the HI recombination line tables of \citet{ost06}. The
computed ionized gas masses are 7.5$\times$10$^{4}$M$_{\sun}$,
9.2$\times$10$^{4}$M$_{\sun}$, 3.5$\times$10$^{5}$M$_{\sun}$ for
galaxies H79a, b and d, respectively. The ionized gas masses for
galaxies H79a and b are within the range of values (between 10$^{3}$
and 10$^{5}$ M$_{\sun}$) found by \citet{mac96} for luminous
elliptical and lenticular galaxies. On the other hand \citet{phi86},
using a larger sample of early-type galaxies, report a mean value
between 10$^{3}$ and 10$^{4}$ M$_{\sun}$. This would place galaxies
H79a and b at the upper end of the distribution of ionized gas
masses for early-type galaxies. Emission line diagnostic diagrams of
\citet{kew06} suggest that much of the gas in H79a may be related to
the AGN rather than to star formation so the already uncertain
estimated mass may be too high. [OIII]/H$\beta$ line ratio of
\citet{coz04} is used.

We used the larger field of view of our PUMA
(10\arcmin$\times$10\arcmin) and Calar Alto
(13\arcmin$\times$5\arcmin) observations to search for H$\alpha$
emission from other galaxies in the field that might represent
previously unknown neighbors of SS. The search was performed two
different ways: using ADHOCw software and IRAF task DAOFIND imposing
a threshold of 4$\sigma$. No candidates were found implying that no
gas rich dwarf systems similar to H79d lie within $\sim$ 90 kpc. We
also looked for discrete H$\alpha$ condensations similar to the ones
found in the debris field of SQ \citep{sul01}. We required
confirmation on at least two independent H$\alpha$ images with
agreement within a few arcsec to allow for field distortion.

\subsection{Optical Spectroscopy}
\subsubsection{H$\alpha$ emission}

H79d is the only galaxy in SS that shows a normal late-type ISM.
Figure 10 compares new Fabry-P\'{e}rot (FP) and slit H$\alpha$ line
of sight velocity curves. They show good agreement (also with R1991,
\citealt{men03} and \citealt{nss00}) when differing spatial/spectral
resolutions are taken into account. The rotation curves are not
strongly distorted suggesting (along with gas content) that H79d is
a relatively recent arrival without evidence of strong perturbation.
Figure 10 suggests a reasonable estimate for the maximum rotation
velocity $v_{max}$ for galaxy H79d is $\sim100$ km s$^{-1}$ after
correction for 80\arcdeg\ inclination. Using $v_{max}$  we derive a
mass within 13\arcsec\ radius ($R\sim3.8$ kpc) of
$\sim9\times10^{9}M_{\sun}$ thus
$(M/L_{r})_{H79d}\sim6~M_{\sun}/L_{\sun}$. R1991 find the mass to
R$_{25}$ for H79d $\sim3\times10^{10}M_{\sun}$ and
$(M/L_{B})_{H79d}\sim3.7~M_{\sun}/L_{\sun}$, using the mass and
luminosity within R$_{25}$, radius corresponding to 25 mag
arcsec$^{-2}$ in B-band.

Figure 11 presents the line of sight velocity curve for H79a. We
plot the heliocentric velocity along the major axis of the H$\alpha$
emission from FP interferometry using CFHT-MOS/SIS data. Our PUMA
measures were indispensable in interpreting the higher resolution
and S/N CFHT ones. PUMA has more than three times the free spectral
range of CFHT and was able to resolve the order overlap that
affected the higher resolution and S/N CFHT measures. The larger
free spectral range of our PUMA data was used to resolve velocity
redundancy inherent in the higher sensitivity CFHT FP data. A
published stellar velocity curve based on long slit observations
\citep{bon99} is superimposed on this plot. In order to match the
long slit data, we mimic a slit through our FP data at
PA=65$^\circ$. We find two H$\alpha$ velocity components: 1) a weak
and compact ($\sim$ 5\arcsec) nuclear component with velocities
($\sim$ 4180 km s$^{-1}$) very similar to stellar values in the
nuclear region (this is part of the NE spatial component in Figure
8) and 2) a more extended (at least $\sim$ 8\arcsec) higher velocity
($\sim$ 4360 km s$^{-1}$) component (part of NE and all of SW
components in Figure 8). The NE spatial component is more intense
because it involves H$\alpha$ emission from both velocity
components. The infalling emission shows almost constant velocity
that, at the center, is $\sim$150 km s$^{-1}$ higher velocity than
the starlight. The match with the stellar velocity curve suggests
that velocity component 1 is nuclear gas and that velocity component
2 can be interpreted as an infalling sheet of gas.

The infalling gas in H79a is either evidence for feedback from
earlier stripping episodes in SS or cross-fuelling from a gas rich
neighbor. H79d is the obvious candidate if one considers the latter
interpretation and it is supported by the apparent H$\alpha$ bridge
between H79a and d that is shown in Figure 8 which shows the sum of
5 velocity channels from the MOS/SIS continuum subtracted image.
This corresponds to velocity range 4350-4395 km s$^{-1}$ in the 0th
order and 4615-4660 km s$^{-1}$ in the first order. These velocities
are consistent with emission from both H79a and H79d. While we see
an apparent bridge (see also R1991) between galaxies H79a/d we do
not find velocity continuity. The northernmost condensations in H79d
can be seen in silhouette on the outskirts of H79a \citep{pal02}.
The rotation curve of H79d shows velocities $\sim$ 4650 km s$^{-1}$
near this overlap zone. This is about 500 km s$^{-1}$ higher than
the nuclear gas in H79a and $\sim$ 300 km s$^{-1}$ higher than the
infalling H$\alpha$ component in that galaxy. If the infalling gas
in H79a originated in H79d then the fuelling was episodic and not
continuous. If H79d did not provide the fuel then feedback of gas
previously stripped from SS members would likely be the source.
Using a velocity dispersion $\sigma_{0}\sim155$ km s$^{-1}$
\citep{bon99} we get a mass for H79a within 14\arcsec(last
concentric isophote) $\sim$10$^{10}$-10$^{11}$ M$_{\sun}$.

Figure 12 presents the line of sight velocity curve for H79b with
our new SPM-PUMA data. In this case we find a well defined H$\alpha$
emission rotation curve that is counter rotating relative to the
stellar velocity curve from \citet{bon99}. This is evidence for a
minor or quiet (lacking a strong MIR/FIR signature) merger as
suggested by \citet{sra94}. Figure 13 shows the counter-rotation of
the H$\alpha$ gas in galaxy H79b relative to \citet{bon99} and our
new stellar velocity curve obtained with ALFOSC. The estimated mass
of H79b within 11\arcsec ($\sim$ the last concentric isophote) is in
the range $9.0\times10^{9}$ --$1.7\times10^{10}M_{\sun}$ using the
\citet{bon99} and ALFOSC rotation curves (Figure 13), respectively.
This is an underestimation of both the current mass and, much more,
the original mass of H79b. Our estimates of the mass to light ratio
range from 1-2. If H79b entered SS as an L$^{\ast}$ galaxy then it
has lost from 50-90\% of its original mass into the common halo. It
has gained the mass of the counter rotating component estimated from
the H$\alpha$ velocity curve to be $\sim3.3\times10^{9}M_{\sun}$.
The latter value suggests that H79b accreted a late-type dwarf
intruder somewhat less massive than H79d. The lack of H$\alpha$
field detections suggest that no new intruders will visit SS in a
significant fraction of the next Gyr.

We attempted to obtain velocity dispersion measures for several of
the galaxies, but our spectral resolution was too low to yield
useful measures except in the case of H79c. Previous measures of
$\sigma$= 155, 130 and 60 km s$^{-1}$ for H79a, b and c respectively
\citep{bon99} all fall close to fundamental plane of elliptical
galaxies (see Figure 7 in \citealt{del01}) and close to a $\sigma$ -
M$_B$ relation defined for dwarf/giant elliptical galaxies, galactic
bulges and dwarf spheroidals \citep{ben92}. H79c, which shows
properties least like an elliptical galaxy, also shows the largest
deviation from $\sigma$ - M$_B$ relation in the sense that the
velocity dispersion is too low. Our corrected estimate for H79c is
slightly higher at $\sigma$=89 km s$^{-1}$.

\subsubsection{H$\alpha$ absorption}

Figure 14 shows the slit spectrum of galaxy H79c. We attempted
several different estimates for the age of galaxy H79c: 1) We fitted
the spectrum with theoretical templates \citep{bru03} computed for
different stellar populations and metallicities. The best fit
template corresponds to a stellar population of $\sim$ 1.5 Gyr with
metallicities of both Z=Z$_{\sun}$=0.02 and Z=0.05. 2) Using the
program \emph{indexf} (Cardiel et al. see
\emph{http://www.ucm.es/info/Astrof/users/ncl/index.html}) we
derived some age indices (e.g. D$_{n}$(4000), H$\delta_{A}$, Fe5015,
etc.) that are commonly used to date galaxy stellar populations
\citep[e.g.][]{kau03,gon05,del07}. We obtained D$_{n}$(4000)=1.56,
H$\delta_{A}$=2.09, Fe5015=3.44, H$\beta$=3.18, D(4000)=1.83,
Fe5406=0.88. We performed also simulations (photon counting) on the
spectrum to get an estimate of the errors in the indices due to S/N:
for example, for D$_{n}$(4000) such errors are $\sim$ 0.01-0.02 and
for H$\delta_{A}$ are $\sim$ 0.5-0.08. Following \citet{kau03} (see
their Figure 2), we estimate that H79c had its last burst of star
formation $\sim$ 1.5-1.8 Gyr ago (if an instantaneous,
solar-metallicity burst model is considered). The estimates are
still consistent if bursts of different metallicities are
considered. 3) The value of D(4000)=1.83 seems to be unreliable in
getting age estimates. While D$_{n}$(4000) and H$\delta_{A}$ indices
are in agreement according to Figure 3 of \citet{kau03}, D(4000) and
H$\delta_{A}$ do not match any of the models illustrated in the
Figure 9 of \citet{gon05}. 4) Figure 3 of \citet{pro04} shows
index-velocity dispersion relations. With its estimated $\sigma=$ 89
km s$^{-1}$ the location of H79c in such plots would infer a spiral
bulge behavior, rather than an S0. Using their Figure 4 the age
estimate is $\sim$ 2-3 Gyr and [Fe/H]$\sim$ -0.75. All these methods
used to estimate the age lead to an average age of $\sim$ 2 Gyr for
H79c. This relatively young age is consistent with the bluer colors
reported earlier for galaxy H79c and with the hypothesis that H79c
is a stripped spiral that intruded into the group near that time. It
was likely the intruder that preceded H79d. At the other extreme a
lower S/N spectrum of H79f is consistent with a much older age
consistent with its being one of the original group members now in a
stage of dissolution.

\subsection{MIR/FIR}

Compact groups generally show depressed levels of FIR emission
because the component galaxies quickly lose their ISMs to encounters
and collisions \citep{sul93}. Only recently formed groups or
transient systems would be expected to show normal or above normal
emission. SQ revealed low levels of star formation igniting in the
tidal debris \citep{sul01} with very little emission in the
component galaxies. The expectation for SS, viewed as significantly
more evolved than SQ, would be that only recent unstripped intruders
might show significant FIR emission with a possible component from
the debris field. Our H$\alpha$ search failed to turn up evidence
for the latter component. The situation is different for MIR
emission where quasi-continuous interactions might efficiently
channel any residual unstripped gas into component nuclei. Thus MIR
emission from a warm ISM (HII regions) would not be expected while
nuclear sources--sometimes connected with AGN could be common. SQ
tells us that warmer emission from large scale shocks or debris
field starbursts can also occur. We must rely upon the IRAS survey
for information about FIR emission in SS while a new ISO map
provides insights into the MIR emission.

SS shows moderate FIR emission \citep{ver98} with estimated
$L_{FIR}\sim8.7\times10^{9}~L_{\sun}$ reflecting IRAS fluxes at 60
and 100 \micron\ S$_{60}$=1.28 Jy and S$_{100}$=2.82 Jy,
respectively. FIR luminosity is computed using
$log~(L_{FIR}/L_{\sun})=logFIR+2logD+19.495$, where
$FIR=1.26\times10^{-14}(2.58S_{60\micron}+S_{100\micron})~W~m^{-2}$
\citep{hel88}, D is the distance in Mpc and S$_{60\micron}$ and
S$_{100\micron}$ are the fluxes at 60 \micron\ and 100 \micron,
respectively expressed in Jy. FIR luminosity is similar to the mean
value for the most isolated luminous Sc spirals in the local
Universe \citep{lis07}. SS is smaller than the IRAS beams at the 25
(3-4$\sigma$), 60 and 100 \micron\ wavelengths where SS is detected.
IRAS processing yields only unresolved detections although various
studies have assigned the FIR emission to one or more members
\citep{hic89b,all96,ver98}. While IRAS resolution was
$\theta$$\geq$1\arcmin\ the pointing accuracy was much higher for
strong detections. The 1$\sigma$ position uncertainty ellipse for
the SS detection was 12\arcsec$\times$4\arcsec\ (PA$\sim$100\arcdeg)
centered E of H79a and S of H79e. This is inconsistent with our
expectation that H79d should dominate the FIR emission. It is almost
certainly a significant contributor however several other sources of
cold dust emission could be present e.g.: 1) H79a: the dust lane and
or the near nuclear H$\alpha$ blobs, 2) H79b: the counter rotating
component and twisted dust lane, 3) H79e: as a luminous background
ScI spiral and 4) possible diffuse dust emission associated with
stripped HI near the IRAS position.

We described earlier new ISO MIR observations of SS which are shown
in Figure 15 as an 11.5 \micron\ ISOCAM map with 6\arcsec\
resolution superimposed on a DPOSS optical image. We resolve SS into
three sources involving H79a, b and e. Photometry was performed with
the IRAF APPHOT package where we interactively derived fluxes for
the three sources present in the field adopting aperture photometry
for our measures. For galaxies H79b and e, that are very close to
each other, we masked galaxy H79b/e when measuring aperture
photometry for galaxy H79e/b. Fluxes are listed in Table 12 and
correspond to a total flux from the group that is roughly a factor
of two below the IRAS upper limit of 90 mJy quoted in \citet{mos90},
while each galaxy flux is again a factor of two/three below the flux
limit estimated from the 12 $\micron$ MaxEnt IRAS map of
\citet{all96}. The lack of an 11.5 $\micron$  detection for H79d is
not surprising because late-type spirals tend to show a colder ISM.
At the same time a massive and 4.5$\times$ more distant ScI spiral
like H79e was detected by ISOCAM. MIR emission might be attributed
to the active nucleus in H79a and to the accretion event in H79b.

The blue and FIR luminosities of the neighbors detected by IRAS are
presented in Table 13. All of them fall on the L$_B$ vs. L$_{FIR}$
correlation defined for isolated galaxies \citep{per97,lis07},
except the pair UGC10197-8 which shows enhanced FIR emission as
expected. If we assume that all of the FIR emission
($log~(L_{FIR}/L_{\sun})=9.93$) belongs to H79d that would imply
that it is about 27$\times$ brighter than the isolated galaxy
expectation ($\sim$ 8.5) for sources of similar morphology and
optical luminosity. There are two possible scenarios: either the
emission in galaxy H79d is enhanced by interaction with the rest of
SS or the FIR emission arises from several of the sources listed
above.

\subsection{Radio Continuum}

A combined analysis of VLA (old and new data), NVSS\footnote{NRAO
VLA Sky Survey (NVSS) is found on the web at
http://www.cv.nrao.edu/nvss/ (see also \citet{con98})} and
FIRST\footnote{Faint Images of the Radio Sky at Twenty-centimeters
(FIRST) is found on the web at http://sundog.stsci.edu/} provides a
clear picture about the sources of radio continuum emission in SS.
VLA (see Figure 16) and NVSS isophotes indicate that the center of
the emission lies closest to H79a. The archival VLA B-array image
(6\arcsec\ resolution) shows that three galaxies (H79a, b and e) are
reasonably compact radio sources with only H79a detected by FIRST.
H79d is only detected with the C-array observations while H79b and e
are blended together (see Figure 16). The photometry shows that the
B-array data is missing some extended flux (see Table 14). This is
not surprising if the emission from H79b, d and e is connected with
star formation in those galaxies

We also detect an apparently unrelated source $\sim$1\arcmin\ north
of the group and near the tidal tail. It shows an integrated C-array
flux of 5.9 mJy in agreement with FIRST, NVSS, \citet{wil91} and
\citet{dic84} measures. The optical counterpart appears to be one
component of a distant interacting galaxy pair that is clearly
resolved on HST images. The $\sim$1 mJy source west of the group
shows no optical counterpart.

We explore the well-known radio-FIR correlation \citep{dis84} in
order to estimate/resolve the FIR flux from galaxies in SS. A
complete sample of 250 normal spiral galaxies brighter than B = 12
\citep{con91} shows the following strong correlation:
$$log~(L_{1.4GHz}~[W~Hz^{-1}]) =
1.29~log~(L_{FIR}~[L_{\sun}])+8.76$$
(we adapted the coefficients for H$_{\circ}$ = 75 km s$^{-1}$
Mpc$^{-1}$). If we assume that radio emission from H79b, d, e is
dominated by star formation related processes then we expect that
H79b and d each account for 25\% of the total FIR flux with galaxy
H79e accounting for 15\%. The residual FIR emission is attributed to
H79a (25-35\%) involving perhaps emission from the AGN, star
formation and the strong dust lane. Using the predicted FIR fluxes
we find that all galaxies fall close to the FIR-optical correlation
found by \citet{lis07} for very isolated galaxies.

We also explore the radio continuum-MIR correlation using ISO MIR
and C-array radio continuum fluxes from Tables 11 and 13,
respectively. Galaxies H79b and H79e closely follow the correlation
found in \citet{gru03}. Galaxy H79a falls somewhat above the
correlation showing excess radio continuum flux likely associated
with the AGN.

\subsection{Radio Line (21cm)}

\citet{wil91} presented HI velocity channel maps for SS. The
emission was centered on H79d with weak emission extending to the E
and NE apparently overlapping H79f. Our new 21cm observations cover
a wider velocity range (4275-4880 km s$^{-1}$) with detections in
the range 4470-4680 km s$^{-1}$ (see Figure 18). While most of the
emission originates in H79d, the new observations (see Figure 17)
clarify the structure of the HI tail towards the east. We do not
confirm an overlap with H79f and find all the eastward emission to
be S of that galaxy. The velocities for the eastward emission are
continuous with the emission from H79d. This is either a coincidence
or evidence that we are seeing the first stages in the ISM stripping
of H79d. The latter interpretation would exacerbate the problem of
the missing gas in SS \citep{ver01} leaving no residual gas from
galaxies H79a, b, c or f.

The velocities in the eastern tidal tail are similar to the
velocities in the southern part of galaxy H79d (see Figure
18--channel map at 4503 km s$^{-1}$). Long one-sided HI tails have
been found in Virgo cluster spirals \citep{chu07}. Simulations of
one of these galaxies, NGC 4654, suggests that ram-pressure
stripping may be the cause of such HI tails \citep{vol03}. These
galaxies are interpreted as recent arrivals into the Virgo cluster
\citep{chu07}. If this interpretation applies to H79d then the tail
may be telling us that it entered the group from the NE. This does
not provide additional support for the hypothesis that H79d is
crossfuelling H79a because the lowest velocity HI that falls near
H79a shows a velocity $\sim$4550 km s$^{-1}$ (see Figure 18) which
is $\sim$200 km s$^{-1}$ higher than the infalling gas.

\citet{ver01} argued that SS as a group was very deficient in cold
gas, traced by HI emission. They estimated the HI deficiency in the
range of 30-50\% depending on whether one assumes that HCG79bc are
intrinsically S0 galaxies or were originally spirals. Similar
numbers were obtained for SQ. The most extreme  estimates are
motivated by the possibility that all E and S0 galaxies in CG are
the product of spiral galaxy ``harassment" (disk destruction and ISM
stripping) as proposed to explain the S0 population in galaxy
clusters \citep{moo96}. This assumption is further supported by the
very high fraction ($\sim$80-85\%) of late-type galaxies found in
low density environments \citep{sul06} that we argue are typical of
CG environments. If we assumed that H79 a and f were also originally
spirals then we would obtain an even more extreme deficiency
($\sim$70\%). Velocity ranges sampled by old and new 21cm
observations of SS unfortunately do not include the full range of
possible velocities implied by optical measures. It is reasonable to
expect HI with velocities as low as $\sim$4000 km s$^{-1}$. The mean
velocity of galaxy H79c is close to 4000 km s$^{-1}$, so stripped
gas might reasonably be expected down to $\sim$3800 km s$^{-1}$.
Comparison of the new VLA HI spectrum with single dish Arecibo
\citep{bie79,gal81} and GBT (Borthakur et al., in preparation) HI
velocity profiles suggest extended emission ($\sim$20\% of total)
might exist between 3900-4800 km s$^{-1}$. An unpublished
position-velocity plot that includes the region north of H79d shows
evidence for another HI component beginning at 4250 km s$^{-1}$ near
the edge of our velocity range. We suggest a conservative estimate
of 30\% HI deficiency for both SQ and SS.

The observed HI mass of galaxy H79d agrees with the mass predicted
for its blue luminosity \citep{ver01}. On the other hand, the HI
mass for H79d is lower than the mean value for Sd galaxies
\citep{rob94}. The position-velocity (PV) plot for H79d (Figure 19)
shows an almost solid-body rotation curve as expected for a low mass
disk galaxy. One can see four fairly equally spaced peaks in the PV
plot, which likely correspond to spiral arms in the disk. The
corresponding local peaks can be seen in the H$\alpha$ image as well
(see Figures 8 and 9).

No molecular gas has been detected in SS although, again, only part
of the plausible velocity range has been sampled. H$_{2}$ masses
traced by CO emission have upper limits of $\sim
5\times10^{8}~M_{\sun}$, $6\times10^{8}~M_{\sun}$,
$6\times10^{8}~M_{\sun}$ for H79a,b and c, respectively
\citep{leo98,ver98}.

\subsection{X-ray}

While the general process producing HI deficiency in a large number
of CG is yet to be clarified, their results suggest that gas heating
may play a role and that very sensitive X-ray maps and flux
measurements could be a way to look for hot gas. In the case of SS
the soft X-ray emission has un upper limit of  L$_{X}<
2.3\times10^{41}$ ergs s$^{-1}$ \citep{pon96}, using ROSAT PSPC
(Position Sensitive Proportional Counter) data. According to
\citet{pil95}, using the same data, this estimate is a 2.6 $\sigma$
detection. Thus, there is no massive soft X-ray emitting halo of hot
gas. This might indicate a dynamical old system where gas has
cooled. On the other hand, N2000a have argued that the 2.6 $\sigma$
photon excess in SS is consistent with an extended halo component
that follows the diffuse optical light.

\section{Discussion}

We report multiwavelength observations of SS that seek to explore
the evolutionary history of the group especially in the context of
previous work on SQ. We assume SS and SQ are representative of the
CG phenomenon \citep[H1982;][]{sul97} and that SS is considerably
more evolved than SQ. The first assumption is based on the fact that
both groups show n=4 or more accordant redshift members with
evidence that virtually all are interacting. The second assumption
is based upon the much higher luminosity fraction contained in the
SS halo (\S\S\S~3.1.3). SQ is unusual at this time because a new
intruder is entering the group at unusually high velocity. Our
inferences about SQ as a typical compact group were only marginally
affected by this transient event. That collision is generating
excess X-ray, optical emission line and radio radiation. The closest
equivalent in the southern CG sample \citep{iov02} would likely
involve the ``Cartwheel'' system where we are observing the
aftermath of a disk penetrating collision. The frequency of
occurrence of such spectacular events depends upon: 1) the density
of high velocity neighbors near compact groups and/or 2) the number
of groups with high enough mass concentration capable of
accelerating neighbors into a group with suitably high velocity.
Most compact groups are not found in regions of high enough density
for high velocity $\Delta$V$\sim$10$^3$ km s$^{-1}$ intruders to be
common. Yet SQ is certainly not the only group with such a potential
intruder \citep[see Figure 2 in][]{sul97}. The important point to
emphasize is that groups like SQ that are currently undergoing a
spectacular event are not less typical because of it.

Specific considerations involve: 1) the age of the group, 2) the
nature of its tidally truncated members, 3) the nature and origin of
its luminous halo, 4) the evidence for merger/accretion events and,
in view of its HI deficiency, 5) the fate of the missing gas. CG can
provide valuable insights into the role of extreme interactions on
galaxy evolution and may be useful for interpreting results at
higher redshift. It was once thought that unstable systems like CG
had very short evolutionary lifetimes that ended in a merger
catastrophe \citep[e.g.][]{mam87,bar89}, however observations do not
support that view \citep{sul87,sul97} and hypothesized dark matter
halos appear able to extend CG lifetimes indefinitely
\citep[e.g.][]{ath97}.

SQ is the most studied compact group with published studies spanning
the electromagnetic spectrum: \citet{sul01} and references therein
plus subsequent: X-ray \citep{tri03,tri05}, UV \citep{xan04,xui05},
optical \citep{gut02,xul03}, MIR \citep{app06}, CO
\citep{yun97,lis02,lis04}, HI \citep{wil02} and radio continuum
\citep{xul03,xan04} observations. If SQ can be viewed as a currently
hyperactive prototype of the compact group phenomenon then we infer
the following:
\begin{enumerate}

\item  Compact groups form by slow sequential acquisition of neighbors
from surrounding larger scale structure. Nucleation points around
which the process might begin could involve dark matter density
fluctuations. Baryonic (primordial merger) fluctuations
\citep{gov96} are disfavored by the rarity ($\sim$10\%) of
first-ranked early-type galaxies in CG \citep{sul97}. Random
continuous formation from more diffuse galaxy aggregates
\citep[e.g.][]{dia94} is also disfavored for several reasons
\citep[see also][]{sul97}: 1) local environmental densities are too
low to make this process important, 2) few of the predicted majority
population of transient groups is observed and 3) observations
indicate that compact group formation is slow and sequential.

\item  New intruders are usually captured and quickly lose most of their
ISM leading to: a) suppression of star formation within the
galaxies: b) possible stimulation of active nuclei due to infall of
residual unstripped gas, c) morphological transformation from
primarily spiral intruders into spiral bulges or into early-type
E-S0 members. While brief episodes of enhanced star formation are
expected and observed in compact groups they are likely associated
with the sequential arrival of new gas rich intruders. This activity
is unlikely to be long lived if new arrivals are also rapidly
stripped. SQ suggests that any star formation activity that arises
in the resultant debris field of stripped gas will be weak and will
not approach the level expected from a single unstripped L$^{\ast}$
spiral galaxy.

\item  In addition to ISM stripping, which can create a multiphase
gaseous debris field, the outer parts of the stellar components are
stripped leading to the growth of a massive diffuse halo akin to
what is now observed in some clusters. Some diffuse X-ray emission
can be associated with this halo but this component is likely to be
overestimated with low resolution observations if: a) high velocity
intrusions take place, b) AGN are common (e.g. HCG16) and c) major
mergers occur (yet rarely).

\item  The major components of the groups persist as discrete (gas poor
early-type) condensations in a diffuse stellar halo for many Gyr. SQ
is mute on the fate of these long lived components but SS, viewed as
a more evolved SQ, suggests that they may slowly dissolve into the
hypergalactic halo before major merging can take place. The rarity
of massive fossil ellipticals outside of cluster cores \citep{sul94}
attests to the rarity of major mergers and the long timescale for
the dissolution process. Groups like SQ and SS disfavor hierarchical
collapse scenarios.

\end{enumerate}

Our new, as well as older published, measures for SS show many
consistencies with what has been inferred from SQ. A complication in
the case of SS may involve an unfavorable orientation where the
principal plane of most components lies near edge-on to our line of
sight. H79a, b, c and d show morphological features, or axial
ratios, consistent with the interpretation that they are highly
inclined disky systems. Following the same numbering/lettering
scheme above we evaluate the empirical clues available for SS. We
refer to relevant subsections of \S~3 whenever our inferences are
supported by particular observations or previously published work.

\textbf{1) CG formation}: Galaxy H79d in SS is an example of the
sequential acquisition process. Such acquisitions are likely to be
rare given the low galaxy surface density environment
\citep{sul87,roo89} in which SS is found. H79d could be regarded
either as a simple projection along the same line of sight or as a
galaxy that is entering the group for the first time. Yet, the low
surface density of accordant and/or discordant galaxies near SS
would justify the latter possibility. This highly inclined late-type
(Sd) spiral shows HI (\S\S~3.6) and H$\alpha$ (\S\S~3.2, 3.3)
distributions consistent with the notion that all or most of the
cold and warm ISM gas components are intact. New and old H$\alpha$
rotation curves (\S\S~3.3) also show little or no evidence for
significant dynamical disruption. The slight ``integral sign'' shape
of the galaxy can be argued to be the result of viewing a disky
spiral at high inclination rather than the product of tidal
deformation. HI emission extending towards NE (see Figure 17; see
also \citealt{wil91}) could be interpreted as debris from earlier
acquisitions although velocity continuity with the gas within H79d
(Figure 18) makes this assumption uncertain and can also be argued
to be the beginning of the stripping process for that intruder.
H$\alpha$ emission appears to link H79d with H79a (Figure 8;
cross-fuelling?), although we do not find velocity continuity
between emission on the side of H79d nearest H79a and the gas
detected in its nuclear region (\S\S~3.3). We cannot unambiguously
establish interaction of H79d with the rest of the group. Typical
timescales for significant dynamical response and/or tidal stripping
are $\sim$10$^8$ years \citep{ion04} coupled with the unequivocal
evidence for lack of tidal perturbation suggests that, if H79d is an
intruder, it must have joined SS within this time frame.

H79c is the best candidate for the intrusion that preceded H79d into
SS. It is the bluest (\S\S\S~3.1.4) member after H79d and also shows
spectroscopic evidence (\S\S\S~3.3.2) for a reasonably young
($\sim$2 Gyr) stellar population perhaps generated just before it
lost its ISM upon joining SS. H79f is the obvious candidate for the
oldest, or one of the oldest members of SS. It is sometimes
described as a tidal tail \citep{nis02} but its morphology
(\S\S\S~3.1.2) and lack of nonstellar component (\S\S~3.2--3.6) are
more consistent with the hypothesis that it is a remnant disk galaxy
slowly dissolving into the common stellar halo \citep[see also
R1991, B1993,][]{bon99}.

\textbf{2a) HI Stripping and SF suppression}:

A statistical study of HI content in CG showed that HI deficiencies
are common \citep{ver01}. Both SQ and SS have estimated deficiencies
in the range 30-70\% depending on how many of the early-type
components entered the groups as spirals (\S\S~3.6). All presently
observed E and S0 components of SQ (3) and SS(4) are candidates
because the environments which supply intruders are spiral rich. Two
recently stripped spirals and, therefore now proto-lenticulars,
\citep[NGC7319 and 7320c;][]{sul01} can be identified in SQ. We
suggest that H79d will likely suffer a similar fate. The reality of
such gas deficiencies virtually guarantees a star formation deficit.
FIR (\S\S~3.4) and H$\alpha$ (\S\S~3.2) observations are regarded as
useful star formation diagnostics. Both are quite consistent with a
normal (for its type and luminosity) level of star formation H79d.
The lack of star formation suppression argues that it is the most
recent SS intruder or a projected field galaxy. If H79d is viewed as
a recent arrival into SS, this is apparently a much slower intrusion
($\Delta V \sim 200~km~s^{-1}$) than the ongoing one observed in SQ
and is closer to the expectation for a low velocity random capture
in a low density environment. A recent and low velocity intrusion
into SS can be also used to argue that no strong star formation
enhancement would be expected at this time. In CG such disk
starbursts are likely short lived if rapid ISM stripping immediately
follows the onset of the enhanced activity. Residual gas channeled
into the nuclei of galaxies might give rise to low level star
formation or AGN activity. Nuclear fuelling might be particularly
efficient in CG because of the quasi continuous nature of the tidal
perturbations. The excess of compact nuclear radio sources
\citep{men85} and low luminosity AGN \citep{coz04} in CG members
supports this suggestion.

Little additional emission related to star formation is observed in
the obviously more evolved members of SS which reflects the observed
deficit of gas in the group. Apart from H79d then SS shows a much
more extreme level of star formation suppression than SQ which is
consistent with it being older and more evolved. H$\alpha$
observations indicate low level star formation is likely  present in
H79b, probably connected with an accretion event and H79a possibly
connected with a feedback event (\S\S\S~3.3.1). Line emission from
H79b shows a reasonably symmetric rotation curve and may originate
entirely from the counter-rotating component. This is consistent
with star formation in a recently accreted dwarf companion
(\S\S\S~3.3.1), although quoted [NII]/H$\alpha$ ratios are close to
those for H79a \citep[N2000a,][see also Table 11]{coz04} and lie
between the HII and Seyfert domains in emission line diagnostic
diagrams \citep{kew06}. The extended H$\alpha$ rotation curve
(Figure 12) argues strongly in favor of star formation in this
galaxy.

H79a shows evidence for both nuclear and near nuclear emission
components. The H$\alpha$ image in Figure 8 and velocity decoupling
(Figure 11) are consistent with the near nuclear gas as a feedback
event, possibly the result of cross fueling of gas stripped from
H79d. Our Fabry-P\'{e}rot measures (\S\S\S~3.3.1) support this
interpretation because the extranuclear component is redshifted
relative to the H79a nucleus by about 150 km s$^{-1}$, which is
consistent with the extranuclear gas viewed as infalling onto the
side of H79a closest (in projection) to H79d (\S\S\S~3.3.1). However
no evidence for velocity continuity is found (\S\S\S~3.3.1) and
emission on the side of H79d closest to H79a shows a velocity
$\sim$300 km s$^{-1}$ higher than that of the extranuclear emission
feature (see Figures 10 and 11). The observations would require a
discrete infall or fueling event rather than a continuous flow of
gas from H79d to H79a. All emission line measures for the gas in
H79a should be treated with caution because the two emission
components are separated by only $\sim$3\arcsec\ (\S\S~3.2). This,
and confusion in the literature between members of the group,
explains the large scatter among redshifts quoted for this galaxy in
NED.

Table 15 provides an interesting comparison of integrated optical,
IR and radio properties for SQ and SS along with corresponding data
for the well studied L$\sim$L* Sb spiral NGC7331. This provides an
alternate demonstration of the HI and star formation deficits. SQ is
$\sim$ two integrated B magnitudes brighter than NGC7331 while SS
shows approximately the same value as that galaxy, but with half of
the luminosity in a diffuse halo (\S\S\S~3.1.3). SQ contains the
same mass of HI as NGC7331 although it originally contained, at
least, three spirals with similar luminosity to that galaxy. SS
currently shows little more than the HI expected from a late-type
spiral like H79d \citep{ver01}. Perhaps 20\% more remains to be
mapped using wider frequency coverage (\S\S~3.6). SQ and NGC7331
show similar integrated star formation measures. However this is an
overestimate of star formation activity in SQ because H$\alpha$ (via
enhanced [NII]) and radio continuum emission in SQ are boosted by a
large scale shock \citep{mol97,sul01}, while those measures and FIR
luminosity are boosted by the luminous AGN NGC7319. SS shows much
lower levels of activity in all measures with H79d contributing most
of the HI and about 60\% of the H$\alpha$ emission (\S\S~3.2). While
composed of less luminous galaxies, if formed from a largely spiral
field, then SS represents an extreme example of star formation
suppression. SQ is clearly evolving in that direction.

Another way to characterize the star formation suppression involves
comparison of integrated colors and/or population synthesis measures
for CG members. Our color measures (\S\S\S~3.1.4) indicate a
dominance of older populations in H79abf and c (with a more recent
component with age $\sim$2 Gyr; \S\S\S~3.3.2). Other recent
spectroscopic synthesis studies \citep{del07,pro04} suggest
systematically older populations (compared to field galaxies) in
larger samples of compact group members. An earlier photometric
study of elliptical members in CG showed typical red colors and
little or no evidence for a recent merger population
\citep{zep91,zep93}.

H$\alpha$ surveys of compact group members \citep{vil98,igl99}
reveal a wide range of inferred star formation activity. Taken
galaxy by galaxy this is unsurprising. However galaxies in the most
evolved CG (like SS) will show very low levels of star formation
activity in most members except for the occasional unstripped new
intruder. These inferences will be confused if triplets are included
in a CG sample. Less evidence of dynamical evolution is expected in
these more intrinsically unstable configurations. This is especially
true for the most studied samples like HCG where hierarchical
systems are rare.

\textbf{2b) AGN}: Gas that is not stripped from CG members will
likely suffer angular momentum dissipation leading to efficient
channeling into component nuclei. There is accumulating evidence
that galaxies in CG show a higher frequency of, generally low level,
AGN activity \citep{coz04,mar06}. This activity can take at least
three forms: Seyfert, LINER and HII nuclei -- the latter not really
an AGN in the classical sense although it has been argued to be a
precursor to Seyfert activity \citep[e.g.][]{bar00}. Recently
stripped NGC7319 in SQ hosts a luminous Seyfert 2 nucleus that is a
strong source from radio to X-ray wavelengths. Both H79 a and b show
evidence for nuclear activity at radio (\S\S~3.5), MIR (\S\S~3.4)
and optical (\S\S~3.1) wavelengths. All H$\alpha$ emission outside
of H79d is found near the centers of these two galaxies. Line ratios
in H79a ([NII]/H$\alpha$ and [OIII]/H$\beta$)
\citep[N2000a,][]{coz04} are consistent with LINER activity
according to emission line diagnostic diagrams of \citet{kew06} if
the derived ratios involve the nuclear gas. Emission from H79b
(NGC6027, VIIZw631) shows well defined counter-rotation in the
emission component (\S\S\S~3.3.1) and is likely dominated by star
forming gas in an accreted neighbor perhaps coupled with some weakly
shocked gas. Radio continuum emission from H79a involves a point
source at all VLA resolutions while H79b shows weaker and more
extended emission (\S\S~3.5) which supports our inference that only
H79a hosts an AGN at this time. Unless CG members host unusually low
mass black holes, the weakness of most AGN activity must be due to
low accretion rates where only residual gas remains to fuel nuclei.
In systems where there is a quasi-continuous perturbation there is
perhaps less opportunity to build up a reservoir of gas that
episodically might fuel more intense AGN activity. The CG
alternative might well involve longer phases of low level activity.

\textbf{2c) Morphological transformation}: SQ contains two recently
stripped spirals and another spiral intruder in the process of being
stripped \citep{sul01}. The remaining two members in SQ are either
elliptical galaxies or remnant bulges of spirals. SS galaxies H79a,
b, c and f show early-type (albeit peculiar) morphologies
(\S\S\S~3.1.2). In analogy with SQ we argue that most of these
galaxies entered SS as spirals because the environment of SS is
spiral rich (\S\S\S~3.1.1); they were subsequently stripped and
transformed into E/S0 systems. In the case of SQ some of the
stripping events must have taken place recently because we can still
recognize spiral structure \citep[NGC 7319 and NGC7320c;][]{sul01}
while in SS the disks and associated spiral structure have been
largely destroyed leaving remnant bulges imbedded in a stellar halo
partly or mostly composed of the stripped disks. We also argue that
the galaxies were considerably more luminous when they first entered
SS. This is self evident because the massive halo must have formed
from stars that originated in existing and/or now disrupted members.
The fraction of early-type members in SS ($\sim$0.7-0.8) is larger
than expected if SS was formed from random infall of galaxies from
its environment. The observed early-type fraction in SS environment
is $\sim$0.3-0.4 (\S\S\S~3.1.2) above the expectation from the low
density environment of extremely isolated galaxies $\sim$0.14
\citep{sul06}.

All known accordant redshift neighbors around SS are larger and
brighter than members of the group (\S\S\S~3.1.2, 3.1.3). Of course
the tabulation is likely biased towards the brightest members in the
neighborhood if for no other reason that we are tabulating neighbors
with measured redshift. All members of SS are smaller and most are
fainter than this neighboring population. Taken at face value SS is
a compact group of sub-L$^{\ast}$ galaxies. Few tabulated compact
groups of dwarf galaxies are observed \citep{hic92,iov02} and there
may be dynamical stability and hierarchical structural reasons for a
paucity of dwarf compact groups. The brightest four accordant
members of most tabulated groups tend to show similar
L$\geq$L$^{\ast}$ luminosities (i.e. they are not hierarchical
systems). Thus we argue SS was formed by sequential accretion of
luminous spiral galaxies that have undergone: 1) gas stripping that
transformed them into E/S0 morphologies and 2) stellar stripping
that has significantly reduced their sizes. This stellar harassment
process might be similar to, but even more efficient (because of
lower systemic velocity dispersions), than the one proposed to
explain the existence of S0 galaxies in clusters \citep{moo96}. In
CG a significant population of spiral bulges may exist as
``early-type'' members.

The early-type members of SS are too bright for their sizes and this
can be explained if disks are stripped leaving the higher M/L bulge
region of each galaxy relatively intact. Comparison of the size and
brightness of bulges in some of the bright neighboring spirals
indeed show values similar to those measured for the early-type
members of SS (\S\S\S~3.1.3). The early-type members of SS are
embedded in a relatively blue massive stellar halo (\S\S\S~ 3.1.3,
3.1.4). It is reasonable to assume that this luminous halo was
generated by stripping/dissipation processes involving the existing
member galaxies. The relatively blue color of the halo supports the
idea that much of it came from a disk stellar population. The excess
brightness for their size (i.e. compactness) of SS members will also
be abetted by minor accretion events of the kind that we observe in
H79b.

If most SS members began as spiral intruders then we are faced with
a serious deficit of stripped hot/warm/cold gas in and around SS
\citep[\S\S~3.2--3.7; see also][]{ver01}. In the case of SQ we found
all three components in the IGM although little was found within the
group members. Part of the HI deficit in SQ is now shocked to X-ray
emitting temperatures \citep{tri03,tri05} some of which may have
cooled and condensed into HII regions \citep{xus99,sul01}. In SS we
observe only logM$_{HI}$(M$_{\sun}$)=9.3 \citep{ver01} most of which
is concentrated in H79d.

The existence of ``passive spirals'' \citep{got03b} could possibly
help solving the problem of ``missing gas'' (gas deficit) in CG. For
example, one could envision a scenario in which the members of SS
entered the group as extremely poor gas spiral galaxies. As
\citet{got03b} suggests, such galaxies seem to prefer environments
density $\sim$1-2 Mpc$^{-2}$, similar to SS neighborhood
(\S\S\S~3.1.1). On the other hand, these gas deficient spiral
galaxies are very rare ($\sim$0.28\%, i.e. they found 73 cases from
a sample of 25813) and quite red (almost as red as elliptical
galaxies). The members of SS seem to be bluer (see our Table 9) than
those ``passive spirals'' \citep[see Figure 9 in][]{got03b}. A
search (dedicated study) for potential ``passive spirals'' among the
neighbors of CG merits further attention.

Compact groups pose a problem for radio line (and often
Fabry-P\'{e}rot) studies because the velocity range over which gas
might be present can be large. In the case of SQ optical emission
and absorption line velocities span almost 2000 km s$^{-1}$ (1200 km
s$^{-1}$ in detected HI). Optical velocities in SS span a range
close to 700 km s$^{-1}$ with at least two condensations in HI at
4400-4700 km s$^{-1}$ (H79d and tail towards E) and $\sim$4200 km
s$^{-1}$ possibly associated with H79a and not well defined in our
new or in previous observations (\S\S~3.6). Earlier single dish
measures \citep{bie79,gal81,gor81} already raised the possibility of
HI at velocities both higher and lower than H79d. We confirm the
latter, but given the scatter among previous single dish mass
estimates it is difficult to estimate how much HI we may have
missed. We cannot rule out the possibility that significant HI
stripped more than 1.5 Gyr ago remains undetected. This is required
in assuming that most of the galaxies H79abcf entered SS as spirals.
A conservative estimate for the total HI content within SS
(logM$_{HI}$(M$_{\sun}$)$\sim$9.4-9.5) is based on the single dish
HI measures. The same comments apply to previous searches for
molecular gas in SS \citep{leo98,ver98}. At the other extreme only a
small hot X-ray emitting component was detected by ROSAT (\S\S~3.7)
which may be associated with the diffuse stellar halo (N2000a).

Either our assumption that SS was formed from spirals is incorrect
or we must account for considerable (30-70\%) missing gas
(\S\S~3.6). We have no evidence for extreme starbursts in less
evolved compact groups that should be viewed as the precursors of an
SS system. In fact as dynamically hyperactive systems CG are
surprisingly deficient in FIR emission \citep[e.g.][]{ver98}. Thus
gas depletion via star formation seems unlikely to explain such a
large deficit. Both HI/CO mapping of a wider spatial/velocity field
and deeper searches for a warm gas component (OSIRIS-GTC?) might
yield interesting results.

\textbf{3) Diffuse halo formation}: After an initial period of
negative results \citep{ros77,sul83,pib95} luminous optical halos
are now found in many compact groups
\citep[][D2005]{sul87,pib95,sul97,whi03,agu06}. Estimated diffuse
light fractions ranges from a few percent up to 50\% suggesting that
this measure might be a useful diagnostic for the dynamical age of a
compact group. Recent studies of the SQ halo show it to be very
complex with an integrated luminosity of $\sim$L$^{\ast}$
($\sim$average luminosity of one L$^{\ast}$ galaxy) \citep{mol98}
and therefore representing about 20\% of the total group luminosity.
A more sensitive study \citep{gut02} suggests that 20\% might be
underestimated. VR/I measures (N2000a) yielded an estimate of
12-13\% for the halo in SS. More recent wavelet processing of deep
CCD images (D2005) suggest that the halo luminosity fraction is
$\sim$46\% in B band which agrees with our own estimate
(\S\S\S~3.1.3). We reduce the D2005 estimate to 40\% because we
cannot confirm an apparent extension towards the east which we
suggest may be an artifact induced by some relatively bright stars
about one group diameter in that direction (see Figure 6 in D2005).
We conclude that the halo fraction in SS is approximately twice that
in SQ. On this basis alone we conclude that SS is much older group
than SQ.

\citet{som06} studied the diffuse light in ``fossil'' groups (FG)
(e.g., hierarchical groups with a very bright central galaxy) and
non-fossil groups (non-FG)(e.g., groups with galaxies of similar
brightness, i.e. non-hierarchical). Their models predict halo
contributions of $\sim$12-45\% with a higher fraction of diffuse
light in FG. Most CG (up to 90\%) tend to be non-hierarchical which
is a long standing argument against a major merger evolutionary
scenario to account for them (\citealt{sul97}; see also
\citealt{pro04,del07} for a different form of evidence). SQ and SS
are certainly non-hierarchical configurations with halo fractions
that appear to exceed the above model expectations. SS provides the
strongest challenges to formation models with a very massive halo
that is also somewhat bluer than the galaxies embedded within it
(see Table 9). It points towards a slow sequential ``dissolution''
process without significant major merger activity. Additional
support for this conclusion comes from the fact that highly evolved
SS is a very weak X-ray emitter compared to many detected compact
and loose groups \citep{hel00} with estimates for any soft halo
component near L$_X$$\sim$10$^{40}$ ergs s$^{-1}$ (N2000a). Before
dismissing SS as atypical it is worth pointing out that the X-ray
luminosity of SQ is strongly boosted by a transient shock component
\citep{tri05}. The X-ray weak halo component in SQ \citep{hel00}
appears to be about 1dex higher than any equivalent component in SS.

\textbf{4) Long lived system}: We earlier estimated a minimum age
for SQ of 2-3 Gyr based upon encounter events that we could identify
over the past 1-2 Gyr \citep{sul01}. Various studies have identified
a sequence of compact group ages, or evolutionary stages, using HI
morphology/deficiency \citep{ver01}, diffuse light fraction
\citep{agu06} and dynamical inferences \citep{rib98}. The fraction
of early-type galaxies normalized by local environmental density
\citep{got03a} offers an additional measure (\S\S\S~3.1.2). All
measures suggest that SQ is reasonably evolved and that SS is
significantly more evolved than SQ. If we assume that the halo age
is directly proportional to the luminosity fraction then we can
assume a minimum age for SS of 4-6 Gyr. The episodic accretion rate
in SS appears to be slower than for SQ and this will slow down the
evolutionary process. If nothing precludes compact group formation
in richer environments one expect many to form but with a large
fraction that are transient \citep[e.g.][]{dia94,dia95}. No more
than 10\% of HCG fall in this category \citep{sul87,roo89}. This
would not be the slow sequential process that characterizes SQ and
SS.

In the case of SS one could identify H79d as the most recent arrival
while H79c likely arrived more than 1.5 Gyr ago. Both the new
intruder and the accreted object in H79b are relatively low mass
galaxies (\S\S\S~3.3.1). Our optical and H$\alpha$ reconnaissance of
the SS neighborhood reveals no other prospective intruders for the
next $\sim$ Gyr (\S\S\S~3.1.1). Apparently after acquiring the
nearest 4-5 neighbors a ``typical'' CG (in lower density
environments) like SQ or SS creates a local void after which it very
rarely acquires additional members.

A possible new way to estimate dynamical ages in CG is provided by
the HST search for young blue star clusters (BSC) thought to form
via interactions. A large number of candidate BSC have been
identified in SQ \citep{gal01} although their true nature remains
unconfirmed. A contradiction is provided by the almost total absence
of BSC candidates in SS \citep{pal02} which at 2/3 the distance
should have been easier to detect. Numerous red star cluster
candidates were found around SS galaxies. Our own high pass
filtering of archival WFPC images of H79a appears to show a halo of
assumed globular clusters surrounding that galaxy. If we assume that
the condensations in SQ are BSC and that they are a signature of
major accretion/interaction events then SQ has undergone significant
activity in the past Gyr which is well confirmed by numerous
observations \citep[and more recent references cited
earlier]{sul01}. HST imaging analysis shows the candidate star
cluster color distributions for SQ peak at B-V$\sim$0.1 and
V-I$\sim$0.4 \citep{gal01} while corresponding average colors in SS
are B-V$\sim$0.5 and V-I$\sim$1.0 \citep{pal02}. This suggests no SS
star cluster population with an age of less than 1 Gyr roughly
consistent with our suggestion that H79c was the last major intruder
about 2 Gyr ago. We interpret galaxies H79abcf as old arrivals and
this is supported by the degree of tidal stripping evidenced through
their surface photometric properties and the luminosity fraction in
the halo. Any BSC associated with these past intrusions have now
become significantly redder.

Galaxies H79abcf are assumed to be the stripped cores of old spiral
intruders. The somewhat bluer color of the diffuse light
(\S\S\S~3.1.4, D2005) is then consistent with the hypothesis that
much of the halo mass came from stripped spiral disks. This bluer
halo color is also evidence that SS formed long ago from a largely
spiral population that has been strongly secularly evolved. Galaxy
H79f is interpreted as the most extreme example of this process. It
has been extensively studied \citep{nis02} and has been interpreted
both as a tidal feature \citep{nis02} and as the remnant of a
disrupted galaxy \citep{bon99}. The central brightness concentration
and concentric luminosity isophotes about the center of this feature
argue for the second interpretation. We suggest that this is the
most evolved member of the group and offers an empirical insight
into the dynamical evolution of compact group members via
dissolution rather than merging. The more remnant cores we identify
in SS, the lower the average initial luminosity of the intruders. If
lower luminosity spirals have a higher average gas fraction then
inclusion of H79f as an intruder remnant actually exacerbates the HI
deficit.

We used rotation curves (H79bd) and velocity dispersion measures
(H79a) to estimate masses for different components of SS
(\S\S\S~3.3.1). Masses for H79c and f are estimated via an assumed
M/L ratio $\sim$10 \citep{bac85}. Galaxies H79b, c and d are each
estimated to be $\sim$10$^{10}$ M$_{\sun}$ with H79a
$\sim$5$\times$10$^{10}$ M$_{\sun}$ and H79f $\sim$5$\times$10$^9$
M$_{\sun}$. We must add almost 10$^{10}$ M$_{\sun}$ for the HI mass.
Given truncated luminosity estimates for the galaxies and perturbed
or truncated rotation curves these values are likely underestimates.
The total mass of discrete components probably lies in the range of
10$^{11}$ M$_{\sun}$ which is similar to our estimate for the halo
mass. While likely not as massive as SQ, SS was not formed from a
dwarf population although most of the members may have entered as
slightly sub L$^{\ast}$ galaxies.

The lack of major mergers and first-ranked ellipticals in compact
groups \citep{men91,sul94} as well as the rarity of fossil
ellipticals in low density galaxy environments \citep{sul94,zab98}
make clear that such groups evolve very slowly and likely require
massive DM halos \citep{ath97,ace02}. Thus hierarchical collapse
models \citep{bau96,kau96,kau98} are strongly disfavored. SQ and SS
further support these conclusions with no evidence for major
mergers. In one sense there is agreement star formation is rapidly
quenched/truncated in compact groups \citep{del07}. This appears to
contradict predictions of monolithic collapse models \citep{chi02}
where star formation continues for a much longer time. It is
important to point out that none of these models makes predictions
about structure at such high spatial frequencies as typified by SQ
and SS. SS offers one striking example of a minor merger while none
has been identified in SQ.

If the majority of SS members are correctly interpreted as cores of
stripped spiral then it is clear that this group, and perhaps most
groups, dynamically evolve via dissolution rather than major
mergers. Rather than the ``beginning of the end'' \citep{pal02} SS
is engaged in a slow process that has already persisted for a
significant fraction of Hubble time. If: 1) fossil ellipticals are
rare \citep{muz99,sul94}, and, concomitantly, 2) major mergers in CG
are rare \citep{sul97} and 3) groups with more than five members are
rare, then we are forced to conclude --without resorting to magic
dark substances--that most CG like SQ and SS that ever formed are
still in existence.

\acknowledgments

One of us (A. D.) acknowledges support from a Graduate Council
Research/Creative Fellowship offered by the Graduate School of the
University of Alabama for the 2006/2007 academic year. A. D. also
thanks R.J. Buta and W.C. Keel for helpful discussions. M. R.
acknowledges grants 46054-F from CONACYT and IN100606 from
DGAPA-UNAM, Mexico. A. dO. and J. P. are partially supported by
Spanish research projects AYA2006-1325, AYA2006-1213 and Junta de
Andaluc\'{\i}a TIC114. We thank C. Da Rocha for providing the CFHT
images of the diffuse light component in SS. We acknowledge V.
Avila-Reese for helpful comments. We also acknowledge P. Amram and
C. Balkowski for making the run to obtain the CFHT H$_{\alpha}$ F-P
images. The new radio data presented in this paper were obtained
with NRAO instrument--VLA. The National Radio Astronomy Observatory
is a facility of the National Science Foundation operated under
cooperative agreement by Associated Universities, Inc.

\clearpage

\begin{deluxetable}{lccccc}

\tablewidth{0pt}

\tabletypesize{\scriptsize}

\tablecaption{Photometric Observations}

\tablenum{1}

\tablehead{\multicolumn{1}{l}{Observations} & \colhead{Telescope} &
\colhead{Filter} & \colhead{N$_{exp}$} & \colhead{T$_{exp}$(s)} &
\colhead{Seeing}}

\startdata
Broad-Band    &   2.5m NOT &    B &     4 &  1620  &  1.1   \\
Images        &   2.5m NOT &    R &     3 &  1330  &  0.9    \\
\enddata

\end{deluxetable}

\begin{deluxetable}{lcccccc}

\tablewidth{0pt}

\tabletypesize{\scriptsize}

\tablecaption{Summary of Long-Slit Spectroscopic Observations at the
NOT 2.5m Telescope}

\tablenum{2}

\tablehead{\multicolumn{1}{l}{Galaxy} & \colhead{PA} &
\colhead{Grism} & \colhead{N$_{exp}$} & \colhead{T$_{exp}$} &
 \colhead{Disper.} & \multicolumn{1}{l}{Spectral} \\
\colhead{Slit direction} & \colhead{(deg)} & \colhead{} & \colhead{}
& \colhead{(s)} & \colhead{({\AA}/px)} & \colhead{Range({\AA})} }

\startdata
H79b/H79b-c    &   81 &    GR4 &    3 &   1800 &  2.97 & 3027-9075 \\
H79b/H79b-c    &   81 &    GR8 &    3 &   1800 &  1.24 & 5816-8339 \\
\tableline
H79c           &   35 &    GR4 &    3 &   1800 &  2.97 & 3027-9075 \\
H79c           &   38 &    GR4 &    2 &   1200 &  2.97 & 3027-9075 \\
H79c           &   38 &    GR8 &    2 &   1200 &  1.24 & 5816-8339 \\
\tableline
H79d           &  179 &    GR4 &    3 &   1800 &  2.97 & 3027-9075 \\
H79d           &  179 &    GR8 &    4 &   3600 &  1.24 & 5816-8339 \\
\tableline
H79f           &   41 &    GR4 &    3 &   2700 &  2.97 & 3027-9075 \\
\enddata

\end{deluxetable}

\begin{deluxetable}{llll}


%\small
\tablenum{3} \tablecolumns{4} \tablewidth{0pt}
\tabletypesize{\scriptsize} \tablecaption{Journal of Fabry-Perot
Observations \label{tbl-1}}

\tablehead{ \colhead{} & \colhead{} & \colhead{}} \startdata

Observations & Telescope & CFHT 3.6m & OAN 2.1m \\

             & Equipment & MOS/FP @ Cassegrain & PUMA @ Cassegrain\\

             & Date & 1996, August, 25 & 2001, March, 24 \\

             & Seeing & 0.6\arcsec & $\sim$ 1.2\arcsec \\

Calibration & Comparison light & $\lambda$ 6598.95 \AA &
 $\lambda$ 6598.95 \AA \\

Perot--Fabry & Interference Order & 1162 @ 6562.78 \AA & 330 @ 6562.78 \AA \\

         & Free Spectral Range at H$\alpha$ & 265~km~s$^{-1}$ & 934~km~s$^{-1}$ \\

         & Finesse at H$\alpha$ & 12 & 24\\

         & Spectral resolution at H$\alpha$ & 13672 \tablenotemark{a} & 7457 \tablenotemark{a}\\

Sampling & Number of Scanning Steps & 24 & 48 \\

     & Sampling Step & 0.24 \AA\ (11~km~s$^{-1}$) & 0.44 \AA\ (19~km~s$^{-1}$)\\

     & Total Field & 220\arcsec$\times $220\arcsec (256$\times $256 px$^2$) & 604\arcsec$\times $604\arcsec (512$\times $512 px$^2$) \\

         & Pixel Size (binned) & 0.86\arcsec & 1.16\arcsec  \\

Detector & &STIS 2 CCD & Site CCD  \\

Exposure Time & & 180s/ch &  120s/ch  \\

Interference Filter &  Central Wavelength & 6665 \AA\  & 6650 \AA\ \\
                    & FWHM                & 19 \AA\    &  30 \AA\ \\
                    & Transmission        & 0.55 @ 4500 km~s$^{-1}$ & 0.70 @ 4500 km~s$^{-1}$ \\

\enddata

\tablenotetext{a}{For a signal to noise ratio of 5 at the sample
step }

\end{deluxetable}

\begin{deluxetable}{l c c r c rr}

\tablewidth{0pt}

\tabletypesize{\scriptsize}

\tablecaption{Properties of Members of SS and Neighbor Galaxies}

\tablenum{4}

\tablehead{\multicolumn{1}{l}{Galaxy} & \colhead{R.A.(J2000)} & \colhead{Dec.(J2000)} & \colhead{v$_r$} & \colhead{Morphological} & \multicolumn{2}{c}{Distance from SS} \\
\cline{6-7}\\ \colhead{} & \colhead{(hh mm ss.s)} & \colhead{(+dd mm
ss.s)} & \colhead{(km/s)} & \colhead{Type} & \colhead{(\arcmin)} &
\colhead{(kpc)} }

\startdata
HCG79 a        &   15 59 11.1 &    +20 45 16.5 &    4292\tablenotemark{a} &   E3    &    &   \\
HCG79 b        &   15 59 12.5 &    +20 45 48.1 &    4446\tablenotemark{a} &   S0    &    &  \\
HCG79 c        &   15 59 10.8 &    +20 45 43.4 &    4146\tablenotemark{a} &   S0    &    &  \\
HCG79 d        &   15 59 11.8 &    +20 44 48.7 &    4503\tablenotemark{a} &   Sd    &    &  \\
HCG79 e        &   15 59 12.9 &    +20 45 35.4 &   19809\tablenotemark{a} &   Sc   &    &  \\
HCG79 f        &   15 59 14.9 &    +20 45 57.3 &    4095\tablenotemark{b} &   SO?   &    &  \\
\tableline
UGC 10127      &   16 00 24.0 &    +20 50 56.9 &    4823\tablenotemark{b} &   Sb    &   17.7  &  309  \\
CGCG 137-004   &   15 56 33.7 &    +21 17 21.0 &    4367\tablenotemark{c} &   S0    &   48.7  &  850  \\
UGC 10117      &   15 59 23.9 &    +21 36 12.7 &    5375\tablenotemark{d} &   Sab   &   50.8  &  887  \\
CGCG 137-019   &   16 02 30.5 &    +21 07 14.3 &    4555\tablenotemark{d} &   E     &   51.2  &  894 \\
NGC 6032       &   16 03 01.1 &    +20 57 21.5 &    4282\tablenotemark{c} &   SBb   &   54.2  &  958 \\
CGCG 108-053   &   16 01 07.1 &    +19 26 54.6 &    4347\tablenotemark{b} &   S0    &   83.1  &  1450 \\
NGC 6052 NED01 &   16 05 12.9 &    +20 32 32.5 &    4500\tablenotemark{e} &   Sc    &   85.4  &  1490  \\
NGC 6052 NED02 &   16 05 13.2 &    +20 32 32.7 &    4541\tablenotemark{e} &   Sc    &   85.5  &  1492   \\
NGC 6008B      &   15 53 08.3 &    +21 04 28.5 &    5119\tablenotemark{c} &   E/SO  &   87.0  &  1518   \\
NGC 6028       &   16 01 29.0 &    +19 21 35.6 &    4475\tablenotemark{b} &   (R)Sa &   89.9  &  1569 \\
NGC 6008       &   15 52 56.0 &    +21 06 01.8 &    4869\tablenotemark{b} &   SBb   &   90.1  &  1572 \\
UGC 10197      &   16 06 04.4 &    +20 48 05.4 &    4771\tablenotemark{c} &   Sd    &   96.5  &  1684 \\
UGC 10198      &   16 06 05.9 &    +20 47 03.3 &    4624\tablenotemark{f} &   Sdm   &   96.8  &  1690 \\
CGCG 137-037   &   16 05 59.5 &    +21 21 37.2 &    4428\tablenotemark{c} &   Sa?   &   101.7 &  1775 \\
NGC 6020       &   15 57 08.1 &    +22 24 16.3 &    4307\tablenotemark{g} &   E     &   102.9 &  1796 \\
NGC 6060       &   16 05 52.0 &    +21 29 05.9 &    4439\tablenotemark{b} &   SBc   &   103.0 &  1798 \\
CGCG 108-085   &   16 03 26.6 &    +19 09 44.0 &    4684\tablenotemark{h} &   Im    &   112.9 &  1970 \\

\enddata

\tablerefs{(a) \citet{hic92}; (b) RC3; (c) \citet{fal99}; (d)
\citet{bee95}; (e) \citet{tar94}; (f) \citet{fre95}; (g)
\citet{mul99}; (h) \citet{bar98}}

\end{deluxetable}

%% Remove the two lines and the last line if you want
%% want to incorporate this table into another LaTex document.
%\documentclass[preprint2]{aastex}
%\begin{document}

%% The values (usually only l,r and c) in the last part of
%% \begin{deluxetable}{} command tell LaTeX how many columns
%% there are and how to align them.
\begin{deluxetable}{c cccc c c c c}

%% Keep a portrait orientation

%% Over-ride the default font size
%% Use Default (12pt)

%% Use \tablewidth{?pt} to over-ride the default table width.
%% If you are unhappy with the default look at the end of the
%% *.log file to see what the default was set at before adjusting
%% this value.

\tablewidth{0pt}

\tabletypesize{\scriptsize}

\tablecolumns{9}

%% This is the title of the table.
\tablecaption{SS Sizes}

%% This command over-rides LaTeX's natural table count
%% and replaces it with this number.  LaTeX will increment
%% all other tables after this table based on this number
\tablenum{5}

%% The \tablehead gives provides the column headers.  It
%% is currently set up so that the column labels are on the
%% top line and the units surrounded by ()s are in the
%% bottom line.  You may add more header information by writing
%% another line between these lines. For each column that requries
%% extra information be sure to include a \colhead{text} command
%% and remember to end any extra lines with \\ and include the
%% correct number of &s.
\tablehead{\colhead{} & \multicolumn{2}{c}{Our data} &
\multicolumn{2}{c}{Our data} & \colhead{H1989a} & \colhead{B1993} &
\colhead{RC3} & \colhead{N2000a}
\\ \cline{2-9} \\  \colhead{HCG79} & \colhead{D$_g$\tablenotemark{a}} &
\colhead{D$_g$\tablenotemark{a}} & \colhead{D$_r$\tablenotemark{a}}
& \colhead{D$_r$\tablenotemark{a}} &
\colhead{D$_R$\tablenotemark{b}}
& \colhead{D$_R$\tablenotemark{c}} & \colhead{D$_B$\tablenotemark{d}} & \colhead{D$_{VR}$\tablenotemark{e}} \\
\colhead{} & \colhead{(\arcsec)} & \colhead{(kpc)} &
\colhead{(\arcsec)} & \colhead{(kpc)} & \colhead{(kpc)} &
\colhead{(kpc)} & \colhead{(kpc)} & \colhead{(kpc)} }

%% All data must appear between the \startdata and \enddata commands
\startdata
a        &  25 &    7.4 &   28  &    8.1 &    14.7    &   6.7     &  11.8     &  4.6   \\
b        &  22 &    6.5 &   22  &    6.5 &    13.6    &   3.3     &   7.6     &  2.0   \\
c        &  18 &    5.3 &   18  &    5.3 &    10.4    &   6.3     &   6.9     &  1.5   \\
d        &  32 &    9.4 &   26  &    7.6 &     4.7    &   \nodata &  15.2     &  4.7   \\
f        &  17 &    5.1 &   17  &    5.1 &    \nodata &   \nodata &  13.9     &  3.7   \\
\enddata

%% Include any \tablenotetext{key}{text}, \tablerefs{ref list},
%% or \tablecomments{text} between the \enddata and
%% \end{deluxetable} commands

\tablenotetext{a}{$D_g$ and $D_r$ are equal to 2 $a_{max}$, where
$a_{max}$ is the semimajor axis of the last concentric isophote in g
and r band, respectively} \tablenotetext{b}{Diameter of $\mu_R$=24
R-mag arcsec$^{-2}$ isophote, where $D_R=\sqrt{A_R/\pi}$ and A$_R$
is the area of the isophote} \tablenotetext{c}{$D_R=2$ $r_e$, where
$r_e$ is the effective radius in R band} \tablenotetext{d}{Diameter
of $\mu_B$=25 B-mag arcsec$^{-2}$ isophote}
\tablenotetext{e}{$D_{VR}=2$ $r_e$ for galaxies a and c, where $r_e$
is the effective radius in VR band; $D_{VR}=2$ $h$ for galaxies b, d
and f, where $h$ is the disk scale length in VR band}

%% No \tablecomments indicated

%% No \tablerefs indicated

\end{deluxetable}
%\end{document}

%% Remove the two lines and the last line if you want
%% want to incorporate this table into another LaTex document.
%\documentclass[preprint2]{aastex}
%\begin{document}

%% The values (usually only l,r and c) in the last part of
%% \begin{deluxetable}{} command tell LaTeX how many columns
%% there are and how to align them.
\begin{deluxetable}{lccccccccc}

%% Keep a portrait orientation

%% Over-ride the default font size
%% Use Default (12pt)

%% Use \tablewidth{?pt} to over-ride the default table width.
%% If you are unhappy with the default look at the end of the
%% *.log file to see what the default was set at before adjusting
%% this value.

\tablewidth{0pt}

\tabletypesize{\scriptsize}

%% This is the title of the table.
\tablecaption{Neighbors Sizes, Magnitudes and Luminosities}

%% This command over-rides LaTeX's natural table count
%% and replaces it with this number.  LaTeX will increment
%% all other tables after this table based on this number
\tablenum{6}

%% The \tablehead gives provides the column headers.  It
%% is currently set up so that the column labels are on the
%% top line and the units surrounded by ()s are in the
%% bottom line.  You may add more header information by writing
%% another line between these lines. For each column that requries
%% extra information be sure to include a \colhead{text} command
%% and remember to end any extra lines with \\ and include the
%% correct number of &s.
\tablehead{\colhead{Galaxy} &  \colhead{D$_{25g}$} & \colhead{D$_{21.5g}$} & \colhead{D$_{21.8g}$} & \colhead{g$_{TC}$} & \colhead{M$_{g}$} & \colhead{r$_{TC}$} & \colhead{M$_{r}$} & \colhead{L$_{rT}$} & \colhead{L$_{rB}$} \\
\colhead{} & \colhead{(kpc)} & \colhead{(kpc)} & \colhead{(kpc)} &
\colhead{} & \colhead{} & \colhead{} & \colhead{} &
\colhead{($10^{10}L_{\sun}$)} & \colhead{($10^{9}L_{\sun}$)} \\
\colhead{(1)} & \colhead{(2)} & \colhead{(3)} & \colhead{(4)} &
\colhead{(5)} & \colhead{(6)} & \colhead{(7)} & \colhead{(8)} &
\colhead{(9)} & \colhead{(10)}}

%% All data must appear between the \startdata and \enddata commands
\startdata
UGC 10127       &  40.6 &      4.5 &    12.4 & 13.0 &    -21.0  &   12.3 &  -21.7 &  3.9 &    11.3  \\
CGCG 137-004    &  20.7 &      5.1 &     5.9 & 13.7 &    -20.1  &   13.0 &  -20.8 &  1.7 &    11.6  \\
UGC 10117       &  28.0 &      6.5 &     7.7 & 13.5 &    -20.8  &   12.8 &  -21.4 &  3.0 &    11.7  \\
CGCG 137-019    &  22.1 &  \nodata & \nodata & 13.7 &    -20.2  &   13.0 &  -20.9 &  1.9 &    \nodata  \\
NGC 6032        &  27.4 &      2.7 &     4.4 & 12.9 &    -20.8  &   12.3 &  -21.4 &  3.0 &    8.7  \\
CGCG 108-053    &  17.8 &      3.9 &     4.6 & 14.3 &    -19.5  &   13.6 &  -20.2 &  1.0 &    6.8  \\
NGC 6008B       &  14.2 &      4.3 &     5.1 & 14.6 &    -19.6  &   13.7 &  -20.4 &  1.2 &    \nodata  \\
NGC 6008        &  28.7 &      4.4 &     8.9 & 13.4 &    -20.6  &   12.8 &  -21.3 &  2.6 &    7.5  \\
UGC 10197       &  24.4 &      0.9 &     1.5 & 14.4 &    -19.6  &   13.9 &  -20.1 &  0.9 &    0.09  \\
UGC 10198       &  22.9 &      1.5 &     2.3 & 14.2 &    -19.8  &   13.8 &  -20.1 &  0.9 &    0.09  \\
CGCG 137-037    &  10.9 &      2.6 &     2.9 & 14.8 &    -19.0  &   14.2 &  -19.6 &  0.6 &    3.0  \\
NGC 6020        &  29.2 &  \nodata & \nodata & 13.4 &    -20.4  &   12.6 &  -21.2 &  2.5 &    \nodata \\
NGC 6060        &  45.2 &      6.5 &    10.7 & 12.3 &    -21.5  &   11.7 &  -22.2 &  5.9 &    5.3  \\
\enddata

%% Include any \tablenotetext{key}{text}, \tablerefs{ref list},
%% or \tablecomments{text} between the \enddata and
%% \end{deluxetable} commands

\tablecomments{Col.(1): Galaxy identification. Note that galaxies
are ordered by distance away from SS as in Table 3. Col.(2):
Diameter of $\mu_g$=25 mag arcsec$^{-2}$ isophote in g band.
Col.(3): Diameter of $\mu_g$=21.5 mag arcsec$^{-2}$ isophote in g
band. Col.(4): Diameter of $\mu_g$=21.8 mag arcsec$^{-2}$ isophote
in g band. Col.(5): Total magnitude in g band corrected for Galactic
and internal extinction and for redshift. Col.(6): Absolute
magnitude in g band. Col.(7): Total magnitude in r band corrected
for Galactic and internal extinction and for redshift. Col.(8):
Absolute magnitude in r band. Col.(9): Total luminosity in r band.
Col.(10): Bulge luminosity in r band.}

%% No \tablerefs indicated

\end{deluxetable}
%\end{document}

%% Remove the two lines and the last line if you want
%% want to incorporate this table into another LaTex document.
%\documentclass[preprint2]{aastex}
%\begin{document}

%% The values (usually only l,r and c) in the last part of
%% \begin{deluxetable}{} command tell LaTeX how many columns
%% there are and how to align them.
\begin{deluxetable}{c cc cc cc c c c}

%% Keep a portrait orientation

%% Over-ride the default font size
%% Use Default (12pt)

%% Use \tablewidth{?pt} to over-ride the default table width.
%% If you are unhappy with the default look at the end of the
%% *.log file to see what the default was set at before adjusting
%% this value.

\tablewidth{0pt}

\tabletypesize{\scriptsize}

\tablecolumns{9}

%% This is the title of the table.
\tablecaption{SS Apparent Magnitudes}

%% This command over-rides LaTeX's natural table count
%% and replaces it with this number.  LaTeX will increment
%% all other tables after this table based on this number
\tablenum{7}

%% The \tablehead gives provides the column headers.  It
%% is currently set up so that the column labels are on the
%% top line and the units surrounded by ()s are in the
%% bottom line.  You may add more header information by writing
%% another line between these lines. For each column that requries
%% extra information be sure to include a \colhead{text} command
%% and remember to end any extra lines with \\ and include the
%% correct number of &s.
\tablehead{\colhead{} & \multicolumn{2}{c}{Our data} &
\multicolumn{2}{c}{Our data} & \multicolumn{2}{c}{H1989a} &
\colhead{R1991} & \colhead{RC3} & \colhead{N2000a} \\ \cline{2-10} \\
\colhead{HCG79} & \colhead{g$_{TC}$\tablenotemark{a}} &
\colhead{r$_{TC}$\tablenotemark{a}} &
\colhead{B$_{TC}$\tablenotemark{b}} &
\colhead{R$_{TC}$\tablenotemark{b}} &
\colhead{R$_I$\tablenotemark{c}} &
\colhead{B$_{TC}$\tablenotemark{d}} &
\colhead{m$_R$($\mu_R$)\tablenotemark{e}} &
\colhead{B$_{T0}$\tablenotemark{f}} & \colhead{VR}\\
}

%% All data must appear between the \startdata and \enddata commands
\startdata
a        &  14.6  &    13.9 &    15.0 &   13.7 &  13.5    &  14.3    &  13.5(22.0) & 14.5    & 13.1 \\
b        &  14.3  &    13.8 &    14.7 &   13.6 &  13.1    &  13.8    &  13.5(21.3) & 14.2    & 12.9 \\
c        &  15.2  &    14.8 &    15.5 &   14.7 &  14.2    &  14.7    &  14.4(21.9) & \nodata & 13.6 \\
d        &  15.4  &    15.8 &    15.5 &   15.8 &  16.7    &  15.9    &  15.7(22.0) & \nodata & 15.8 \\
f        &  16.6  &    16.1 &    16.8 &   16.5 &  \nodata &  \nodata &  15.4(21.6) & \nodata & 17.2 \\
\enddata

%% Include any \tablenotetext{key}{text}, \tablerefs{ref list},
%% or \tablecomments{text} between the \enddata and
%% \end{deluxetable} commands

\tablenotetext{a}{magnitude enclosed by the last concentric isophote
in g  and r band respectively corrected for Galactic and internal
extinction and for redshift} \tablenotetext{b}{calculated using the
transformation equations given in Lupton (2005)-
http://www.sdss.org/dr4/algorithms/sdssUBVRITransform.html:
$B=g+0.3130 (g-r)+0.2271$ and $R=r-0.1837 (g-r)-0.0971$}
\tablenotetext{c}{R magnitude within $\mu_R$=24 R-mag arcsec$^{-2}$
isophote} \tablenotetext{d}{$B_T$ asymptotic magnitude corrected for
internal and external extinction}\tablenotetext{e}{$m_R$ extending
to R surface brightness $\mu_R$} \tablenotetext{f}{total "face-on" B
magnitude, corrected for Galactic and internal extinction and for
redshift}

%% No \tablecomments indicated

%% No \tablerefs indicated

\end{deluxetable}
%\end{document}

%% Remove the two lines and the last line if you want
%% want to incorporate this table into another LaTex document.
%\documentclass[preprint2]{aastex}
%\begin{document}

%% The values (usually only l,r and c) in the last part of
%% \begin{deluxetable}{} command tell LaTeX how many columns
%% there are and how to align them.
\begin{deluxetable}{ccccc}

%% Keep a portrait orientation

%% Over-ride the default font size
%% Use Default (12pt)

%% Use \tablewidth{?pt} to over-ride the default table width.
%% If you are unhappy with the default look at the end of the
%% *.log file to see what the default was set at before adjusting
%% this value.

\tablewidth{0pt}

\tabletypesize{\scriptsize}

%% This is the title of the table.
\tablecaption{SS Absolute Magnitudes and Luminosities}

%% This command over-rides LaTeX's natural table count
%% and replaces it with this number.  LaTeX will increment
%% all other tables after this table based on this number
\tablenum{8}

%% The \tablehead gives provides the column headers.  It
%% is currently set up so that the column labels are on the
%% top line and the units surrounded by ()s are in the
%% bottom line.  You may add more header information by writing
%% another line between these lines. For each column that requries
%% extra information be sure to include a \colhead{text} command
%% and remember to end any extra lines with \\ and include the
%% correct number of &s.
\tablehead{\colhead{HCG79} & \colhead{M$_{g}$} & \colhead{L$_{gT}$} & \colhead{M$_{r}$} &\colhead{L$_{rT}$} \\
\colhead{} & \colhead{} & \colhead{($10^{9}L_{\sun}$)} & \colhead{}
& \colhead{($10^{9}L_{\sun}$)} }

%% All data must appear between the \startdata and \enddata commands
\startdata
a        &   -19.3 &  4.2 &  -20.0 &    7.9   \\
b        &   -19.6 &  5.5 &  -20.1 &    8.7   \\
c        &   -18.7 &  2.4 &  -19.0 &    3.5   \\
d        &   -18.5 &  2.0 &  -18.1 &    1.4   \\
f        &   -17.3 &  0.7 &  -17.3 &    0.7   \\
\enddata

%% Include any \tablenotetext{key}{text}, \tablerefs{ref list},
%% or \tablecomments{text} between the \enddata and
%% \end{deluxetable} commands

%% No \tablecomments indicated

%% No \tablerefs indicated

\end{deluxetable}
%\end{document}

%% Remove the two lines and the last line if you want
%% want to incorporate this table into another LaTex document.
%\documentclass[preprint2]{aastex}
%\begin{document}

%% The values (usually only l,r and c) in the last part of
%% \begin{deluxetable}{} command tell LaTeX how many columns
%% there are and how to align them.
\begin{deluxetable}{lccc}

%% Keep a portrait orientation

%% Over-ride the default font size
%% Use Default (12pt)

%% Use \tablewidth{?pt} to over-ride the default table width.
%% If you are unhappy with the default look at the end of the
%% *.log file to see what the default was set at before adjusting
%% this value.

\tablewidth{0pt}

\tabletypesize{\scriptsize}

%% This is the title of the table.
\tablecaption{SS Colors}

%% This command over-rides LaTeX's natural table count
%% and replaces it with this number.  LaTeX will increment
%% all other tables after this table based on this number
\tablenum{9}

%% The \tablehead gives provides the column headers.  It
%% is currently set up so that the column labels are on the
%% top line and the units surrounded by ()s are in the
%% bottom line.  You may add more header information by writing
%% another line between these lines. For each column that requries
%% extra information be sure to include a \colhead{text} command
%% and remember to end any extra lines with \\ and include the
%% correct number of &s.
\tablehead{\colhead{} & \multicolumn{2}{c}{Our data} &
\multicolumn{1}{c}{H1989a} \\ \cline{2-3} \cline{4-4 \space} \\
\colhead{HCG79} & \colhead{(g-r)$_{\circ}$} &
\colhead{(B-R)$_{\circ}$} &\colhead{B-R}}

%% All data must appear between the \startdata and \enddata commands
\startdata
 a        &    0.6 &  1.3-1.8 &  1.6      \\
 b        &    0.5 &  1.3-1.5 &  1.4      \\
 c        &    0.4 &  1.2-1.3 &  1.3      \\
 d        &   -0.1 &  0.7-0.9 &  0.8      \\
 f        &    0.4 &  1.3-1.4 &  \nodata  \\
halo           &    0.4 &  0.8-1.3 &  \nodata \\
NW tail        &\nodata &  1.2-1.3 & \nodata \\
\enddata

%% Include any \tablenotetext{key}{text}, \tablerefs{ref list},
%% or \tablecomments{text} between the \enddata and
%% \end{deluxetable} commands

%\tablenotetext{}{}
%% No \tablecomments indicated

\tablecomments{Col.(1): Galaxy identification. Col.(2): g-r color
within the last concentric isophote (for galaxies abcdf) in r-band
(SDSS). The color doesn't change if we use the last concentric
isophote in g-band. Col(3): B-R color (NOT). Col(4): B-R color
within the $\mu_{B}$=24.5 mag arcsec$^{-2}$ isophote \citep{hic89a}.}

%% No \tablerefs indicated

\end{deluxetable}
%\end{document}

%% Remove the two lines and the last line if you want
%% want to incorporate this table into another LaTex document.
%\documentclass[preprint2]{aastex}
%\begin{document}

%% The values (usually only l,r and c) in the last part of
%% \begin{deluxetable}{} command tell LaTeX how many columns
%% there are and how to align them.
\begin{deluxetable}{c rcr cccc}

%% Keep a portrait orientation

%% Over-ride the default font size
%% Use Default (12pt)

%% Use \tablewidth{?pt} to over-ride the default table width.
%% If you are unhappy with the default look at the end of the
%% *.log file to see what the default was set at before adjusting
%% this value.

\tablewidth{0pt}

\tabletypesize{\scriptsize}

\tablecolumns{8}

%% This is the title of the table.
\tablecaption{SS H$\alpha$ Fluxes}

%% This command over-rides LaTeX's natural table count
%% and replaces it with this number.  LaTeX will increment
%% all other tables after this table based on this number
\tablenum{10}

%% The \tablehead gives provides the column headers.  It
%% is currently set up so that the column labels are on the
%% top line and the units surrounded by ()s are in the
%% bottom line.  You may add more header information by writing
%% another line between these lines. For each column that requries
%% extra information be sure to include a \colhead{text} command
%% and remember to end any extra lines with \\ and include the
%% correct number of &s.
\tablehead{\colhead{} & \colhead{Our data\tablenotemark{1}} &
\colhead{Our data\tablenotemark{2}} &
\colhead{I\&V1999\tablenotemark{a,1}} &
\colhead{Our data\tablenotemark{2}} & \colhead{C2004\tablenotemark{b,2}} & \colhead{N2000a\tablenotemark{c,2}} &  \colhead{S2000\tablenotemark{d,3}} \\
\cline{2-4}  \cline{5-8 \space \space}\\
\colhead{HCG79} & \multicolumn{3}{c}{log $f_{H\alpha+[NII]}$} &
\multicolumn{4}{c}{log $f_{H\alpha}$} \\
\colhead{} & \multicolumn{3}{c}{(erg $s^{-1}$ $cm^{-2}$)} &
\multicolumn{4}{c}{(erg $s^{-1}$ $cm^{-2}$)}}

%% All data must appear between the \startdata and \enddata commands
\startdata
 a   &   -13.65 &  \nodata &    -13.58 &  \nodata  &   -13.99  &    -14.60  & (-15.04) \\
 b   &   -13.56 &  -13.54 &    -13.56 &  -13.71  &   -13.62  &    -14.00  & (-14.20) \\
 c   &  $<$ -16.53 &  abs &   $<$ -16.91 &  abs  &   \nodata   &  abs &  abs \\
 d   &   -13.64 & -13.93 &   -14.05 &   -13.90  &   -15.33  &    -14.75  & (-15.25) \\
\enddata

%% Include any \tablenotetext{key}{text}, \tablerefs{ref list},
%% or \tablecomments{text} between the \enddata and
%% \end{deluxetable} commands

\tablenotetext{1}{from reduced H$\alpha$ interference filter image}

\tablenotetext{2}{from long-slit spectra}

\tablenotetext{3}{from nuclear spectra}

%% No \tablecomments indicated

\tablecomments{abs denotes absorption}

%% No \tablerefs indicated
\tablerefs{(a) \citet{igl99}; (b) \citet{coz04}; (c) \citet{nis00};
(d) \citet{shi00}}

\end{deluxetable}
%\end{document}

\begin{deluxetable}{c ccccc}

\tablewidth{0pt}

\tabletypesize{\scriptsize}

\tablecolumns{6}

\tablecaption{SS Spectroscopic Measures}

\tablenum{11}

\tablehead{\colhead{HCG79} &
\colhead{$\frac{[NII]\lambda6584}{H_{\alpha}}$} &
\colhead{$\frac{[SII]\lambda6717,31}{H_{\alpha}}$} &
\colhead{$\frac{[SII]\lambda6717}{[SII]\lambda6731}$} &
\colhead{$EW(H_{\alpha})$} & \colhead{$EW([NII])$} \\
\colhead{} & \multicolumn{5}{c}{}}

\startdata
 a   & \nodata & \nodata & \nodata & \nodata & \nodata \\
 b   & 0.47$\pm$0.01 & 0.33$\pm$0.01 & 1.25$\pm$0.01 & 12.74$\pm$0.64 & 5.80$\pm$0.25 \\
 c   & abs & abs & abs & abs & abs \\
 d   & 0.15$\pm$0.02 & 0.38$\pm$0.02 & 1.37$\pm$0.06 & 48.50$\pm$3.00 & 3.50$\pm$0.30 \\
\enddata

\tablecomments{abs denotes absorption; all the values in the table
are obtained from our long-slit spectra}

\end{deluxetable}

%% Remove the two lines and the last line if you want
%% want to incorporate this table into another LaTex document.
%\documentclass[preprint2]{aastex}
%\begin{document}

%% The values (usually only l,r and c) in the last part of
%% \begin{deluxetable}{} command tell LaTeX how many columns
%% there are and how to align them.
\begin{deluxetable}{cc}

%% Keep a portrait orientation

%% Over-ride the default font size
%% Use Default (12pt)

%% Use \tablewidth{?pt} to over-ride the default table width.
%% If you are unhappy with the default look at the end of the
%% *.log file to see what the default was set at before adjusting
%% this value.

\tablewidth{0pt}

\tabletypesize{\scriptsize}

%% This is the title of the table.
\tablecaption{SS MIR Fluxes}

%% This command over-rides LaTeX's natural table count
%% and replaces it with this number.  LaTeX will increment
%% all other tables after this table based on this number
\tablenum{12}

%% The \tablehead gives provides the column headers.  It
%% is currently set up so that the column labels are on the
%% top line and the units surrounded by ()s are in the
%% bottom line.  You may add more header information by writing
%% another line between these lines. For each column that requries
%% extra information be sure to include a \colhead{text} command
%% and remember to end any extra lines with \\ and include the
%% correct number of &s.
\tablehead{\colhead{HCG79} & \colhead{S$_{11.5\micron}$ (mJy)} }

%% All data must appear between the \startdata and \enddata commands
\startdata
a        &   18.2  \\
b        &   12.7  \\
e        &   12.5  \\
\enddata

%% Include any \tablenotetext{key}{text}, \tablerefs{ref list},
%% or \tablecomments{text} between the \enddata and
%% \end{deluxetable} commands

%% No \tablecomments indicated

%% No \tablerefs indicated

\end{deluxetable}
%\end{document}

%% Remove the two lines and the last line if you want
%% want to incorporate this table into another LaTex document.
%\documentclass[preprint2]{aastex}
%\begin{document}

%% The values (usually only l,r and c) in the last part of
%% \begin{deluxetable}{} command tell LaTeX how many columns
%% there are and how to align them.
\begin{deluxetable}{lccrrr}

%% Keep a portrait orientation

%% Over-ride the default font size
%% Use Default (12pt)

%% Use \tablewidth{?pt} to over-ride the default table width.
%% If you are unhappy with the default look at the end of the
%% *.log file to see what the default was set at before adjusting
%% this value.

\tablewidth{0pt}

\tabletypesize{\scriptsize}

%% This is the title of the table.
\tablecaption{Neighbors Blue and FIR Luminosities}

%% This command over-rides LaTeX's natural table count
%% and replaces it with this number.  LaTeX will increment
%% all other tables after this table based on this number
\tablenum{13}

%% The \tablehead gives provides the column headers.  It
%% is currently set up so that the column labels are on the
%% top line and the units surrounded by ()s are in the
%% bottom line.  You may add more header information by writing
%% another line between these lines. For each column that requries
%% extra information be sure to include a \colhead{text} command
%% and remember to end any extra lines with \\ and include the
%% correct number of &s.
\tablehead{\colhead{Galaxy} &  \colhead{B$_{T0}$} & \colhead{log L$_B$} & \colhead{S$_{60\micron}$} & \colhead{S$_{100\micron}$} & \colhead{log L$_{FIR}$} \\
\colhead{} & \colhead{} & \colhead{($L_{\sun}$)} & \colhead{(Jy)} &
\colhead{(Jy)} & \colhead{($L_{\sun}$)}}

%% All data must appear between the \startdata and \enddata commands
\startdata
UGC 10127       &  13.5\tablenotemark{a} &     10.06 &    1.25 &  3.57 &    9.98  \\
UGC 10117       &  14.5\tablenotemark{a} &      9.66 &    0.29 &  0.91 &    9.37  \\
NGC 6032        &  13.7\tablenotemark{a} &      9.98 &    0.42 &  0.98 &    9.47  \\
NGC 6008        &  13.6\tablenotemark{a} &     10.02 &    0.51 &  2.07 &    9.68  \\
UGC 10197       &  14.8\tablenotemark{b} &      9.54 &    0.24 &  1.09 &    9.38  \\
UGC 10198       &  14.5\tablenotemark{b} &      9.65 &    0.24 &  1.09 &    9.38  \\
NGC 6020        &  13.7\tablenotemark{a} &      9.98 & $<$0.03 & $<$0.19 &  $<$8.59  \\
NGC 6060        &  13.1\tablenotemark{a} &     10.21 &    1.52 &  4.91 &   10.10  \\
CGCG 108-085    &  15.4\tablenotemark{a} &      9.30 &    0.61 &  1.44 &    9.63  \\
\enddata

%% Include any \tablenotetext{key}{text}, \tablerefs{ref list},
%% or \tablecomments{text} between the \enddata and
%% \end{deluxetable} commands

\tablecomments{Col.(1): Galaxy identification. Col.(2): $B_T$
asymptotic magnitude corrected for Galactic and internal extinction
and K-corrected. Col.(3): Logarithm of the blue luminosity. Col.(4):
IRAS flux density at 60 \micron. Col.(5): IRAS flux density at 100
\micron. Col.(6): Logarithm of FIR luminosity, obtained as explained
in \S\S~3.4. Upper limits are indicated with $<$ in front of the
value. }

\tablenotetext{a}{Total blue magnitudes from RC3}
\tablenotetext{b}{calculated from g an r modelMag from SDSS using
the transformation equations given in Lupton (2005)-
http://www.sdss.org/dr4/algorithms/sdssUBVRITransform.html:
$B=g+0.3130 (g-r)+0.2271$.}

%% No \tablerefs indicated

\end{deluxetable}
%\end{document}

\begin{deluxetable}{lccl}
\tablecolumns{4}

\tablewidth{0pt}

\tabletypesize{\scriptsize}

\tablecaption{Summary of 1.4 GHz Radio Continuum Sources in SS}

\tablenum{14}

\tablehead{\colhead{} & \colhead{B-array} & \colhead{C-array} &
\colhead{Notes} }

\startdata
HCG79a    & $4.9\pm0.7$ mJy   & $6.9\pm1.0$ mJy & \\
HCG79b+e  & $1.1\pm0.1$ mJy   & $3.4\pm0.5$ mJy & \\
~~~HCG79b    &~~ $0.41\pm0.07$ mJy &                 & \\
~~~HCG79e    &~~ $0.69\pm0.10$ mJy &                 & \\
HCG79c    & $<0.2$ mJy        & $<0.6$ mJy       & $3\sigma$ limits \\
HCG79d    & $<0.2$ mJy        & $1.4\pm0.2$ mJy &  \\
Source W  & $<0.2$ mJy        & $1.0\pm0.2$ mJy & no optical counterpart \\
Source N  & $5.7\pm0.9$ mJy   & $5.9\pm0.9$ mJy & a compact optical counterpart \\
\enddata

\tablecomments{Uncertainties quoted are either a $1\sigma$ noise
estimate in the continuum map or the 15$\%$ absolute calibration
uncertainty, which ever is larger.}
\end{deluxetable}

% Remove the two lines and the last line if you want
%% want to incorporate this table into another LaTex document.
%\documentclass[preprint2]{aastex}
%\begin{document}

%% The values (usually only l,r and c) in the last part of
%% \begin{deluxetable}{} command tell LaTeX how many columns
%% there are and how to align them.
\begin{deluxetable}{lcccccc}

%% Keep a portrait orientation

%% Over-ride the default font size
%% Use Default (12pt)

%% Use \tablewidth{?pt} to over-ride the default table width.
%% If you are unhappy with the default look at the end of the
%% *.log file to see what the default was set at before adjusting
%% this value.

\tablewidth{0pt}

\tabletypesize{\scriptsize}

%% This is the title of the table.
\tablecaption{Basic Properties for SS, SQ and NGC 7331}

%% This command over-rides LaTeX's natural table count
%% and replaces it with this number.  LaTeX will increment
%% all other tables after this table based on this number
\tablenum{15}

%% The \tablehead gives provides the column headers.  It
%% is currently set up so that the column labels are on the
%% top line and the units surrounded by ()s are in the
%% bottom line.  You may add more header information by writing
%% another line between these lines. For each column that requries
%% extra information be sure to include a \colhead{text} command
%% and remember to end any extra lines with \\ and include the
%% correct number of &s.
\tablehead{\colhead{Name} & \colhead{log M$_{HI}$} & \colhead{log
L$_{H\alpha}$} & \colhead{log L$_{FIR}$} &
\colhead{log L$_B$} & \colhead{M$_B$} & \colhead{log L$_{1.4GHz}$} \\
\colhead{} & \colhead{(M$_{\sun}$)} & \colhead{($L_{\sun}$)} &
\colhead{($L_{\sun}$)} & \colhead{($L_{\sun}$)} & \colhead{} &
\colhead{($L_{\sun}$)} }

%% All data must appear between the \startdata and \enddata commands
\startdata

HCG 79   &   9.3\tablenotemark{a} & 6.6\tablenotemark{c} &  9.9\tablenotemark{e} & 10.3\tablenotemark{g} & -21.0\tablenotemark{g} & 4.3\tablenotemark{i}  \\
HCG 92   &  10.0\tablenotemark{a} & 7.4\tablenotemark{d} & 10.2\tablenotemark{e} & 11.2\tablenotemark{g} & -23.2\tablenotemark{g} & 5.7\tablenotemark{j}  \\
NGC 7331 &  10.0\tablenotemark{b} & 7.8\tablenotemark{b} & 10.3\tablenotemark{f} & 10.5\tablenotemark{h} & -21.5\tablenotemark{h} & 4.5\tablenotemark{j}  \\

\enddata

%% Include any \tablenotetext{key}{text}, \tablerefs{ref list},
%% or \tablecomments{text} between the \enddata and
%% \end{deluxetable} commands

\tablecomments{Col.(1): Galaxy/Group identification. Col.(2):
Logarithm of the HI mass. Col.(3): Logarithm of the H$_{\alpha}$
luminosity. Col.(4): Logarithm of FIR luminosity. Col.(5): Logarithm
of the total blue luminosity. Col.(6): Total blue absolute
magnitude. Col.(7): Logarithm of 1.4 GHz radio continuum luminosity.
}

\tablerefs{(a) \citet{ver01}; (b) \citet{ken94}; (c) \citet{coz04};
(d) \citet{sev99}; (e) \citet{ver98}; (f) \citet{san03}; (g)
\citet{hic93b}; (h) RC3; (i) our VLA data; (j) NVSS. }

%% No \tablerefs indicated

\end{deluxetable}
%\end{document}

\clearpage

\begin{figure}
\figurenum{1} \epsscale{0.9} \plotone {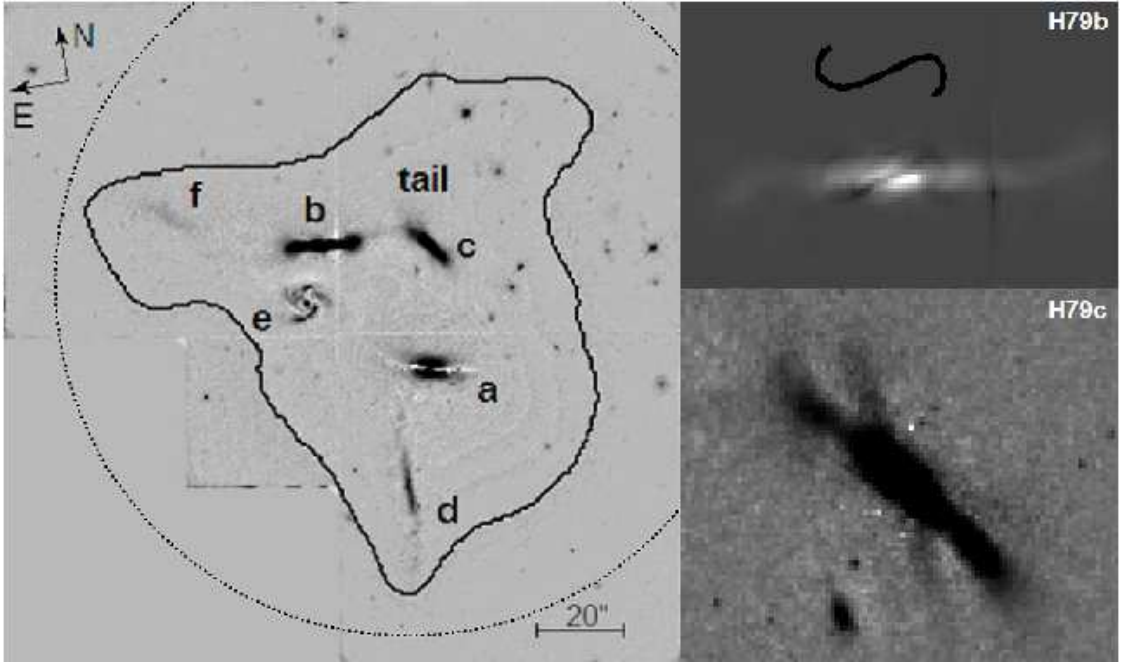} \caption {{HST/WFPC2
images of SS. ($left$) High pass median filter image. ($right$) Low
pass median filter image (5.5X zoom in).}\label{fig1}}
\end{figure}

\begin{figure}
\includegraphics[scale=0.7,angle=-90]{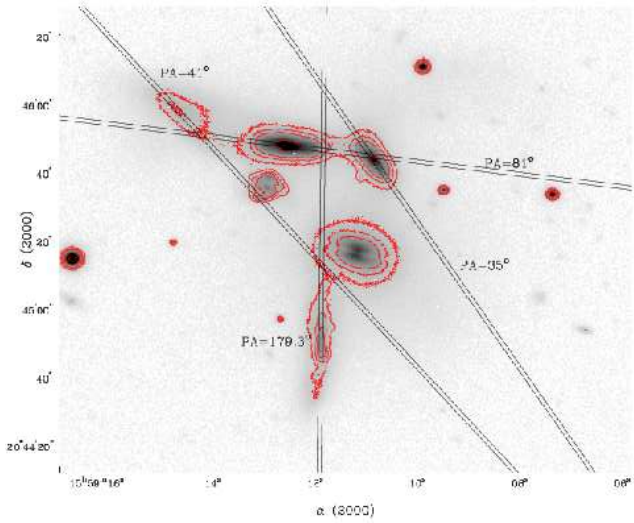}
\figurenum{2} \caption{{SS: The direction of the slits for the
long-slit spectra obtained with ALFOSC at the NOT telescope.}
\label{fig2}}
\end{figure}

\begin{figure}
\figurenum{3} \epsscale{0.9}
\plottwo {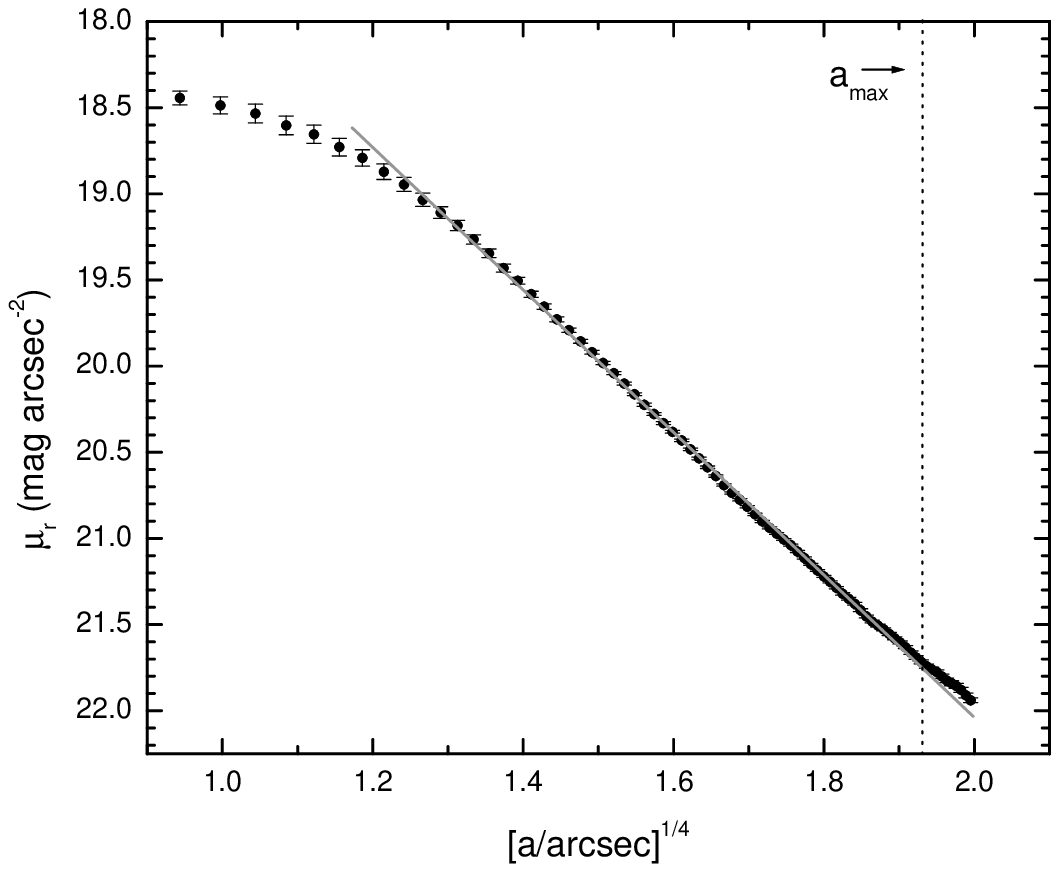}{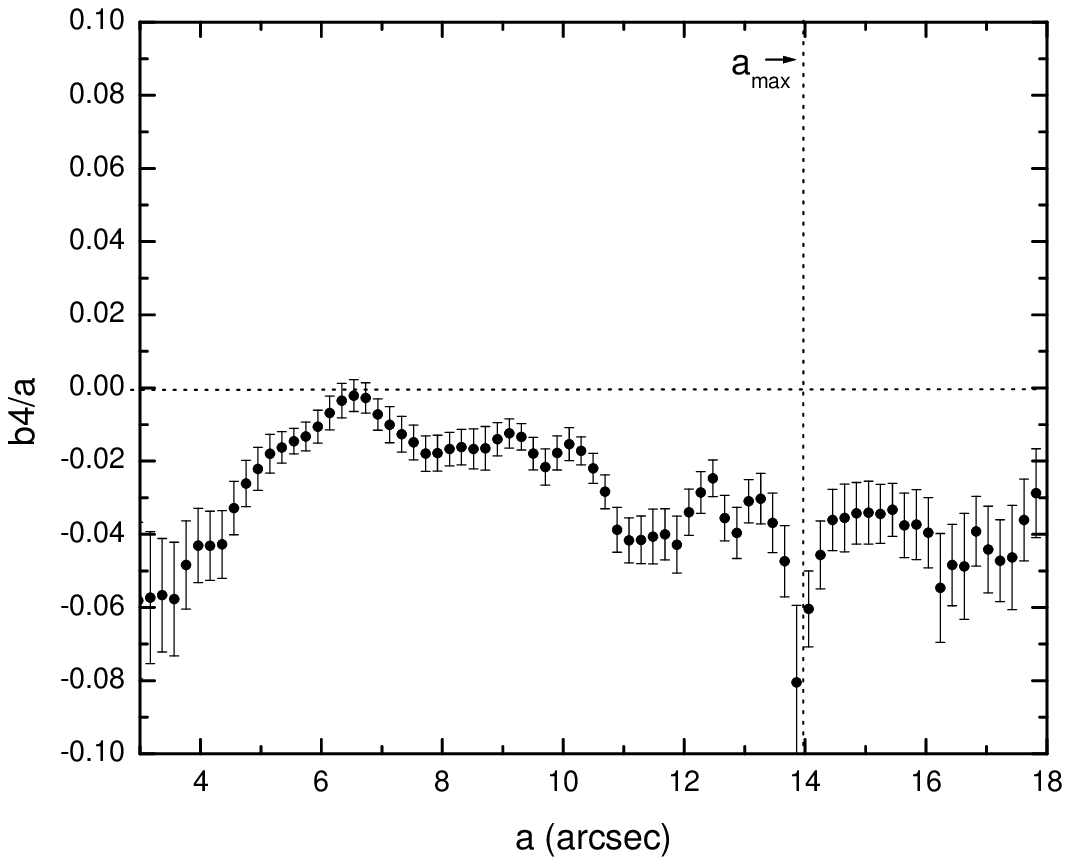} \caption {{H79a: ($left$) Elliptically
averaged surface brightness profile in r-band. ($right$) b4
coefficient profile. The dotted horizontal line corresponds to a
fourth cosine Fourier coefficient of 0 (perfect ellipse). We
truncated the profile from 3 arcsec because of the dust lane.}}
\label{fig3}
\end{figure}

\begin{figure}
\figurenum{4} \epsscale{0.9}
\plottwo {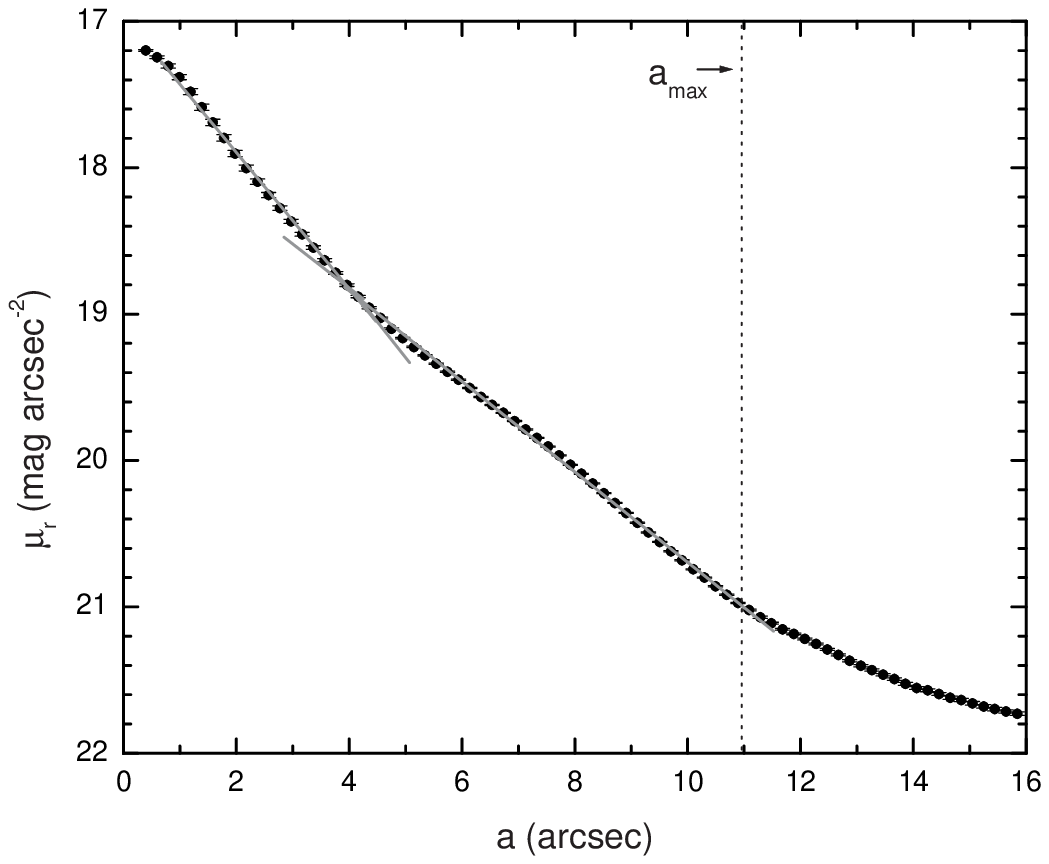}{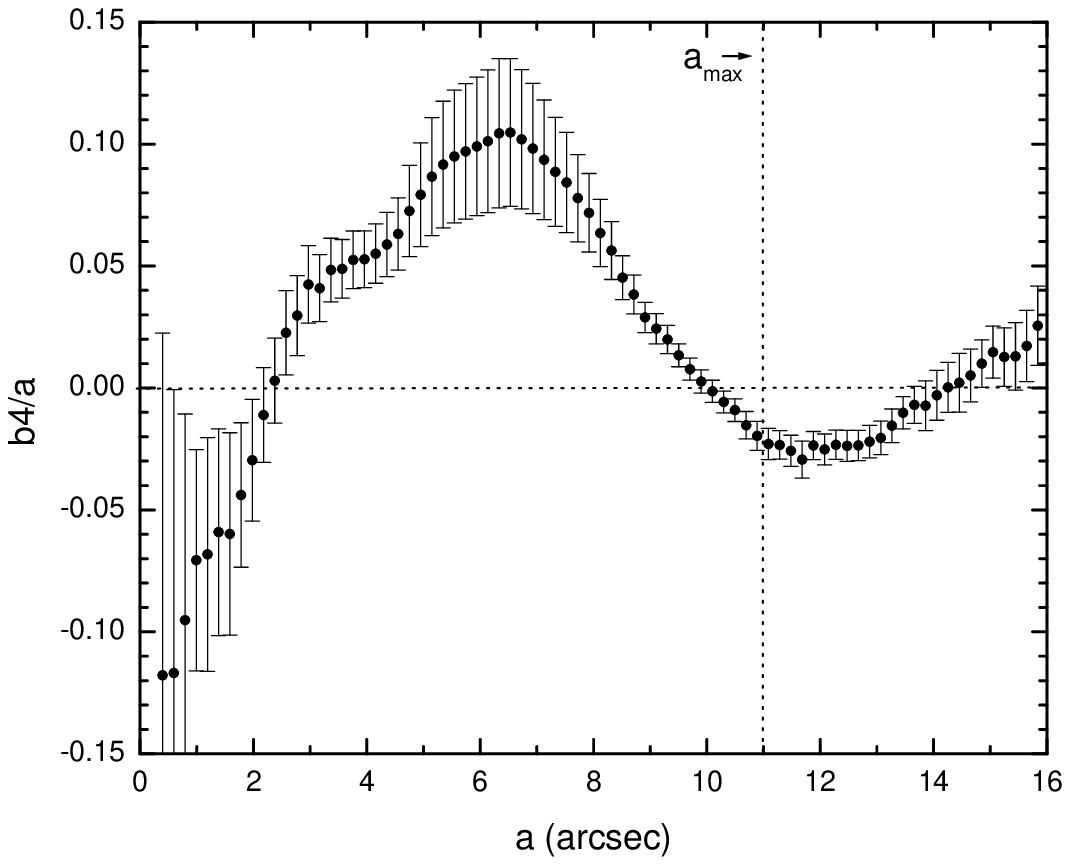} \caption {{H79b: ($left$) Elliptically
averaged surface brightness profile in r-band. ($right$) b4
coefficient profile. The dotted horizontal line corresponds to a
fourth cosine Fourier coefficient of 0 (perfect ellipse).}
\label{fig4}}
\end{figure}

\clearpage

\begin{figure}
\figurenum{5} \epsscale{0.9}
\plotone {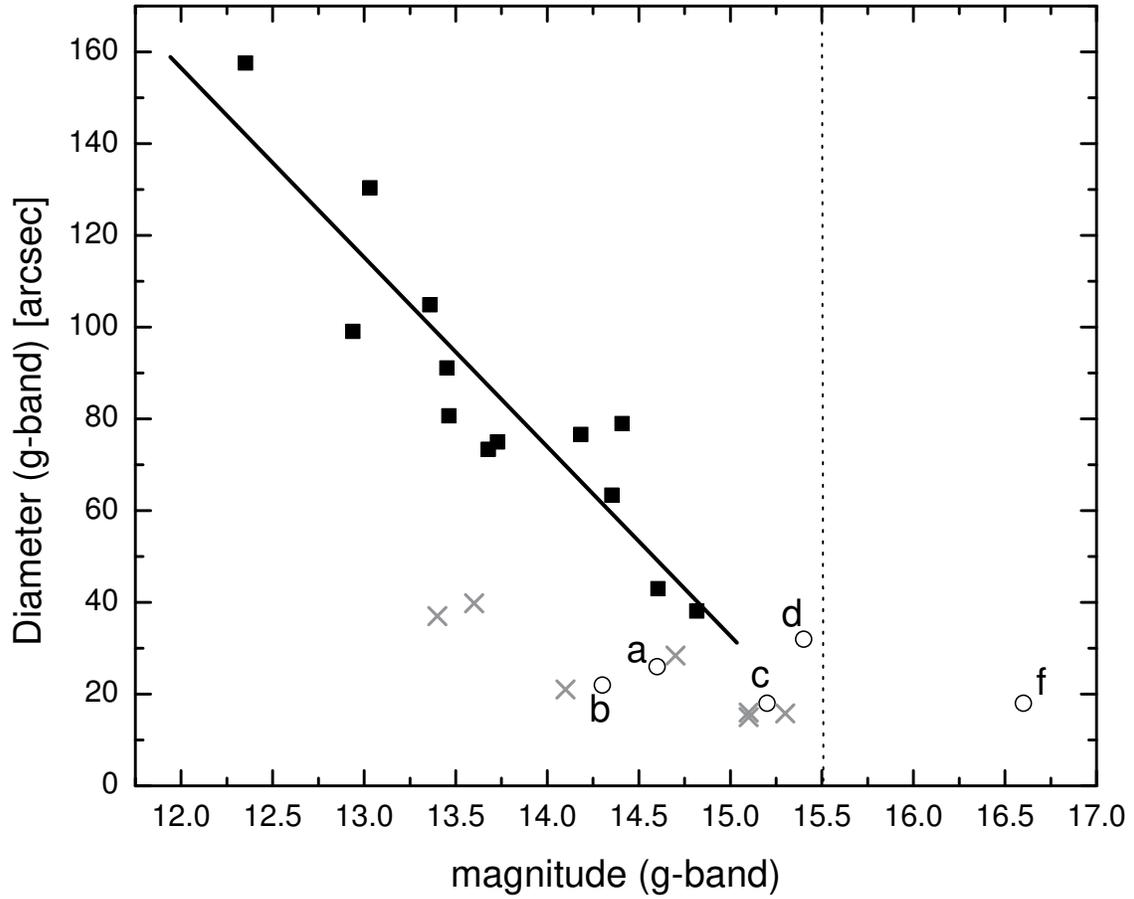}\caption {{Size versus magnitude for neighbors
(filled squares), bulges of SO and Sb-Sc neighbors (crosses) and
galaxies in SS (open circles)} \label{fig5}}
\end{figure}

\begin{figure}
\figurenum{6} \epsscale{1.0}
\plotone{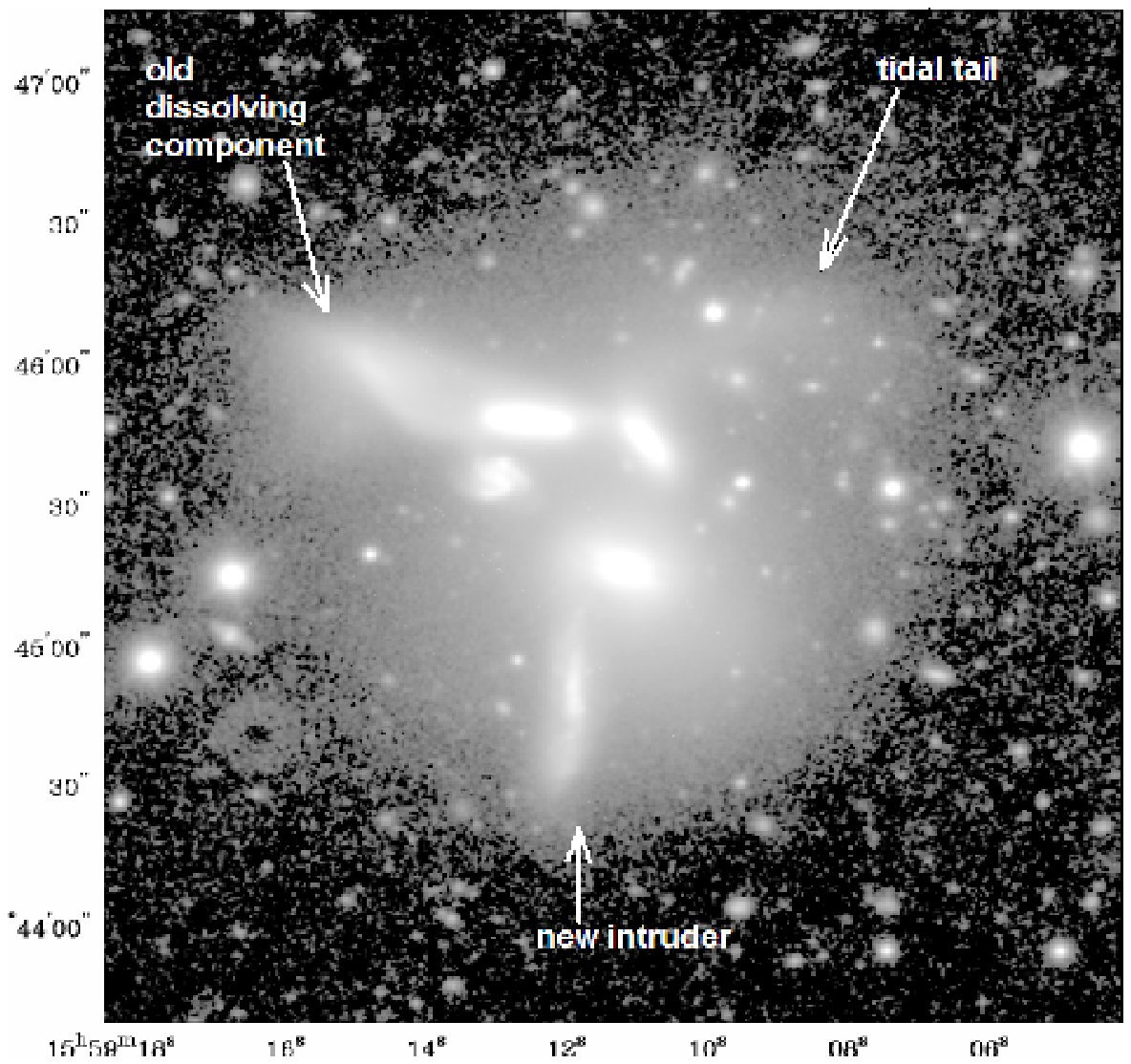}\caption{{B+R color image of SS}\label{fig6}}
\end{figure}

\begin{figure}
\figurenum{7} \epsscale{1.0}
\plotone {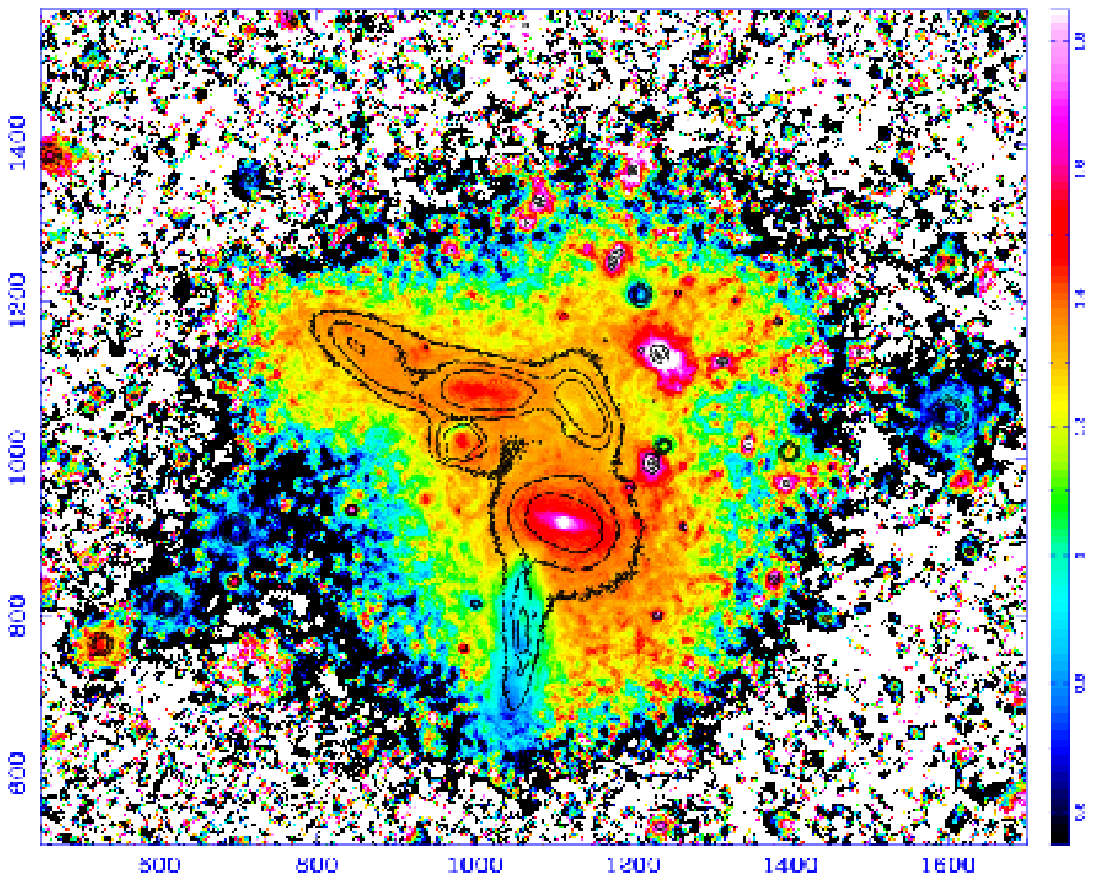}\caption {{B-R color image of SS}\label{fig7}}
\end{figure}

\begin{figure}
\figurenum{8} \epsscale{0.5}
\plotone {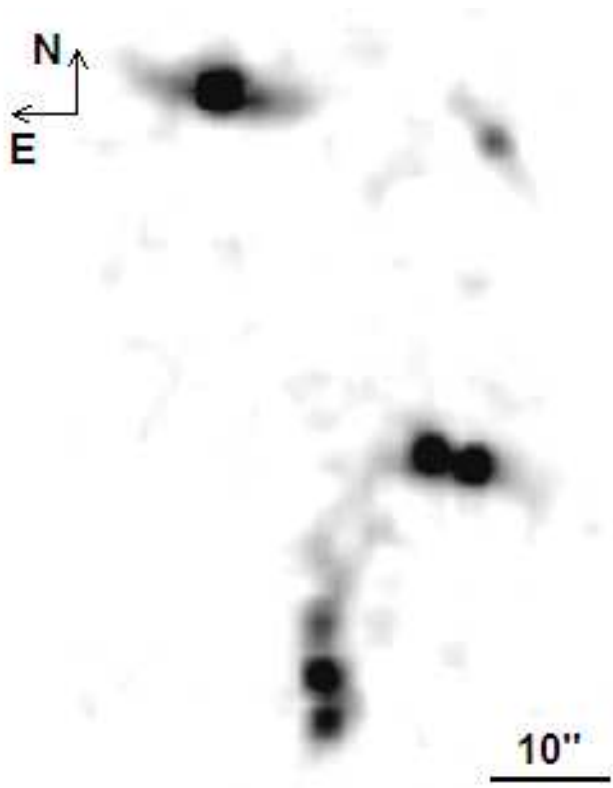}\caption {{MOS/SIS Fabry-P\'{e}rot (CFHT 3.6m)
continuum-subtracted $H_{\alpha}$ image: Sum of 5 velocity channels,
corresponding to velocities between 4350-4395 km/s (galaxy H79a) in
0 interference order and 4615-4660 km/s (galaxy H79d) in +1
interference order. }\label{fig8}}
\end{figure}

\begin{figure}
\figurenum{9} \epsscale{0.7}
\plotone {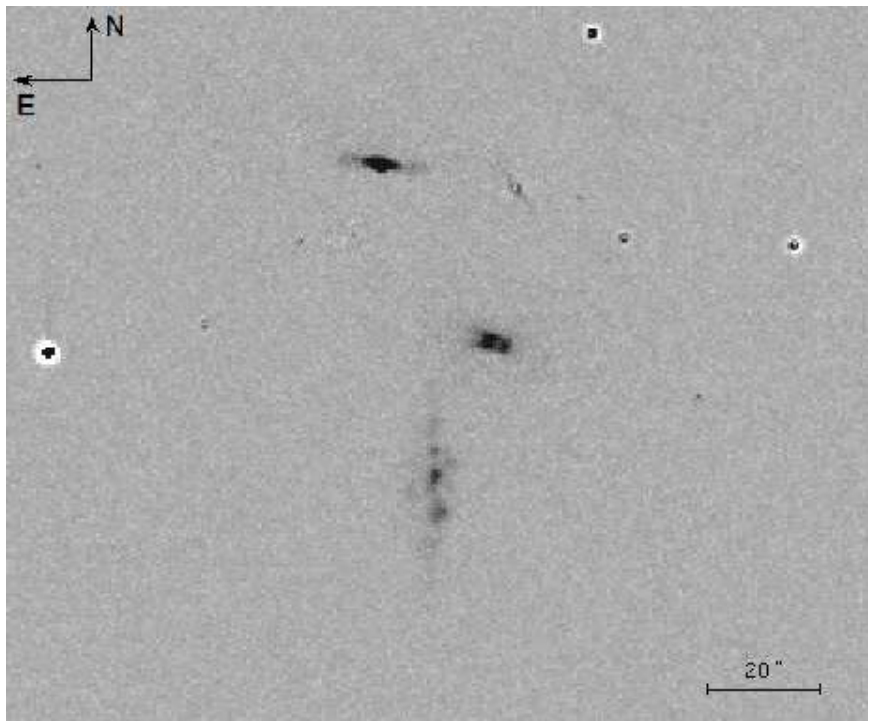}\caption {{Continuum-subtracted $H_{\alpha}$
interference filter image centered at 6667 {\AA} of
SS.}\label{fig9}}
\end{figure}

\begin{figure}
\figurenum{10} \epsscale{0.9} \plotone {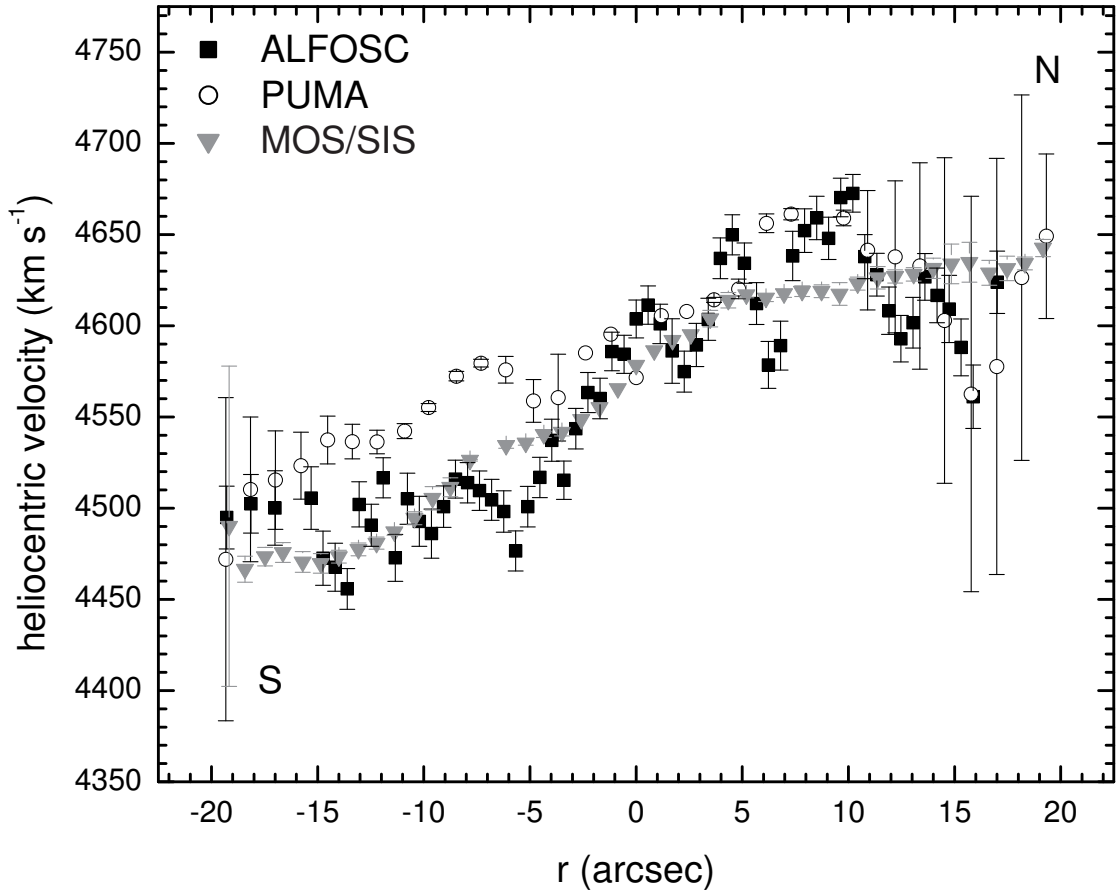}\caption {{H79d line
of sight velocity curve along major axis (PA=180\arcdeg - PUMA \&
MOS/SIS; PA=179.3\arcdeg - ALFOSC).}\label{fig10}}
\end{figure}

\clearpage

\begin{figure}
\figurenum{11} \epsscale{0.9}
\plotone {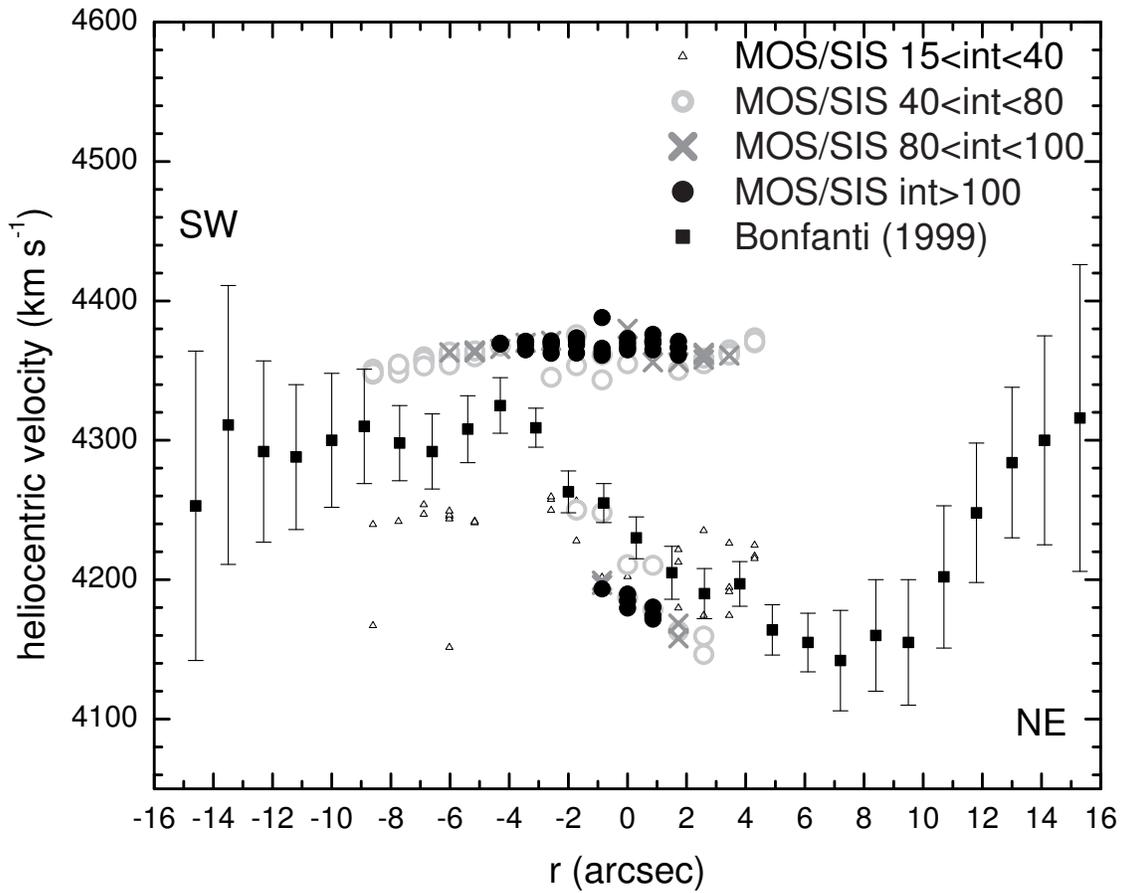}\caption {{H79a line of sight velocity curve along
major axis (PA=65\arcdeg). "int" corresponds to the intensity of the
peak of each component.}\label{fig11}}
\end{figure}

\begin{figure}
\figurenum{12} \epsscale{0.9}
\plotone {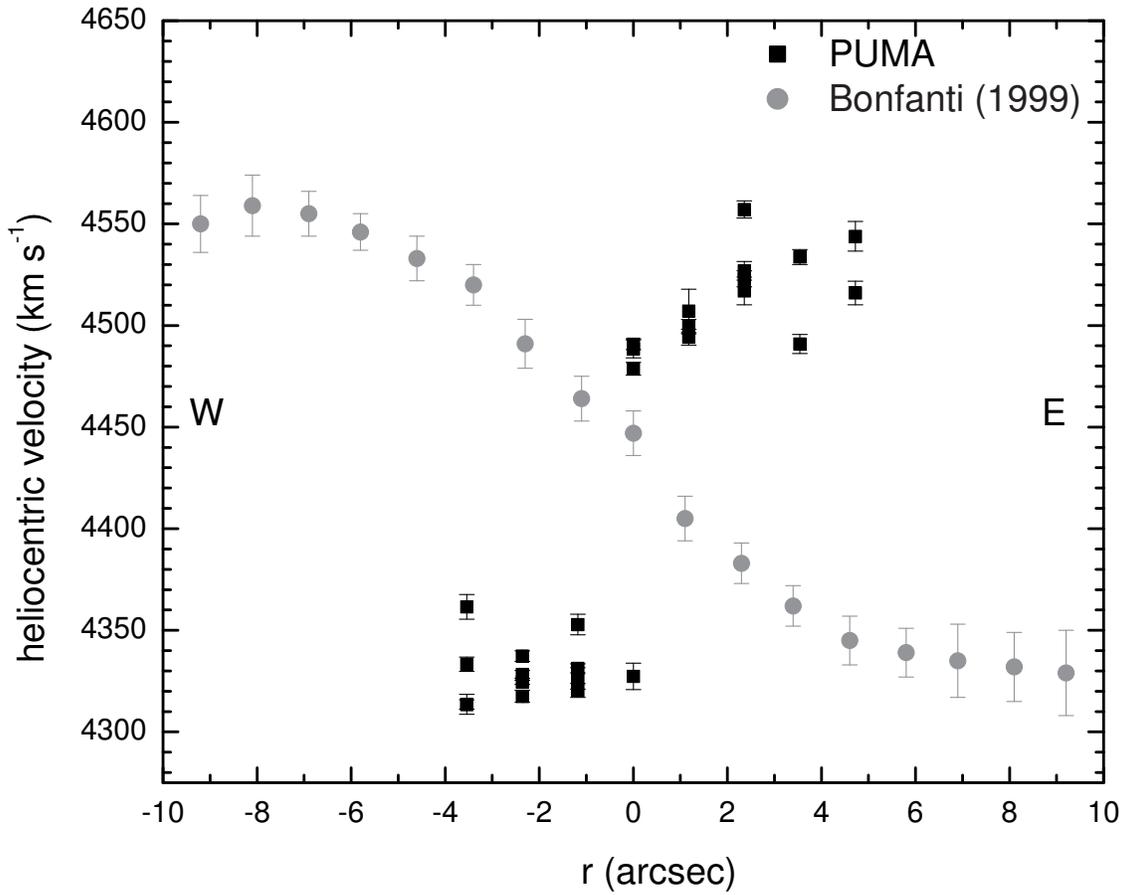}\caption {{H79b line of sight velocity curve along
the direction connecting the nuclei of galaxies H79bc
(PA=79\arcdeg).}\label{fig12}}
\end{figure}

\begin{figure}
\figurenum{13} \epsscale{0.9} \plotone {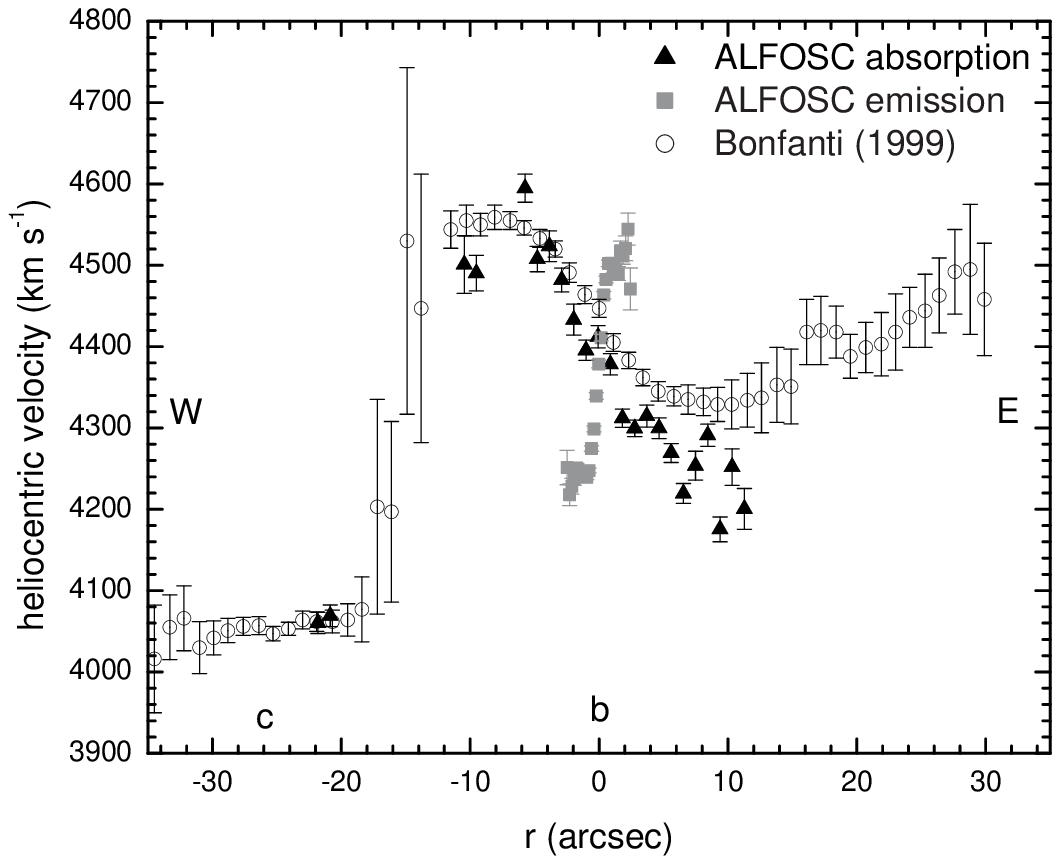}\caption {{Line of
sight velocity curve along the direction connecting the nuclei of
galaxies H79bc (PA=79\arcdeg- \citealt{bon99}; PA=81\arcdeg-
ALFOSC).}\label{fig13}}
\end{figure}

\clearpage

\begin{figure}
\figurenum{14} \epsscale{0.9} \plotone {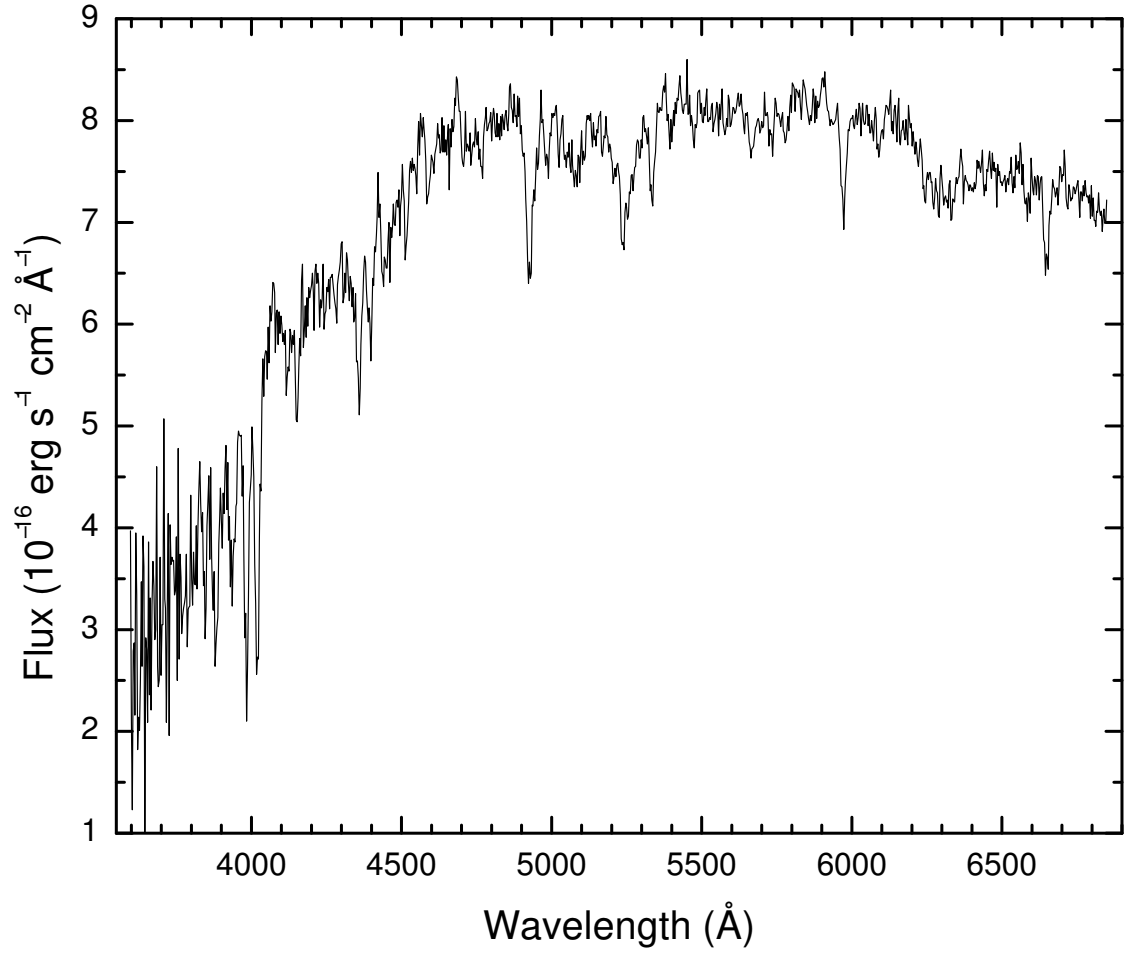}\caption {{Spectrum
along the slit of galaxy H79c corresponding to an aperture of
12\arcsec (PA=35\arcdeg).}\label{fig14}}
\end{figure}

\begin{figure}
\figurenum{15} \epsscale{0.9} \plotone {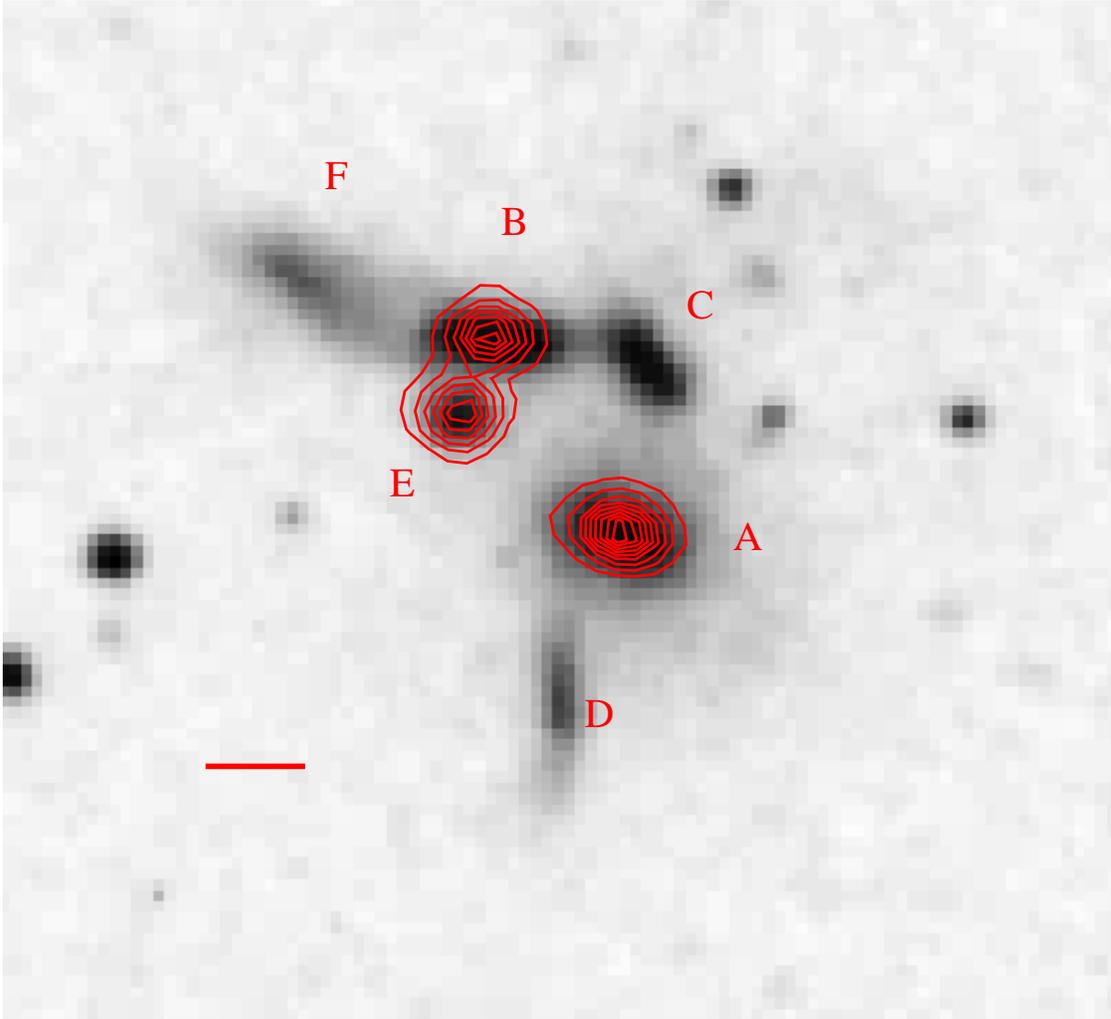}\caption
 {{Isophotal contours of the LW10 ISOCAM image of the
SS overimposed on a DPOSS image of the same group. The contours
drawn go from 4 to 5.5 mJy per pixel, in steps of 0.15 mJy. The
background level is around 3.8 mJy per pixel. The horizontal bar on
the left shows the aperture radius adopted to measure the fluxes
reported in Table 12.}\label{fig15}}
\end{figure}

\begin{figure}
\figurenum{16} \epsscale{1.0} \plotone {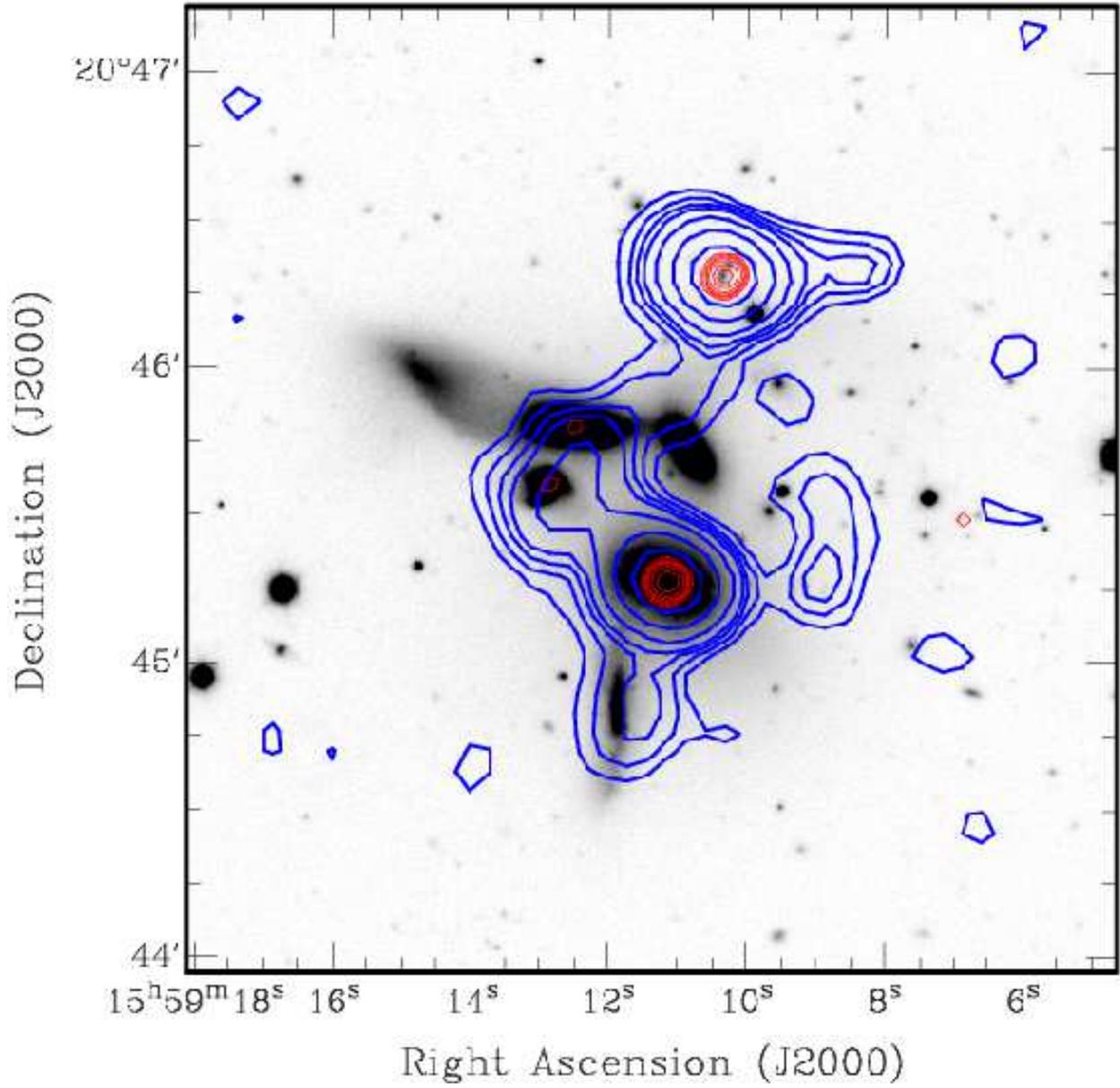} \caption
 {{VLA: (Bold contours) C-Array 1.4 GHz radio continuum. (Thin contours:) B-Array 1.4 GHz radio continuum
 }\label{fig16}}
\end{figure}

\clearpage

\begin{figure}
\figurenum{17} \epsscale{1.0} \plotone {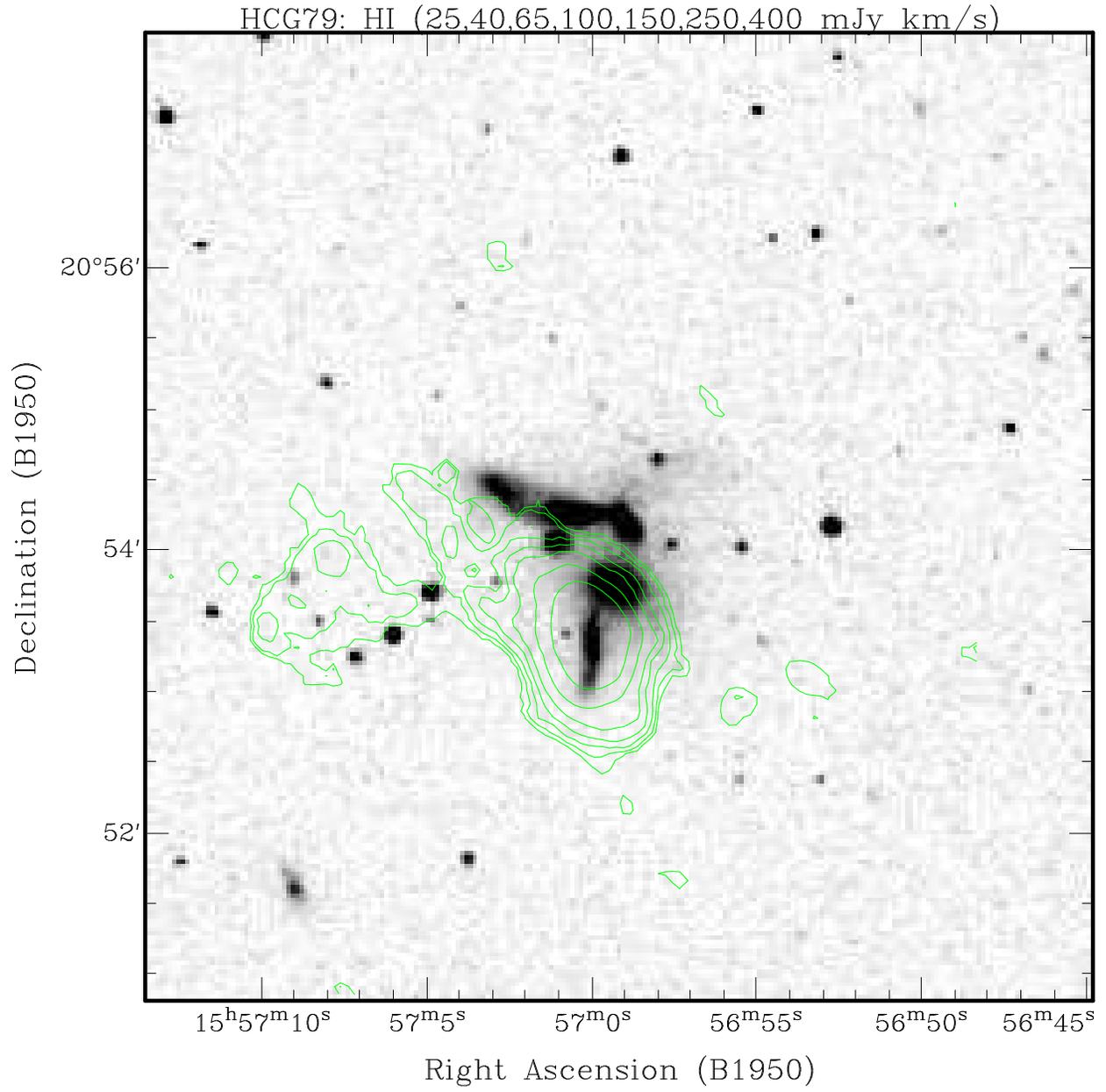} \caption
 {{VLA: HI contours.}\label{fig17}}
\end{figure}

\begin{figure}
\figurenum{18} \epsscale{1.0} \plotone {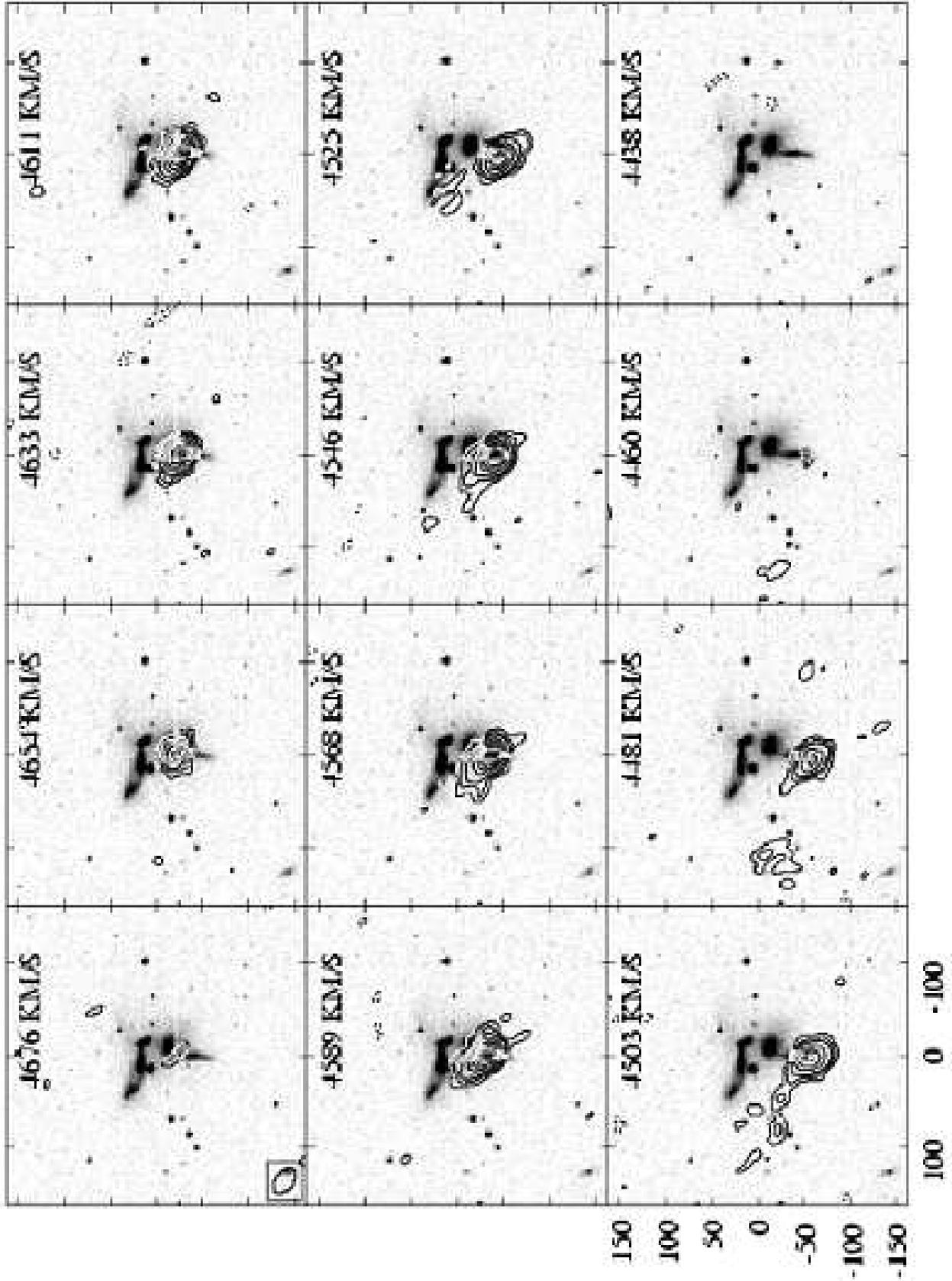} \caption
 {{VLA: HI channel maps.}\label{fig18}}
\end{figure}

\begin{figure}
\figurenum{19} \epsscale{1.0} \plotone {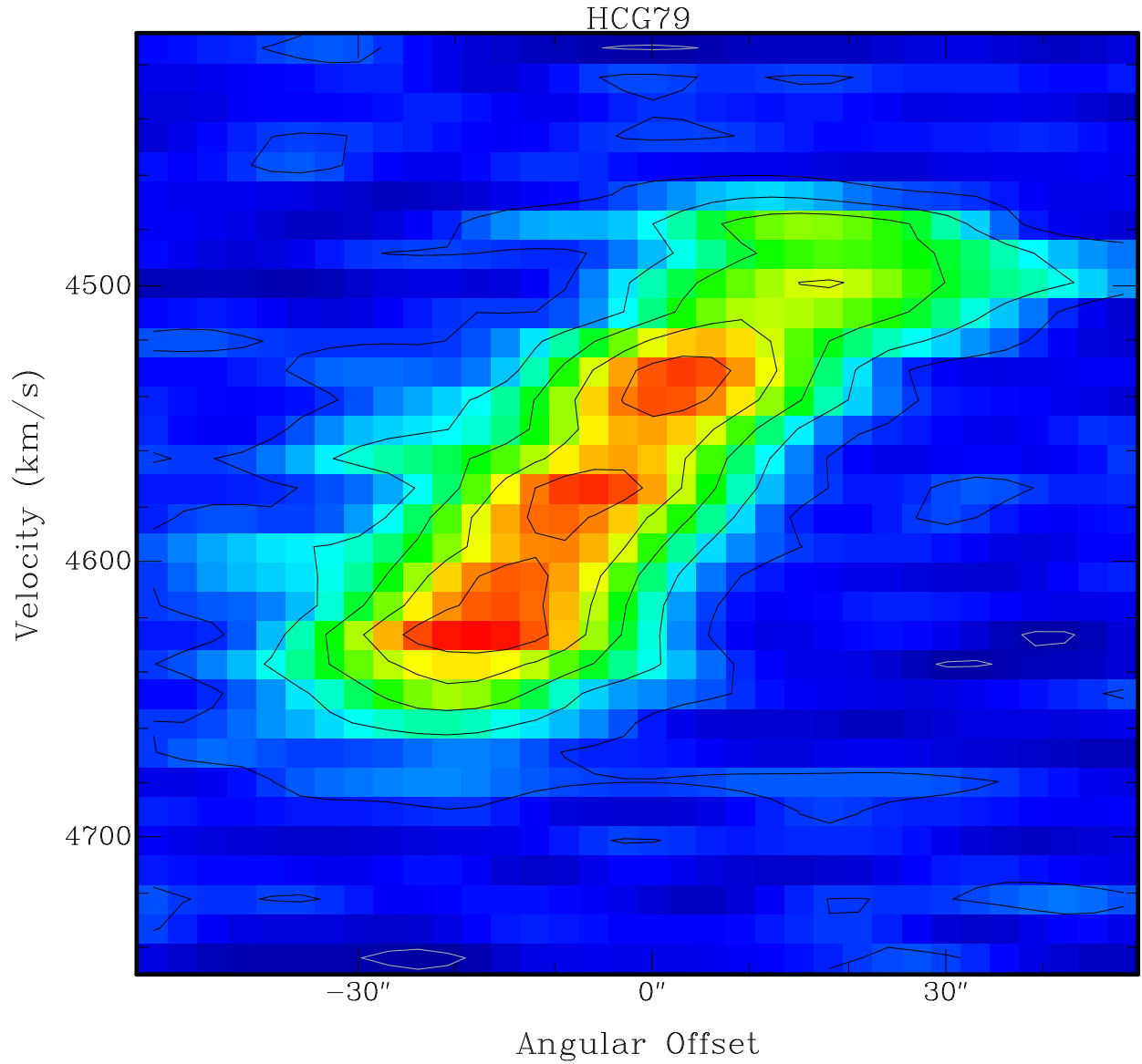} \caption
 {{VLA: HI Position-velocity plot of H79d along major axis
 (PA=180\arcdeg).}\label{fig19}}
\end{figure}

\end{document}